\documentclass[10pt,journal,compsoc]{IEEEtran}
  \pdfoutput=1\relax                   % create PDFs from pdfLaTeX
  \usepackage{graphicx}                % allow us to embed graphics files
  \DeclareGraphicsExtensions{.pdf,.png,.jpg,.jpeg} % for pdflatex we expect .pdf, .png, or .jpg files
  \usepackage{graphicx}                % allow us to embed graphics files
  \DeclareGraphicsExtensions{.eps}     % for pure latex we expect eps files
\graphicspath{{figures/}{pictures/}{images/}{./}} % where to search for the images

\usepackage{microtype}                 % use micro-typography (slightly more compact, better to read)
\PassOptionsToPackage{warn}{textcomp}  % to address font issues with \textrightarrow
\usepackage{textcomp}                  % use better special symbols
\usepackage{mathptmx}                  % use matching math font
\usepackage{times}                     % we use Times as the main font
\usepackage{cite}                      % needed to automatically sort the references
\usepackage{tabu}                      % only used for the table example
\usepackage{booktabs}                  % only used for the table example
\usepackage{color}
\usepackage{xcolor}
\usepackage{multirow}
\usepackage{url}
\usepackage{enumitem}
\usepackage{wrapfig}
\usepackage{caption}
\usepackage{etoolbox}
\usepackage{soul}
\BeforeBeginEnvironment{wrapfigure}{\setlength{\intextsep}{0pt}}
\BeforeBeginEnvironment{wrapfigure}{\setlength{\columnsep}{3pt}}

\newcommand{\rv}[1]{{\color{black}{#1}}}
\newcommand{\rb}[1]{{\color{black}{#1}}}
\begin{document}
%% Paper title.
\title{How Does Automation Shape the Process of Narrative Visualization: A Survey \rv{of} Tools}

%\title{Automatic Visual Narrative \\ Techniques and Tools : A Survey}

%% This is how authors are specified in the journal style

\author{
Qing Chen,
        Shixiong Cao,
        Jiazhe Wang,
        and~Nan Cao% <-this % stops a space
\IEEEcompsocitemizethanks{
\IEEEcompsocthanksitem Qing Chen, Shixiong Cao, and Nan Cao are with Intelligent Big Data Visualization Lab, Tongji University. \\Email: \{qingchen,caoshixiong,nan.cao\}@tongji.edu.cn. \\Nan Cao is the corresponding author. 
\IEEEcompsocthanksitem Jiazhe Wang is with Ant Group. E-mail: jiazhe.wjz@antgroup.com.
}% <-this % stops an unwanted space
}

\IEEEtitleabstractindextext{%
\begin{abstract}
In recent years, narrative visualization has gained much attention. Researchers have proposed different design spaces for various narrative visualization \rv{genres} and scenarios to facilitate the creation process. As users' needs grow and automation technologies advance, increasingly more tools have been designed and developed. \rv{In this study, we summarized six genres of narrative visualization (annotated charts, infographics, timelines \& storylines, data comics, scrollytelling \& slideshow, and data videos) based on previous research and four types of tools (design spaces, authoring tools, ML/AI-supported tools and ML/AI-generator tools) based on the intelligence and automation level of the tools. We surveyed 105 papers and tools to study how automation can progressively engage in visualization design and narrative processes to help users easily create narrative visualizations.} This research aims to provide an overview of current research and development in the automation involvement of narrative visualization tools. We discuss key research problems in each category and suggest new opportunities to encourage further research in the related domain.

% 这里应该写105个paper+工具呢还是91个paper+工具呢？包括最后的conclusion

% In this paper, we surveyed \rv{104} papers and tools to study how automation can progressively engage in the visualization design and narrative process. By investigating the narrative strengths and the drawing efforts of various visualizations, we created a two-dimensional coordinate to map different visualization types. Our resulting taxonomy is organized by the seven types of narrative visualization on the +x-axis of the coordinate and the four automation levels (i.e., design space, authoring tool, ML/AI-supported tool, and ML/AI-generator tool) we identified from the collected work. The taxonomy aims to provide an overview of current research and development in the automation involvement of narrative visualization tools. We discuss key research problems in each category and suggest new opportunities to encourage further research in the related domain.% to encourage targeting the narrative visualization genre that will help researchers enhance their understanding and continue to develop the field.
\end{abstract}
%our goal is to sort out the design spaces and tools for various narrative genres and provide researchers and practitioners with an overview of the development and research for narrative visualization tools.
\begin{IEEEkeywords}
Data Visualization; Automatic Visualization; Narrative Visualization; Design Space; Authoring Tools; Survey
\end{IEEEkeywords}
}

\maketitle
%\firstsection{Introduction} %for journal use above 
\section{Introduction} 

\IEEEPARstart{D}ata visualization has been broadly applied to communicate data and information in an effective and expressive manner. Recently, an emerging trend has been to combine narrative and storytelling with visualization~\cite{kosara2013storytelling}. 
\rv{The norms of communicative and exploratory information visualization are used in narrative visualizations in order to tell the desired story~\cite{hullman2011visualization}.} 
However, creating visualizations with narrative information is a challenging and time-consuming task. Such a creation usually requires data analytic skills and visualization design expertise. Even experts need to spend much time and effort creating an ideal visualization for a specific design scenario.
Therefore, by summarizing the experience in practice, researchers specify various design spaces and visualization scenarios for distinct narrative genres, which are used to guide users to create narrative visualizations. 

With the emergence of new user requirements and the advancement of automation technology, an increasing number of intelligent tools have been created to assist the visual creative process. 
Authoring tools offer rich interactions that allow \rv{users} to adequately control the creation process. However, such tools still require users to decide on each visualization element manually. To further weaken the barriers and reduce the burdens of creation, \rv{researchers have developed} ML/AI-supported tools and ML/AI-generator tools to support a more automatic process. ML/AI-supported tools usually provide recommendations \rb{as} part of the narrative visualization creation process. \rb{Normally}, users need to make their own design choices to achieve the design outcome. \rb{However}, ML/AI-generator tools do not require user expertise in visualization and can generate a complete set of visualization designs without user intervention.

\rv{Over the past few years, related surveys of automated techniques have focused on the automation of traditional statistical charts \cite{wang2021survey, wu2021ai4vis, zhu2020survey}. Automatic tools that support various genres of narrative visualizations have not been sufficiently \rb{investigated}. However, systematic reviews on how (and to what extent) automation shapes visual design and visual narrative processes are generally lacking. The narrative process describes the primary responsibilities and actions of data visualization storytellers and the types of artifacts that come from these activities \cite{lee2015more}.}
%Relevant visualization automation surveys in recent years have primarily focused on one type of automation. Automatic tools which support various genres of narrative visualizations have not been sufficiently touched. There is a lack of a systematic review of how (and to what extent) automation shapes the processes of visual design and visual narrative tools.
In addition, most previous studies aim at the creation process from the visual design level. % Few investigations have focused on the automation tools to support narrative visualization. 
Advances in artificial intelligence and human-computer interaction have brought more opportunities and challenges to this field. Therefore, a state-of-the-art survey is required to provide a better understanding of automation involvement in narrative visualization creation tools.

\rv{To fill this gap, we collected 91 design spaces and tools covering the six genres of narrative visualization and classified them into four automation levels, allowing us to describe how automatic techniques could be progressively used in visualization design and visual narrative, further allowing users to create data visualizations. By analyzing the tools of each narrative visualization genre, we compared the focus of the four levels of tools in each narrative genre in order for users to easily choose the appropriate tool to create according to different scenarios. Furthermore, we identified both mature and less-explored research directions for automated visual narrative tools and presented new research problems and future work to assist researchers in advancing their grasp of the subject matter and pursuing their investigations. In addition to the state-of-art survey, we developed an interactive browser to facilitate the exploration and presentation of the collected design spaces and tools at \textbf{http://autovis.idvxlab.com/.}}
\renewcommand\arraystretch{1.8}
\begin{table*}

\textbf{\caption{The design \rb{spaces} and tools of major narrative visualization \rv{genres}.}}
%描述table的名字
\label{table1}

\begin{tabular}{|m{2.8cm}<{\centering}|m{2.8cm}<{\centering}|m{2.8cm}<{\centering}|m{2.8cm}<{\centering}|m{2.8cm}<{\centering}|m{1.3cm}<{\centering}|}
\hline
\multicolumn{1}{|l|}{} & \textbf{Design Space} & \textbf{Authoring tool} & \textbf{ML/AI-supported tool} & \textbf{ML/AI-generator tool} & \textbf{SUM} \\ \hline

% \textbf{Dashboard}
% &\cite{sarikaya2018we} \cite{vazquez2019extending} \cite{froese2016lessons} \cite{javed2012exploring} \cite{chen2020composition} \cite{qu2017keeping} \cite{gramazio2014relation} \cite{suprata2019data} \cite{fernandez2022beyond}
% & \cite{tableau2006} \cite{ahlberg1996spotfire}  \cite{elias2011exploration} \cite{elias2012annotating}
% & \cite{belo2014restructuring} \cite{elshehaly2020qualdash} \cite{dabbebi2017towards}  \cite{ma2020ladv}
% & \cite{datapine2012} \cite{SisenseFusionAnalytics2019} \cite{thoughtspot2012} \cite{QlikSense2010}
% & 21  \\ \hline

\textbf{Annotated Chart}
&\cite{borkin2013makes} \cite{borkin2015beyond} \cite{kong2017internal} 
&\cite{ren2017chartaccent} %后面跟着这3个全是编程工具 \cite{gomez2017ggplot2} \cite{swoopyDragjs2016}\cite{wongsuphasawatlabella2015} 
&\cite{chen2010touch2annotate} \cite{chen2010click2annotate}  \cite{kandogan2012just} \cite{bryan2016temporal} \cite{latif2021kori} \cite{fan2022annotating} \cite{kong2012graphical} \cite{srinivasan2018augmenting} \cite{subramonyam2018smartcues}
&\cite{hullman2013contextifier} \cite{liu2020autocaption}  
& 15 \\ \hline

\textbf{Infographic} 
& %\cite{artacho2008influence} \cite{santos2018influence}  \cite{albers2014infographics} 
\cite{cmeciu2016beyond} \cite{harrison2015infographic} \cite{lyra2016infographics} \cite{lan2021smile} \cite{diakopoulos2011playable} \cite{dunlap2016getting}  %\cite{naparin2017infographics}
&\cite{kim2016data} \cite{cui2021mixed}  \cite{wang2018infonice} \cite{zhang2020dataquilt} \cite{coelho2020infomages} \cite{kim2019dataselfie} \cite{xia2018dataink} \cite{lee2013sketchstory}
&\cite{lu2020exploring} \cite{tyagi2021user} \cite{yuan2021infocolorizer} % \cite{visme2013} \cite{infogram2012}   \cite{CANVA2018} 
&\cite{cui2019text} \cite{qian2020retrieve} \cite{chen2019towards} 
& 20 \\ \hline

\textbf{Timeline \& Storyline}
&\cite{brehmer2016timelines} \cite{lan2021understanding} \cite{bach2015time} 
\cite{di2020storyline} \cite{tanahashi2012design} \cite{gronemann2016crossing}
&\cite{kim2017visualizing} \cite{brehmer2019timeline}  
%\cite{webalon2011tiki} \cite{dukes2010dipity} \cite{TimelineJS2013} \cite{TimelineSetter2011} 
&\cite{nguyen2016timesets} \cite{liu2013storyflow} \cite{tang2018istoryline} 
\cite{tang2020plotthread}  \cite{satyanarayan2014authoring} \cite{fulda2015timelinecurator}
&
& 14 \\ \hline

\textbf{Data Comics} 
&\cite{zhao2015data}    \cite{wang2019teaching} \cite{bach2017emerging} \cite{wang2019comparing} \cite{bach2018design} \cite{wang2020data} \cite{zhao2019understanding} \cite{hasan2022playing}
& \cite{kim2019datatoon} \cite{kang2021toonnote} \cite{wang2021interactive} \cite{suh2022codetoon}
&\cite{zhao2021chartstory}
&\cite{wang2019datashot} \cite{shi2020calliope}
& 15 \\ \hline

\textbf{Scrollytelling \& Slideshow} 
& \cite{seyser2018scrollytelling} \cite{godulla2017digitale} \cite{elias2018towards} \cite{hullman2013deeper} \cite{roth2021cartographic}
&\cite{conlen2018idyll} \cite{sultanum2021leveraging}
%\cite{powerpoint2016} \cite{keynote2003} %\cite{bocklandt2021sandslide} 
&\cite{winters2019automatically}
&\cite{lu2021automatic}
& 9 \\ \hline

\textbf{Data Video}
&\cite{amini2015understanding} \cite{cao2020examining} \cite{xu2022wow} \cite{thompson2020understanding} \cite{sallam2022towards} \cite{wang2016animated} \cite{li2020improving} \cite{shu2020makes} \cite{shi2021communicating} \cite{tang2020design}     
&\cite{amini2016authoring} \cite{lan2021kineticharts} \cite{chen2021augmenting}
&\cite{thompson2021data} \cite{wang2021animated} \cite{kim2021gemini} \cite{ge2021cast}
&\cite{shi2021autoclips}
& 18  \\ \hline

\textbf{SUM} & 38 & 20 &24  &9  & 91 \\ \hline

\end{tabular}
\vspace{-2mm}
\end{table*}

% 20230301: 先把文章中删掉的句子的cite去掉了，还需要把商业类工具的cite去掉 
%\input{Table/table2-0413}
%\input{Table/table2-0329}
%\input{Table/table3-0413}

%\input{Table/table2-0329-edit}
%\input{Table/table2-0228-edit}
%\input{Table/table2-0228}
%\input{Table/table3-0228-edit}
%\input{Table/table3-0228}

% \begin{figure*}
%     \centering
%     \includegraphics[width=18.3cm]{Figure/20220228.png}
%      %\textbf{\caption{}{fig2. Listed by year of publication of narrative visualization and tool type}}
% \end{figure*}

 %\begin{figure}
     %\centering
     %\includegraphics[height=15cm]{Figure/02.png}
     %\caption{Narrative Visualization Genre Development Timeline}
 %\end{figure}

\section{Related Survey and Taxonomy}
%briefly describe why we need this survey - 1.no previous survey focus on the the automation of AI involvement in the scope of narrative visualization. 2.recent advances in AI and HCI brings more opportunities and challenges in this domain. 3.we need a state-of-the-art survey on xx could facilitate xx (benefits of this survey can bring)

\begin{figure}[htp]%[hbt!]
\setlength{\abovecaptionskip}{0.2cm}
\setlength{\belowcaptionskip}{-0.1cm}
    \centering
    \includegraphics[width=9cm]{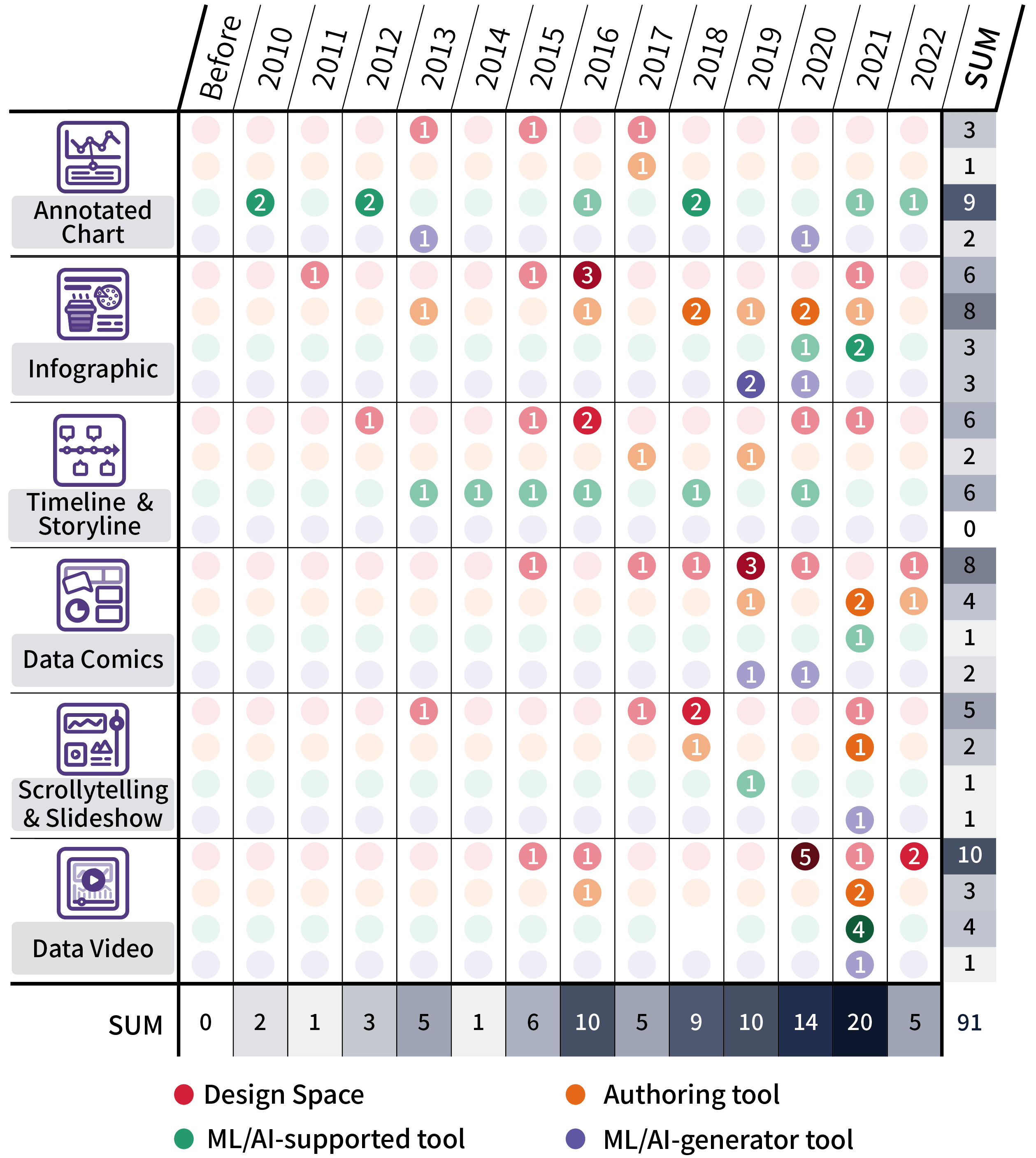}
    \caption{Number of relevant research publications or tools in different genres for narrative visualization in chronological order.}
   \label{fig:figures1} %交叉引用的标签
\end{figure}

% \textcolor{blue}{1.remove dashboard (put into discussion); 2.AI-supported -> ML/AI-supported, AI-generator -> ML/AI-generator (TBD); 3.remove coordinate \& rewrite 2.4.2}

In this section, we first perform a literature review on narrative visualization. Then, we introduce papers that are most relevant to our work. Then, we present our survey scope and methodology. Finally, we describe the taxonomy of this survey.

%In this section, we first review the development of narrative visualization from previous studies. Then, we introduce several survey papers that are most relevant to our work. Next, we present our survey scope and methodology. Finally, we describe the taxonomy of our survey based on the classification of automation levels and narrative genres. 

% \begin{figure*}[htp]%[hbt!]
% \setlength{\abovecaptionskip}{-0.1cm}
% \setlength{\belowcaptionskip}{-0.1cm}
%     \centering
%     \includegraphics[width=18cm]{Figure/20220418-table2.png}
%     \caption{Number of relevant research publications on different types of tools for narrative visualization in different years}
%   % \label{fig:figures2.11} %交叉引用的标签
% \end{figure*}

 %The theories around data storytelling and visual narratives have been extensively studied in the visualization domain.
%Lee et al. \cite{lee2015more} argued that visual data stories need to contain a set of story segments supported by data that should be presented in a meaningful order. 
%Kosara \& Mackinlay~ \cite{kosara2013storytelling} suggested that ``narrative'' promises to open up a whole new path for visualization research. 

\subsection{Narrative Visualization and Storytelling Process}

Our research is influenced by the emergence of narrative visualization theories and visual storytelling technologies.
According to \rv{Segel et al.}~\cite{segel2010narrative}, narrative visualization comprises three essential components: narrative genres, narrative structure (methods for organizing plot or information), and visual narrative (visual methods for generating story experiences and transmitting messages). 

Hullman et al.~\cite{hullman2013deeper} provided a summary of how automated sequencing might be included in design systems to assist users in making organized choices when developing narrative visuals. According to Lee et al.~\cite{lee2015more}, aiming to achieve the goal of message delivery, visual data stories must have a collection of narrative segments backed by data and presented in a coherent order. \rv{In addition, the process of creating narrative visualizations is not always 
linear. This approach can be roughly categorized into three steps: investigating the data, making a story, and telling the story.}
%When visualization is used for decision-making, the narrative structure is even more critical because it allows for the concise and logical representation of essential facts. %A large number of researchers have studied how to create appropriate narrative structure. %Lee et al.\cite{lee2015more} state that creating narrative visualizations is not necessarily linear but can be roughly divided into three steps: exploring data, making a story, and telling a story. 
%Yang et al. \cite{yang2021design} proposed a design space for narrative patterns, data flow, and visual communication by extracting stories from 103 data videos. %The design space provides a practical framework for participants to select, organize, and populate data facts into the different stages of the ``Freytag Pyramid'' while supporting visual content design throughout the structure. %Lan et al. \cite{lan2021understanding} proposed six narrative sequencing models (chronology, trace-back, trailer, recurrence, halfway-back, anchor). The findings show that nonlinear narratives are more likely to increase user engagement and make stories more expressive without hindering comprehension. 
Tong et al. \cite{tong2018storytelling} surveyed the literature on storytelling in visualization, covering the logical concepts of who is the subject of the narrative visual (creation tool and audience), how the story is told (narration and transition), and why we can use visual narratives (memory and interpretation). %They also classified storytelling into four types of narrative orders: linear, user-directed path, parallel, random access or other. %In this way, a matrix was constructed to study the literature.
%In this paper, our goal is to sort out the design spaces and tools for various narrative genres and provide researchers and practitioners with an overview of the development and research for narrative visualization tools.
The goal for this research is to provide researchers and practitioners with an overview of the development and research for various narrative visualization tools.

\subsection{Related Survey}
This section outlines the surveys related to automated visualization techniques and tools.
%Khan\cite{khan2011data} discusses the definition of visualization, the different steps of the visualization process, the problems faced, the classification of visualization techniques based on different perspectives, standard interaction methods, and interaction processes. Mei et al.\cite{mei2018design} reviews declarative specifications and user interfaces used for visualization design and proposes a design space to describe different aspects of information visualization construction efforts. 
Wang et al. \cite{wang2021survey} surveyed 88 papers on ML4VIS and explained seven main processes of machine learning techniques applied to visualization: 
 \textit{Data Processing4VIS, Data-VIS Mapping, Insight Communication, Style Imitation, VIS Interaction, VIS Reading, and User Profiling.} %and answered the questions, "What ML can assist with visualization processes?" and "How can ML techniques be used to solve visualization problems?" Two main questions. 
Wu et al. \cite{wu2021ai4vis} reviewed recent advances in artificial intelligence techniques applied to visual data, examining a number of key research questions related to the development and management of visual data and the support provided by artificial intelligence for these operations. The study by Zhu et al. \cite{zhu2020survey} is the most relevant to us, in which they investigated automated visualization techniques for infographics. %However, the researchers only focused on three visualization aspects: network, annotation, and storytelling, and only investigated automated tools. %Moreover, this study will sort out the different narrative visualization design spaces, authoring tools, ai-supported tools, and ML/AI-generator tools from the perspective of narrative visualization so that practitioners can guide their work through that investigation.
However, no previous work has thoroughly analyzed different levels of automation and how those tools help the design and creation process of visual storytelling in different narrative forms. Our effort seeks to give an overview of available design tools that may assist a variety of users in various design situations. Moreover, through the analysis, \rv{we identify directions that remain undeveloped for future research}.

\subsection{Survey Scope and Methodology}
%This study aimed to obtain an overview of design space, authoring tools, AI-supported tools, and ML/AI-generator tools for narrative visualization types. 
\rv{Our research focuses on narrative visualization tools. Tong et al.~\cite{tong2018storytelling} emphasized in their research that narrative visualization focuses more on information visualization than scientific visualization. \rb{In addition, studies on narrative scientific visualization have been limited; therefore, scientific visualization was excluded from our study.}}

%\hl{Moreover, due to the complexity of heterogeneous data and specific research challenges, we exclude scientific visualization in this review} \cite{xu2020survey}.}

%此外，关于叙述性科学可视化的研究还没有那么多，因此在我们的研究中排除掉了科学可视化。

%\cite{ynnerman2018exploranation, chai2020crowdchart}.
%And some tools that simply generate images and videos are not part of the data visualization field. Therefore, these two parts are excluded from our study.} 

%In our investigation of narrative visualization, we found a gradual increase in research related to immersive visualization. The content related to immersive visualization will be covered in the \ref{sec:discussion} chapter and will not be the main research content in this study. Therefore, these two parts are excluded from our study.
%Since our focus is narrative visualization tools, we limited the scope in information visualization. %Therefore, our study does not include scientific visualization and pure image/video-generated visualizations. %Tong et al.\cite{tong2018storytelling}  pointed out through his research that narrative visualization focuses more on information visualization than scientific visualization. 
%The purely generated images and videos are also not part of the data visualization domain. Therefore these two parts are excluded from our study.

To create the corpus of articles, we gathered from visualization journals and conferences by using reference-driven and search-driven methods. We started with a collection of references on the categorization of narrative visualization in this area for the selection of reference-driven, and we then broadened the focus \rb{by looking up both citing and cited publications.} We completed two rounds of article gathering for the search-based choices. A preliminary search for narrative visualizations, relevant design tools, and best practices was conducted in the first round by using high-impact visualization conferences and publications. In particular, \rv{we selected five conferences} (ACM CHI, IEEE InfoVis, IEEE VAST, IEEE PacificVis and IV) and five journals (IEEE TVCG, IEEE CGA, and ACM Transactions on Graphics, Computer Graphics Forum, Visual Informatics). We gathered a variety of publications by using two search terms (``visualization'' and ``design space/design guide,'' ``visualization'' and ``authoring tool'') and then evaluated abstracts and full texts to narrow down our \rb{sample}. 

%具体的论文数目与分类方法
After this round of article selection, 348 papers and tools were obtained. To achieve a more precise review of the literature about narrative visualization, we used narrative visualization \rv{genres} and tools (e.g., ``data comics'' and ``design space/authoring tool,'' ``infographics'' and ``design space/authoring tools,'' etc.) to \rb{categorize} the papers. Furthermore, we \rb{removed} programming tools and domain-specific application tools, as they are beyond the scope of this research. Finally, 91 narrative visualization papers and tools are summarized in Table 1 and Figure~\ref{fig:figures1}. \rb{In Table 1 and Figure~\ref{fig:figures1}, we excluded \rb{14} commercial software mainly because most of them do not have a definite publication date, and commercial software tends to have frequent updates and additional features, which makes it difficult to fix a specific year.}

\subsection{Taxonomy}
%In this section, we first sort out the common types of narrative visualizations and then summarize the narrative structures and interactions of different narrative genres based on the collected tools and design spaces. This is followed by a discussion of the difficulty of traditional visualization types in the context of existing research. To clarify the scope of this study, we finally constructed a two-dimensional axis, where the horizontal axis is an exploration to explanation, and the vertical axis is simple to complex. The purpose of making this axis is to more clearly represent the difficulty of the different visualization types and the strength of the narrative and thus clarify the specific scope of this study. Second, four types of authoring tools are summarized based on the degree of involvement of automated machines in the visual design process.

In this section, we will first describe the four levels of automation and then introduce the detailed classification of narrative visualization in our survey.
%工具分类
\subsubsection{Tool Classification Method}
In this section, we categorize the visualization tools into four groups based on their automation and intelligence\rv{\cite{zhu2020survey,tyagi2021user}}.

\begin{wrapfigure}{l}{0.055\textwidth}
  %\begin{center}
    \includegraphics[width=0.055\textwidth]{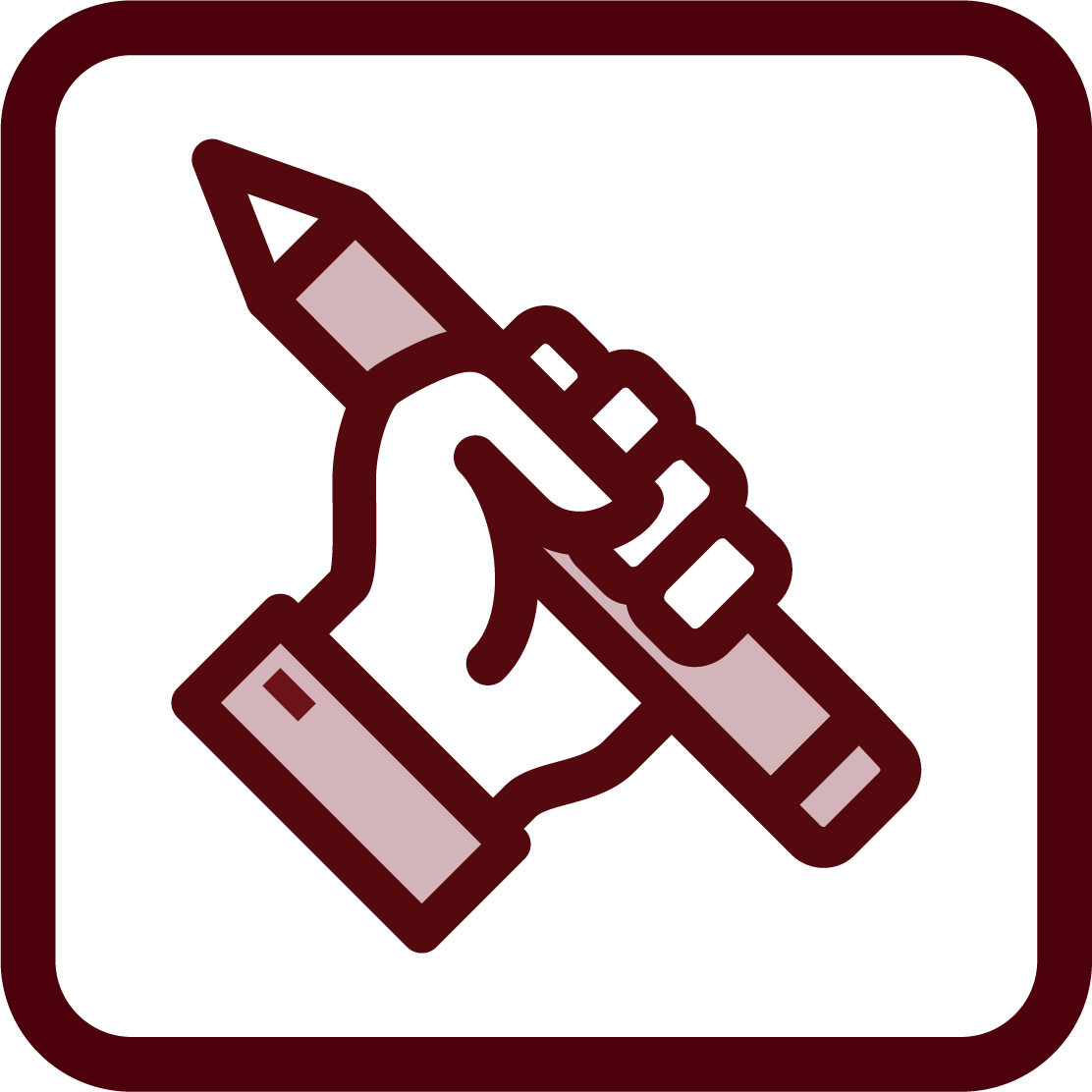}
  %\end{center}
\end{wrapfigure} 
\noindent
\rv{\textbf{Design space} is a conceptual set of possibilities rather than a software tool~\cite{botero2010expanding}.} Design space stresses the ability to choose from a variety of possibilities and investigate alternatives \cite{fischer2006meta, westerlund2005design}. 
% \rv{Botero et al.\cite{botero2010expanding} propose that the design space should be conceptualized as the space of possibilities for realizing a design, which extends beyond the concept design stage into the design in use activities of people.}
% Botero et al. \cite{botero2010expanding} refers to design space as a potential space that is not ``pre-existing.'' %It is realized through the presence of different stakeholders, tools, technologies, materials, and social processes and protocols. 
%In this space, all designers make choices that ultimately result in content beyond conceptual design throughout the design process. 
The design space is a description of all potential design options throughout the design process. Utilizing basic design principles from current visualization techniques is the most preferred method for building a design space \cite{schulz2010explorative}. Moreover, visual design spaces allow us to capture some implicit knowledge of graphic designers \cite{schulz2010design}. 
%Once defined, the design space can be used to access an otherwise scattered and extensive set of visualization technologies \cite{schulz2010design}.
We believe that the visualization design space is an attempt to understand how visualizations are created by designers in a systematic process, that is, to decompose a design work into several design elements and arrange them properly. A straightforward design space makes the design more structured and disciplined, allowing designers to create designs without relying on pure feelings. \rv{It is also the basis for computers to understand the design and eventually create tools to facilitate the design process.}

%Following the guidelines and considerations in design spaces, automatic tools could better understand designers and help generate ideal visualizations. %It is the basis for computers to read designs and eventually automate them.

\begin{wrapfigure}{l}{0.055\textwidth}
  %\begin{center}
    \includegraphics[width=0.055\textwidth]{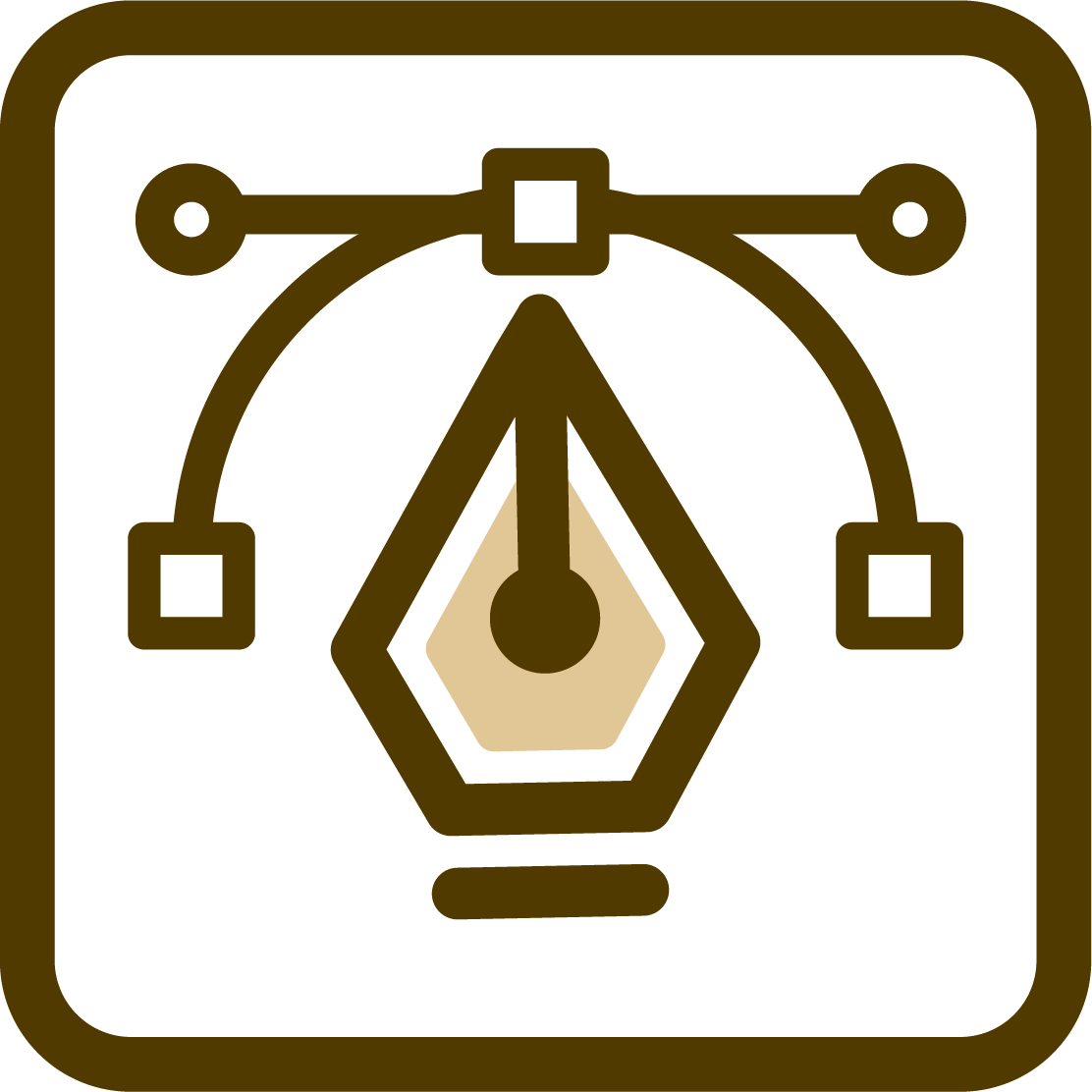}
  %\end{center}
\end{wrapfigure} 
\noindent
{\textbf{Authoring tool}} encapsulates key software functionalities and features for content creation \cite{zikas2020immersive}.  It is an application or tool designed for a specific design purpose. Authoring tools allow users to create visualizations freely with interactive features. They usually require designers to design starting from scratch, allowing designers to have major control of the creation process. On the users’ side, authoring tools allow them to understand the creation framework in advance and eventually interact with the system.

\begin{wrapfigure}{l}{0.055\textwidth}
  %\begin{center}
    \includegraphics[width=0.055\textwidth]{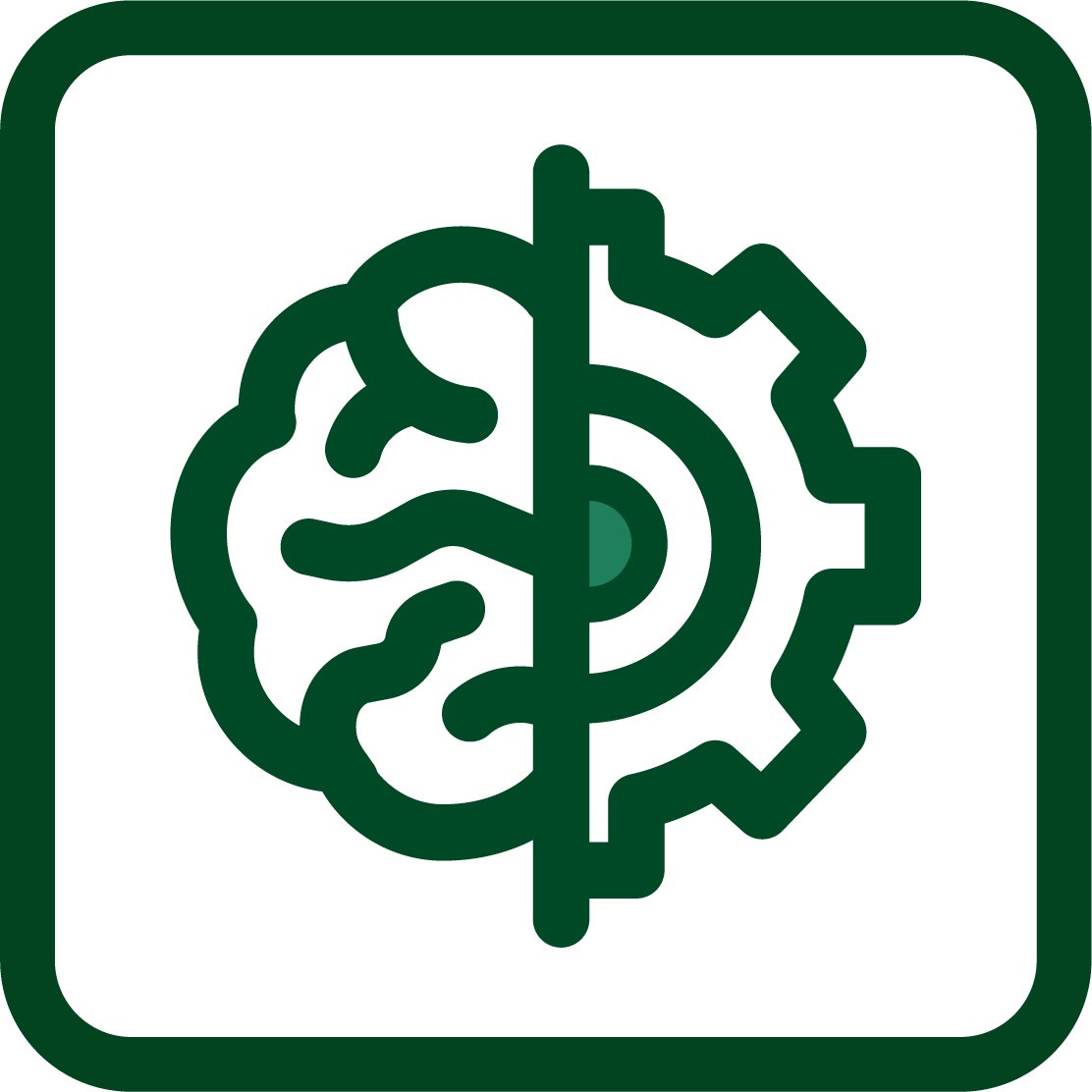}
  %\end{center}
\end{wrapfigure} 
\noindent
{\textbf{ML/AI-supported tools}} apply intelligent algorithms to facilitate visualization creation. Such tools ease visual generation while ensuring a certain degree of control for user in the creation process. \rv{ML/AI-supported tools focus on automatically providing some steps or automatically visualizing some elements, while users need to make decisions on some important steps to create the visualization. A recommended solution is usually provided for a particular part of the visualization. Eventually, users can organize the design content to form the final visualization outcome.}

% Most tools only support specific categories of visualizations or certain parts of visual design. Recommended solutions are usually provided in some steps of visualization creation. Users still need to make decisions on design choices and \rv{be} involved in the authoring process. %The users themselves will combine the final design content to form a perfect visualization design solution.

\begin{wrapfigure}{l}{0.055\textwidth}
  %\begin{center}
    \includegraphics[width=0.055\textwidth]{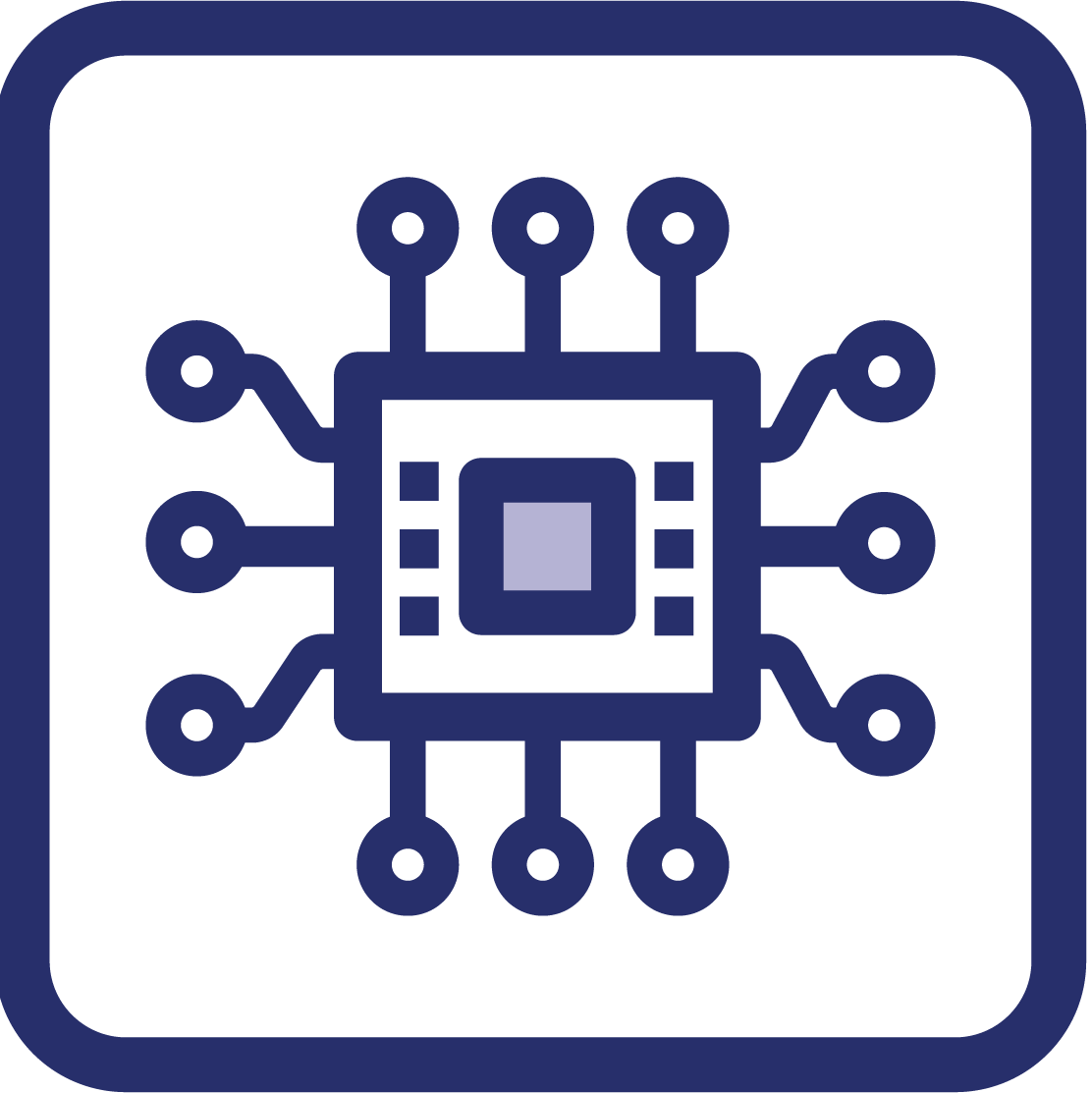}
  %\end{center}
\end{wrapfigure} 
\noindent
{\textbf{ML/AI-generator tool}} is even more intelligent, as users no longer need to participate decision making in the authoring process. The ML/AI-generator tool is designed to reduce barriers for amateurs to create visualizations automatically and ease the burdens for experts to search and select without manually specifying  all elements \cite{zhu2020survey}. When the user uploads the data, \rv{this type of tool} automates the process and analysis of the data and can generate a complete set of visual design solutions without user intervention. %This part of the tool is designed mainly for users who do not know how to program or design visualizations. 

\subsubsection{Visual Classification Method}

\begin{figure}[htp]%[hbt!]
\setlength{\abovecaptionskip}{0.2cm}
\setlength{\belowcaptionskip}{-0.1cm}
    \centering
    \includegraphics[width=8.5cm]{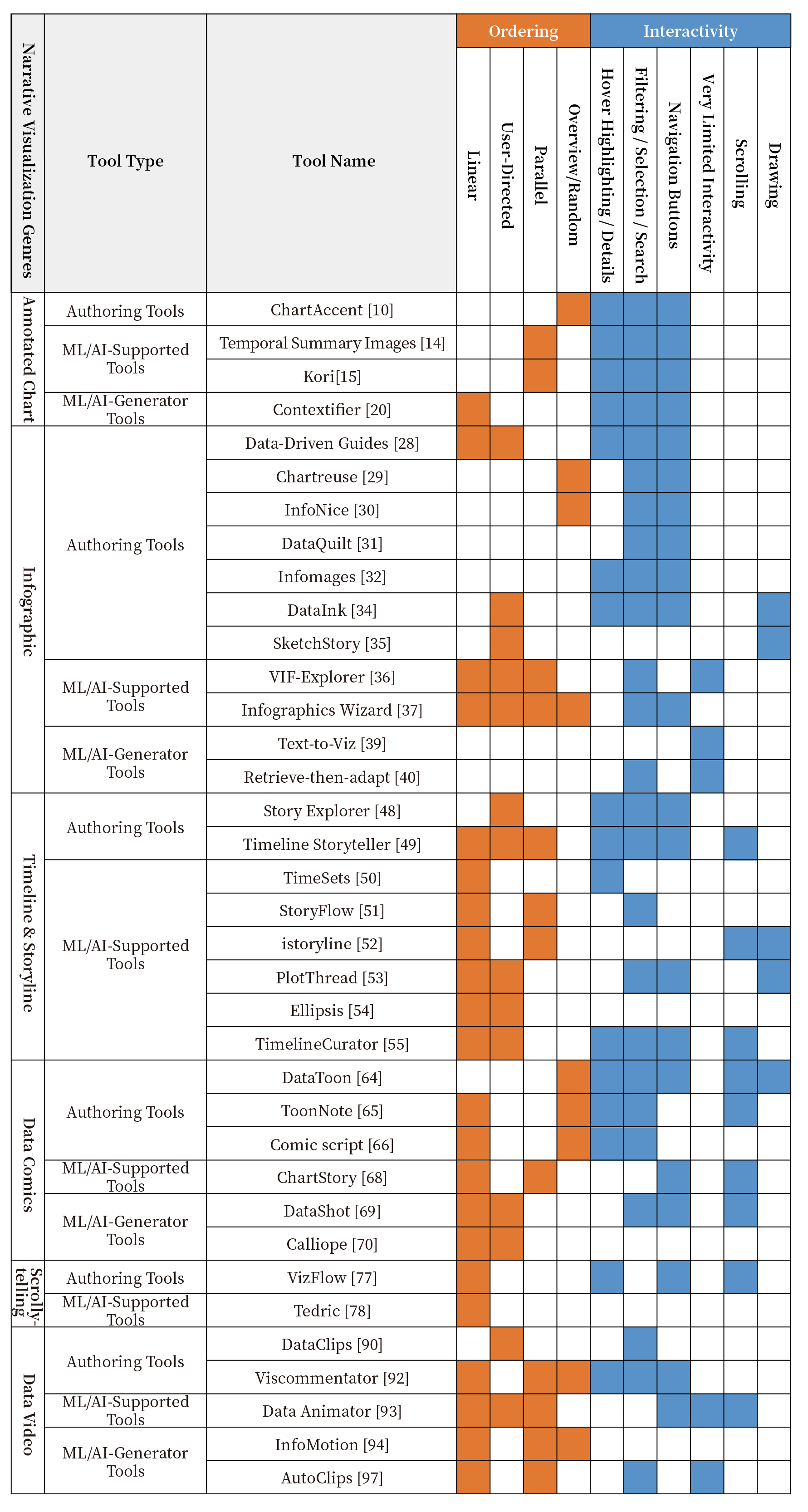}
    \caption{\rv{A summary of interactive tools in each narrative genre, with the supported narrative orderings and interactions of each tool.}}
   \label{fig:intera} %交叉引用的标签\
\end{figure}
%提出了叙事性可视化的分类
\textls[-15]{\rv{Segel et al.} \cite{segel2010narrative} presented seven genres of narrative visualization: magazine-style, annotated chart, partitioned poster, flowchart, comic strip, slideshow, film/video/animation. Recently, Roth \cite{roth2021cartographic} classified visual storytelling into seven \rv{genres}: static visual stories, long-form infographics, dynamic slides, narrative animations, multimedia visual experiences, personalized story maps, and compilations \rv{(compilations provide a ``visual abstract'' that typically links to further text)~\cite{roth2021cartographic}.  %rvq
\rv{On the basis of their findings and the presentation outcomes, we reclassified the narrative visualization genres (Table 2).
In this research, magazine style, partitioned posters, and static visual stories were jointly studied and then collectively referred to as infographics. Film/video/animation, narrated animations, and multimedia visual experiences are called data videos. Slideshow, compilations, and long-form infographics are collectively called ``scrollytelling \& slideshow''. In the literature review, we found only a few works about flowcharts. However, many works on timelines \& storylines are presented in the form of flowcharts; thus, we jointly utilized flowcharts and timelines. Roth et al.\cite{roth2021cartographic} found that personalized story maps are similar to adding annotations to maps; in this study, we classified them as annotated charts. In conclusion, we focused on six genres of narrative visualizations in this survey: \textit{annotated charts, infographics, timelines \& storylines, data comics, scrollytelling \& slideshow, and data videos.}}}}

% In this study, three visualization \rv{genres}, magazine-style, static visual stories, and long-form infographics are studied together and collectively referred to as infographics. \rv{Segel et al.}  \cite{segel2010narrative} pointed out that the presentation of the partitioned poster is multi-view visualization, whose layout is similar to a dashboard and an infographic. %Although both dashboards and infographics could present multiple visualizations on the same page, dashboards are mainly used to monitor interactive and real-time information. 
% Infographics are often presented in a static form and require designers to organize the information, while dashboards are mainly used to monitor interactive and real-time information.% It does not have the property of real-time updates of information \cite{cmeciu2016beyond}. 
%我们自己的分类
% Therefore, we categorize the partitioned poster with interactive properties and real-time updated data into a dashboard and vice versa in the infographic. Comic strips are referred to as data comics for a clearer presentation. The narrated animations and multimedia visual forms are classified into data videos. Slideshow, scrollytelling and compilations are similar in a narrative style. Therefore, we put these three categories together for our study. There is less literature about the flow chart type in the collected research. However, we found quite a few works on timeline and storyline, which are also presented in a flow formation. In this study, flow charts and timeline-related contents are studied together.

%叙事结构的定义+本节中要分析的工具的范围 
{\rv{We surveyed the literature ~\cite{segel2010narrative,tong2018storytelling} to further summarize various tools with different narrative orderings and interactivity. Segel et al.\cite{segel2010narrative} summarized three kinds of ordering for narrative visualization: linear (the author specifies this path), random access (no path is specified), and user-directed (users may choose a route from various available pathways or design their own). Tong et al.~\cite{tong2018storytelling} added another ordering type called parallel (multiple paths can be displayed simultaneously). %Linear means the author specifies this path; random access means no way is specified; user-directed means the user can choose a path among multiple paths or create his own path. Parallel means multiple paths can be displayed at the same time. 
Apart from the 38 references listed in the Design Space category in Table 1, some studies proposed techniques or algorithms without developing a fully functional visualization tool with \rb{appropriate interfaces}. Therefore, we selected the 36 visualization tools that include interactive functions and support the creation of narrative structures for each narrative genre. Their narrative orderings and interactivity are also marked in Figure 2.}}

% \usepackage{multirow}
\renewcommand\arraystretch{1.3}

\begin{table}[ht]
\textbf{\caption{Segel et al.\cite{segel2010narrative} and Roth et al.\cite{roth2021cartographic} proposed seven genres of narrative visualization, respectively, and we combined their findings to obtain new genres of narrative visualizations.}}
\begin{tabular}{|m{2.4cm}<{\centering}|m{2.4cm}<{\centering}|m{2.7cm}<{\centering}|}
\hline
\textbf{Segel \& Heer\cite{segel2010narrative}} & \textbf{Roth et al.\cite{roth2021cartographic}} & \rb{\textbf{Our work}} \\ \hline
magazine style & \multirow{2}{*}{static visual stories} & \multirow{2}{*}{infographics} \\ \cline{1-1}
partitioned poster & & \\ \hline
\multirow{3}{*}{film/video/animation} & narrated animations & \multirow{3}{*}{data videos} \\ \cline{2-2}
 & multimedia visual experiences          & \\ \hline
\multirow{4}{*}{slideshow}           & dynamic slideshows & \multirow{4}{*}{\begin{tabular}[c]{@{}c@{}}scrollytelling \\  \& slideshow\end{tabular}} \\ \cline{2-2}
& compilations & \\ \cline{2-2}
& longform infographics & \\ \hline
annotated chart & personalized story maps & annotated chart \\ \hline
comic strip & - & data comics                              \\ \hline
flow chart & - & \begin{tabular}[c]{@{}c@{}}timeline \& storyline\end{tabular}                        \\ \hline
\end{tabular}
\vspace{-1mm}
\end{table}

{\rv{As shown in Figure 2, most tools support linear ordering, and relatively few support random ordering. On average, the tools for annotated charts support the fewest narrative ordering types, while the tools for timelines and data videos support the most narrative ordering types. Segel et al.\cite{segel2010narrative} proposed six types of interactions for narrative structures, of which \textit{\rb{hover highlighting}, filtering/selection/search, and navigation buttons} are the three most common interaction types. As we explored the selected tools, we discovered two standard interaction types: \textit{scrolling} which includes landscape and portrait scrolling, and \textit{drawing,} which supports \rb{``touch+pen'' interaction.}}}

\section{Annotated Chart}

\begin{figure*}[htp]%[hbt!]
\setlength{\belowcaptionskip}{-0.1cm}
    \centering
    \includegraphics[width=16.5cm]{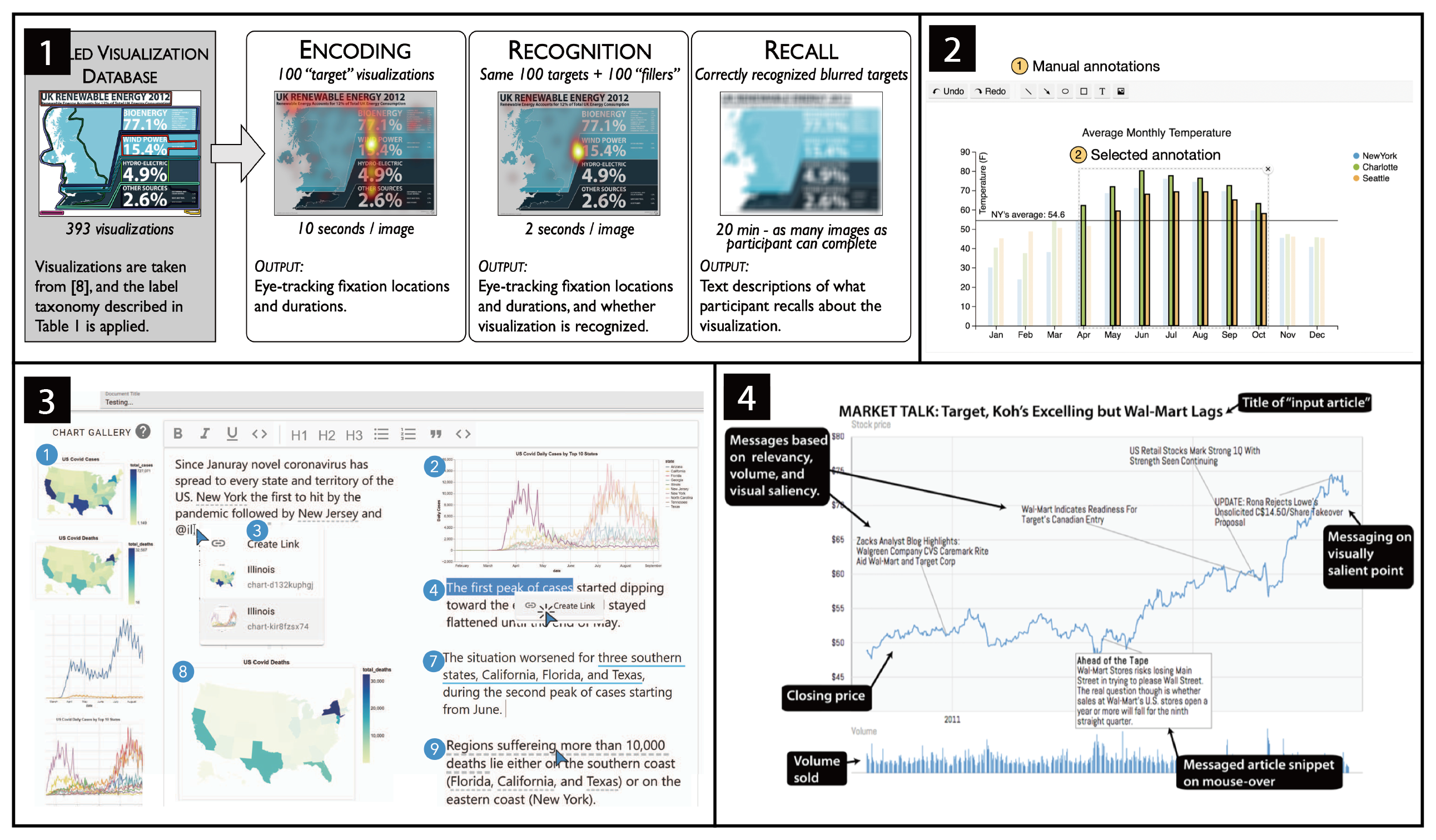}
    \caption{Selected examples of annotated charts' design spaces and tools. (1) Design space: an experiment on how visualizations are recognized and recalled. The study shows that adding captions and annotating text improves user attention and recall \cite{borkin2015beyond}. (2) Authoring tool: ChartAccent \cite{ren2017chartaccent}, which allows one to manually and interactively generate data annotations. (3) ML/AI-supported tool: Kori's \cite{latif2021kori} Tools viewport. As the user enters text, Kori automatically prompts for potential references (gray underlining). Simple interactions to manually create links are also supported. (4) ML/AI-generator tool: annotations generated by Contextifier \cite{hullman2013contextifier}.}% Graphical labels and arrows point out interface functions.}
   % \label{fig:figures2.11} %交叉引用的标签
   %\vspace{-2mm}
\end{figure*}

\begin{wrapfigure}{l}{0.055\textwidth}
    \includegraphics[width=0.055\textwidth]{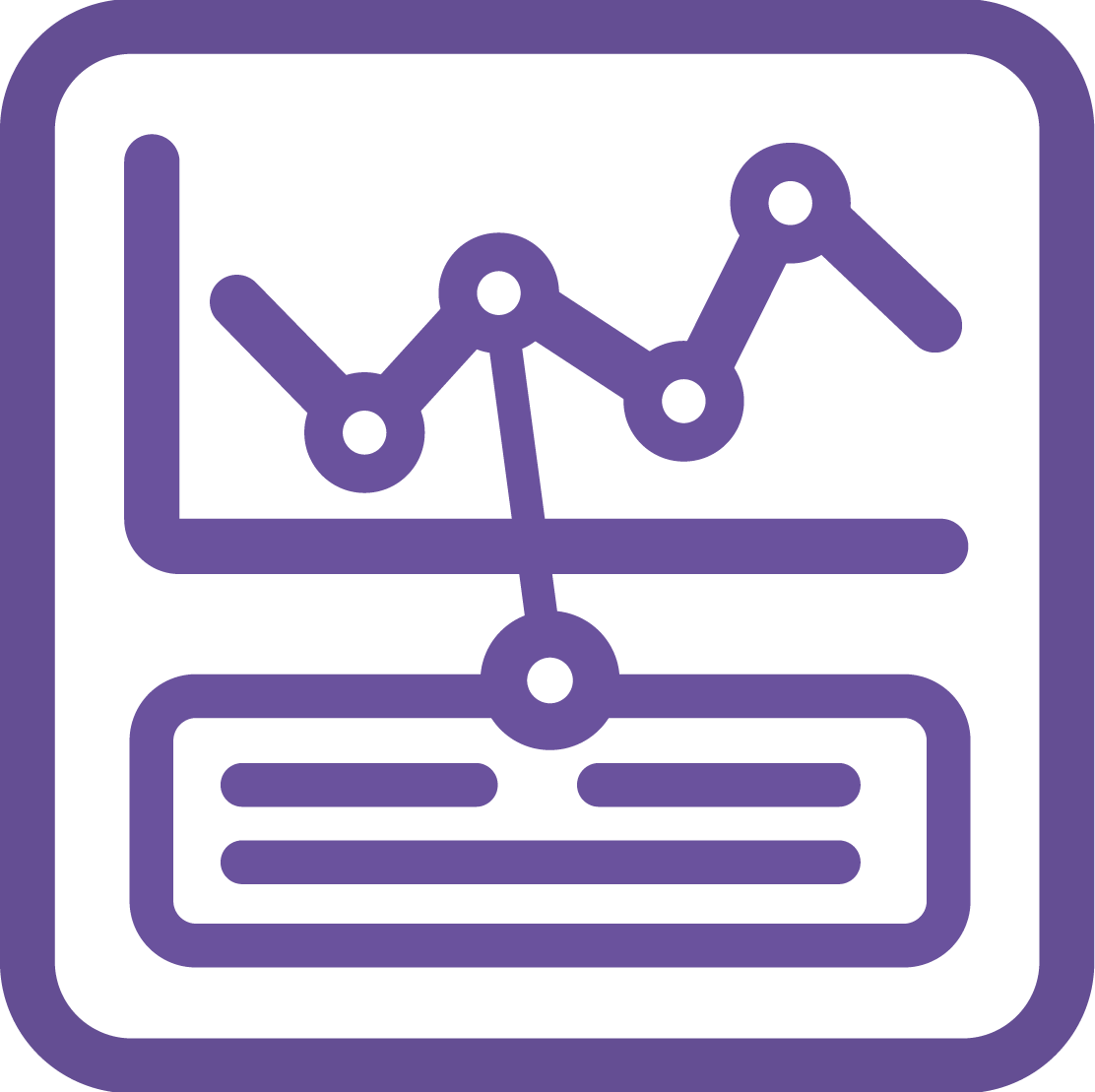}
\end{wrapfigure} 

Annotated charts use graphics (arrows or trend lines) or text (data values or commentary) to supplement information, adding contextual information to a visualization to supplement or introduce the meaning of the data. Annotations allow audiences to focus on specific content or critical information while retaining complete details of the contextual data \cite{brath2018automated,marshall1997annotation}.

{\textbf{Design space:}}
Adding annotations to visualizations makes the detailed information more accessible to users and improves the memorability of the images \cite{borkin2015beyond}. Borkin et al. \cite{borkin2015beyond} applied eye movement studies and \rv{cognitive experimental techniques} to verify that adding captions and annotated text to charts can communicate visual information more effectively. 
%That is, infographics with low data-to-ink ratios and high visual density (i.e., more chart junk and ``clutter'') are more likely to be remembered than ``clean'' infographics.A specific form of annotation applied to a specific target ideally achieves a specific effect that adds value to annotated \cite{ren2017chartaccent}.
% In some instances, the act of annotation is self-serving, such as when a student adds an annotation to a textbook to aid learning or when someone adds an annotation to a calendar as a personal reminder. 
When annotating charts, both the form of the annotation (text, shapes, highlights, and pictures) and the kind of desired annotation (data items, coordinate spaces, structural chart components, and previous annotations) must be considered~\cite{ren2017chartaccent}. In addition, Kong et al. \cite{kong2017internal} defined annotation as a visual cue. \rb{They divided the annotations into two categories: internal cues that modify the existing image by highlighting or downplaying the focus area (i.e., the context) and external cues that add supplementary elements (e.g., outlines, annotations and glyphs) to the existing image to emphasize the focus. They showed that internal cues are often more effective in directing attention than external cues.} 
%\rv{They separated annotations into internal cues\rb{(modify the existing image by highlighting or downplaying the focus area (i.e., the context))} and external cues \rb{(add supplementary elements (e.g., outlines, annotations, glyphs) to the existing image to emphasize the focus point)} and demonstrated that utilizing internal cues is often preferable \rb{to} using external cues. 
Internal cues affect the current picture by highlighting the focal region or de-emphasizing the rest of the visualization.
%Researcher divided them} annotations into internal cues \rv{(modify the existing image by emphasizing the focus area or de-emphasizing the remainder of the image (i.e., the context).) } and external cues \rv{(append additional components (e.g., outlines, annotations, glyphs) to the existing image to emphasize the focus point.)} and finally proved that using internal cues is usually better than external cues.}

{\textbf{Authoring tool:}}
\textls[10]{Researchers have developed a range of visual programming libraries and packages for diagram annotations~\cite{gomez2017ggplot2, bostock2011d3}. These tools require users to have programming skills, while programming tools can only provide asynchronous feedback to designers. To help create chart annotation more easily, researchers have developed authoring tools that have \rb{appropriate interfaces} and can provide feedback to users, which significantly facilitates the annotating process without requiring specialized programming knowledge.} Tableau \cite{tableau2006} provides several basic options for annotating charts. For example, the \rv{tool} allows users to add trend lines to charts. User-created annotations via text can be data-driven but are limited to some standard forms of annotation. ChartAccent \cite{ren2017chartaccent} is an interactive tool that allows users to generate data annotations manually. \rv{It offers many functions, such as highlighting markers, which are more straightforward and flexible than Tableau \cite{tableau2006}. Selected markers can be highlighted directly without affecting unselected markers. Although these tools can easily create annotations, they still rely largely on the designer's expertise to create manually.}

{\textbf{ML/AI-supported tool:}}
{ML/AI-supported tools of annotated charts reduce manual operations by automatically providing annotated suggestions via user interactions. 
\rv{SmartCues \cite{subramonyam2018smartcues}, which provides multitouch interaction, is a library that supports details-on-demand via dynamic computational overlays to assist users in building queries and generating data-aware annotations.
Touch2Annotate \cite{chen2010touch2annotate} and Click2Annotate \cite{chen2010click2annotate} are early semi-automatic annotation generators. 
Touch2Annotate~\cite{chen2010touch2annotate} is a tool for adding annotations to multidimensional data visualizations on a multitouch interface. The tool provides annotation templates and allows users to create high-quality chart annotations by simply highlighting the data and selecting the appropriate annotation template according to the annotated content. 
Click2Annotate \cite{chen2010click2annotate} allows simple data analysis and generates easy-to-understand annotations. The semantic information encoded in its annotations can be browsed and retrieved. %Click2Annotate reduces the need for manual operations.
Similarly, Kandogan~\cite{kandogan2012just} introduced the idea of just-in-time descriptive analysis. In this scheme, when a user interacts with a diagram, the diagram is automatically annotated in response to that interaction.}} %Just-in-time descriptive analysis is based on (a) identifying visual features, such as clusters, outliers, and trends, that the user may observe automatically in the visualization and (b) determining the semantics of these features by performing statistical analysis as the user interacts, and (c) enriching the visualization with annotations that not only describe the semantics of the visualized features but also facilitate the interaction to help the user understand the deep structure of the data. However, it can only annotate point-based multidimensional visualizations. 

{Latif et al. \cite{latif2021kori} developed Kori based on a design space analysis of textual and graphical references and added visualization \rv{genres}, such as line charts, pie charts, and maps. When users create \rb{visualizations} with the tool, the system automatically provides annotation suggestions using natural language and enables combining text and graphs via manual interaction. 
\rv{Kong et al.~\cite{kong2012graphical} proposed an automated system that overlays user-selected graphics onto existing chart bitmaps and allows users to customize published visualizations by identifying visual markers and attributes of axes of encoded data to better assist users with chart reading tasks. Srinivasan et al. \cite{srinivasan2018augmenting} explored the potential applications of interactive data facts for visual data exploration and communication. The researchers also developed the Voder system to demonstrate how users can use interactive data facts to suggest optional visualizations and modifications, which helps users interpret the visualizations and convey their findings.} % When the user hovers over a data fact, the portion of the visualization corresponding to the fact is dynamically highlighted. This helps users to interpret the visualization and communicate their findings.}
Bryan et al. \cite{bryan2016temporal} focused on narrative visualizations for multivariate, time-varying datasets. They proposed a method called Temporal Summary Images (TSI) consisting of temporal layout, data snapshots in the form of comic strips, and textual annotations. %When users explore the visualization interactively, the system provides annotations for three types of data attributes: digital vector, storyline, and impact map. 
\textls[-5]{Moreover, researchers have noted that line graphs are the most common type of visualization in daily life~\cite{lee2021viral}. However, some line charts are deceptive with exaggeration, understatement, and message reversal. \rb{For example, exaggerating or minimizing the effect size via aspect ratio manipulation in line charts leads to deceptive representation~\cite{fan2022annotating}.} %论文原句，还未修改
To address this problem, Fan et al. \cite{fan2022annotating} introduced a tool for detecting and annotating line graphs in the wild that reads line graph images and outputs text and visual annotations to assess the truthfulness of line graphs and help readers understand faithful line charts.}}

Compared with authoring tools, ML/AI-supported tools further simplify the difficulty of creating annotated visual diagrams and reduce manual operations by automatically providing annotation suggestions. Furthermore, ML/AI-supported tools allow users to promptly add annotations to the diagram while interacting with the visualization based on AI assistance. %The disadvantage of these tools is that they support fewer types of charts. Tools that support more types of visual charts also tend to require more manual operations by the user.

{\textbf{ML/AI-generator tool:}}
As annotations are essential in visualization design, researchers have explored annotation approaches for different visualization \rv{genres}. The Contextifier \cite{hullman2013contextifier} provides an algorithm for selecting annotations that automatically creates a stock timeline graph and matches the appropriate annotation to the line graph by referring to the content in the news article.
Liu et al. \cite{liu2020autocaption} developed AutoCaption to build a scheme to accomplish the task of diagram title generation by using deep neural networks. One-dimensional residual neural network is used to analyze the relationships between visualization elements, identify essential features of the visualization diagram, and generate a complete description. Both tools create the appropriate information for the diagram without user intervention.

\textbf{Summary:}
Annotations are informative additions to visual diagrams and are an essential part of visual design, helping audiences quickly understand diagram information and helping analysts revisit and reuse analysis processes conducted in the past~\cite{shrinivasan2009connecting}. \rv{Researchers have verified the importance of annotation at the visual memory level \cite{borkin2013makes} and at the cognitive level \cite{latif2021kori}, which both indicate that annotations are an integral part of visualization design.} Although researchers have studied the layout problem of annotated charts and the distraction caused by repeatedly switching views by using interactive highlighting \rb{\cite{kittivorawong2020fast}}, solutions to occlusion problems, such as annotations blocking the charts, have not yet been addressed. Therefore, more advanced techniques and tools are required to improve the efficiency of the automatic layout. Moreover, for tools to become more intelligent and accurate, the extraction of the existing annotated diagram corpuses and the research related to the identification and correction of incorrect annotations must both be enhanced. \rv{Researchers have also developed various tools based on annotated design spaces. Just-in-time annotations and automated annotations provide a new method for users to promptly update and convey visual information \cite{brath2018automated}. In the future, automated annotations can focus more on internal annotations with the option of rich and aesthetically appealing visual cues \cite{zhu2020survey}.}

\section{Infographic}

\begin{figure*}[htp]%[hbt!]
\setlength{\belowcaptionskip}{-0.1cm}
    \centering
    \includegraphics[width=16.5cm]{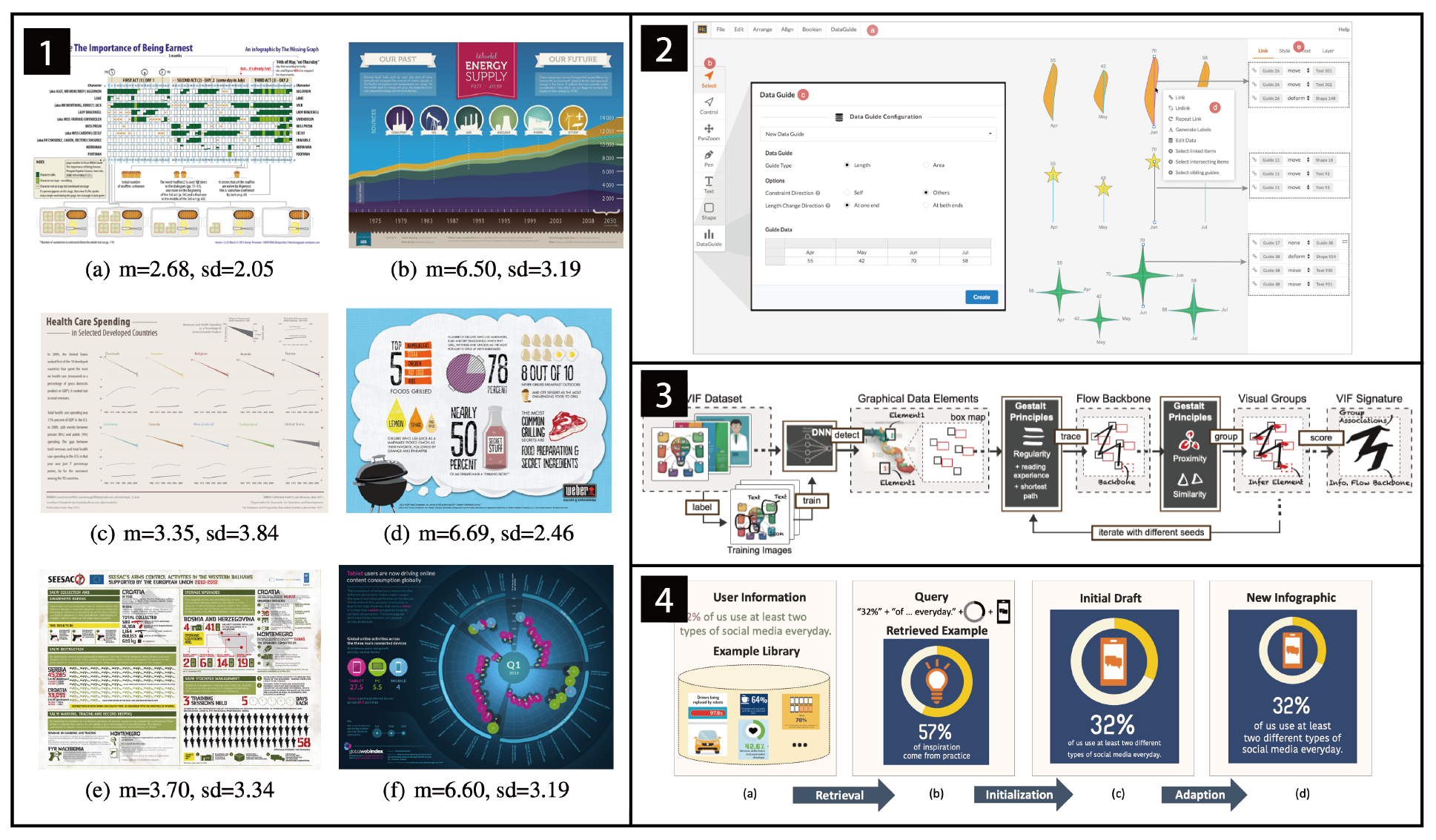}
    \caption{Selected examples of infographic's design spaces and tools. (1) Design space: different types of infographics have different levels of appeal to users, \rv{the scores are the means and standard deviations from user experiments using 9-point Likert scale}) \cite{harrison2015infographic}. (2) Authoring tool: DDG vector drawing tool which can be used to bind vector graphics to data \cite{kim2016data}. (3) ML/AI-supported tool: utilizes a deep neural network using manually labeled infographics as training data to find visual data items while ignoring creative aspects \cite{lu2020exploring}.
    (4) ML/AI-generator tool: infographics are automatically generated by simulating online examples in two main steps: retrieval (indexing of online instances based on visual elements) and matching (replacement with personal user data) \cite{qian2020retrieve}.}
   % \label{fig:figures2.11} %交叉引用的标签
   %\vspace{-2mm}
\end{figure*}

\begin{wrapfigure}{l}{0.055\textwidth}
  %\begin{center}
    \includegraphics[width=0.055\textwidth]{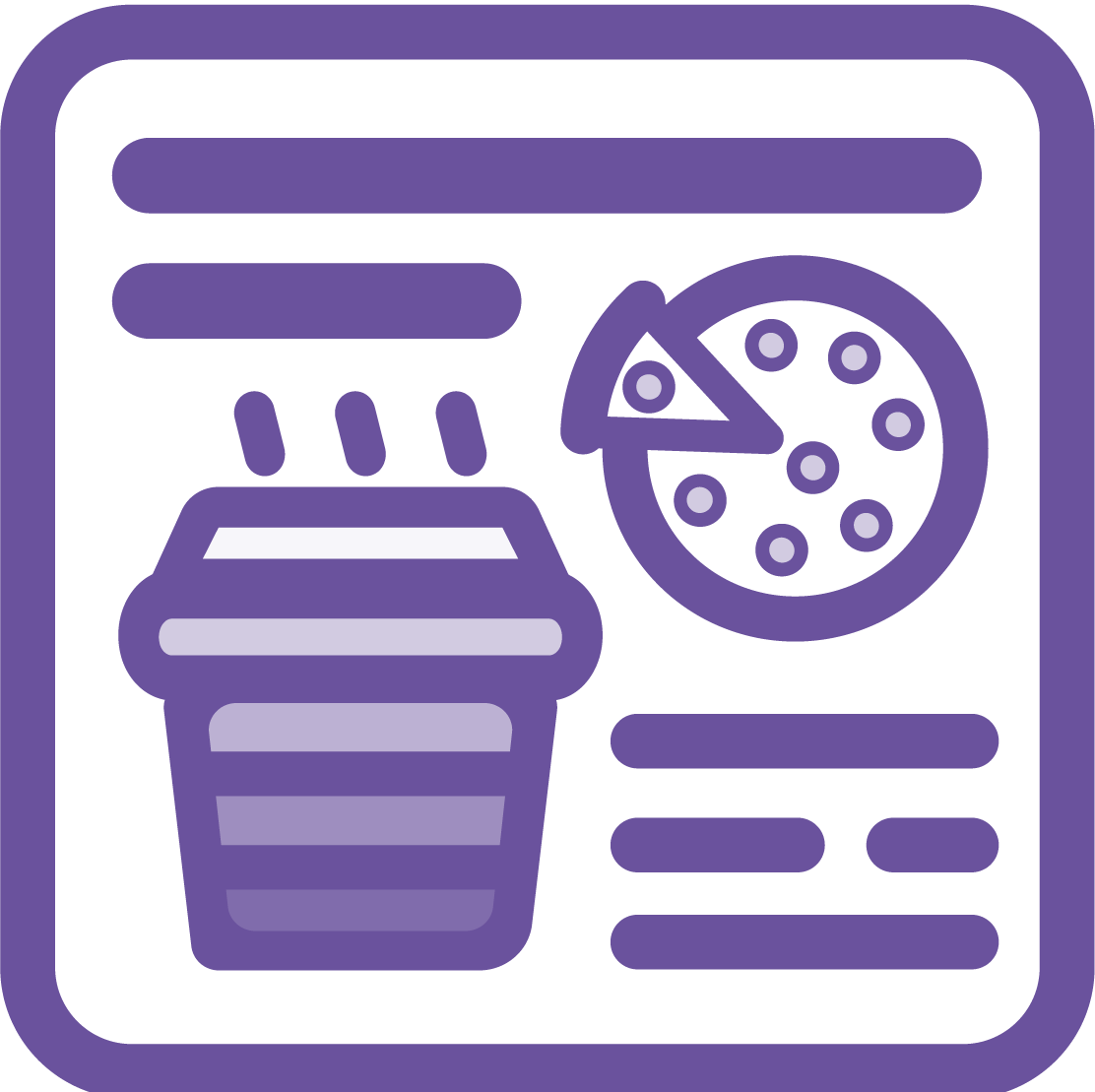}
  %\end{center}
\end{wrapfigure} 

The term infographics, which stands for informational graphics, refers to a type of visualization that focuses on the use of graphically designed icons, images, colors, and other elements to illustrate data and textual information. Otten et al. \cite{otten2015infographics} defined infographics as ``to convey a particular set of information to a specific audience by transforming complex and abstract concepts into visual components.'' %Although both infographic and dashboard could present multiple visual graphs on the same page, the essence of an infographic lies in the communication and exchange of information, while the dashboard lies in the exploration of data \cite{moere2011role, cmeciu2016beyond}.

{\textbf{Design space:}}
%infographic的分类方式
Infographics are frequently utilized in a variety of sectors because they are simple to comprehend and can improve the viewer's visual working memory \cite{harrison2015infographic, naparin2017infographics}.
% Infographics are widely used in social media, politics \cite{cmeciu2016beyond, amit2018digital}, medicine \cite{oliveira2020visual, balkac2018role}, business presentations \cite{harrison2015infographic}, journalism \cite{wang2018infonice}, and education \cite{noh2015use, naparin2017infographics} because they are easy to understand and enhance the viewer's visual working memory.
Different categories of infographics, information units, and presentation formats have been studied by researchers. Albers et al.\cite{albers2014infographics} summarized four types of infographics, including bullet list infographics, \rb{snapshot} infographics, flat information infographics, and information flows. 
%\rb{\st{Only the last three are suitable for infographics.}} The first two types do not support the complex communication of information. The first two types usually do not have a clear audience; the latter two types are more suitable for representing complex information. 
%\rb{\st{According to the information units, infographics can be divided into four types: statistical-based, timeline-based, process-based, and location-based}} \cite{santos2018influence}.
According to the presentation forms, infographics can be divided into static, dynamic, and interactive categories. 
%\rb{\st{Since different media produce different effects on users, it is essential to be aware of the differences between printed and digital infographics at the beginning of the design process. Especially for interactive infographics, visibility, feedback, constraints, and consistency must be considered.}}\cite{albers2014infographics}.

%Infographic的具体design guidelines
\rb{A good infographic should be attractive, easy to understand, and easy to remember}\cite{cmeciu2016beyond}. Studies have found that audiences usually form a primary impression of an infographic within the first 500 milliseconds. This impression depends heavily on the color and visual complexity of the page. Therefore, to increase the appeal of infographics, designers should display them by increasing the \rv{contrast} between colors or selecting a limited number of images with text~\cite{harrison2015infographic}. However, an infographic is not only a simple combination of graphics and text. Infographics affect how well audiences remember information; when audiences are pleased by infographics, they are more likely to remember it \rb{over} a longer time period \cite{lyra2016infographics}. The studies by Lan et al. \cite{lan2021smile} showed that adding emotional factors to visual designs can create better infographics. Other researchers point out that embedding games into infographics encourages user interaction and improves their exploration experience \cite{diakopoulos2011playable}.
In addition, several specific design guidelines for infographics are proposed.
Dunlap \& Lowenthal~\cite{dunlap2016getting} gave design recommendations on four levels: overall design, structure, content, and infographics visuals. %Namely, a compelling infographic usually has an unexpected element (e.g., humor, metaphor); has no more than two pages of content; has an apparent visual focus and purpose of communication; uses images and elements relevant to the topic, etc. Effective infographics typically use simple visuals rather than high-fidelity visuals to maintain effective delivery of important information \cite{dunlap2016getting, cui2019text}.

% \rb{\st{Naparin \& Saad  asserts that infographics must adhere to three requirements: knowledge translation, color scheme, and narrative. In addition to effectively conveying information to the audience, it is critical to consider how to include narrative components or emotional factors to increase the infographic's readership and memorability.}\cite{naparin2017infographics}}

%Naparin \& Saad \cite{naparin2017infographics} stated that infographics need to satisfy three dimensions of knowledge translation, choosing the right colors and storytelling. Apart from effectively delivering information to the audience, it is necessary \rv{to consider how involve} emotional considerations or narrative attributes to enhance the audience's understanding and memorability of an infographic.

{\textbf{Authoring tool:}}
Infographics have many advantages, but designing infographics can be laborious for \rv{amateurs} and time-consuming even for experts. %Researchers have proposed a large number of design spaces and guidelines for infographics. However, they have also developed many infographic creation tools to reduce the difficulty of creating infographics for non-professionals and increase the speed and efficiency of infographics for professionals.
Numerous tools can be used to create infographics in the design field, including Adobe Illustrator \rb{\cite{adobeIllustrator}}, Sketch \rb{\cite{sketch2010}}, and other vector drawing tools. However, these tools do not support associating data with graphics, suggesting the complexity involved in matching data with graphics when used together to create data-driven infographics. 
Researchers have developed specialized tools to solve this problem by binding data to vector graphics. For example, designers can manually draw graphics and associate data with the created graphics by using Data-Driven Guides (DDG)~\cite{kim2016data}. This tool relieves the burdens of designers to manually code data into custom graphics. Chartreuse~\cite{cui2021mixed} and InfoNice~\cite{wang2018infonice} help users create \rb{evocative} bar graphs with custom markers that convert new bars into infographics with visual elements. Both tools are integrated with Microsoft Office as plug-ins, lowering the barrier to creating infographics. In addition to associating data with vector graphics, DataQuilt \cite{zhang2020dataquilt} and Infomages \cite{coelho2020infomages} are tools for binding data to bitmap images. %DataQuilt \cite{zhang2020dataquilt} can borrow the stylistic features of natural images (e.g., paintings, photographs, sketches) to encode and decorate data. Infomages \cite{coelho2020infomages} embeds commonly used data charts into thematic images and support designers in annotating data.
In addition, certain tools are integrated with the sketch functions, allowing users to create designs more freely~\cite{kim2019dataselfie,xia2018dataink,lee2013sketchstory}. DataInk~\cite{xia2018dataink} provides ``pen+touch'' interactions enabling designers to express their creative thinking by drawing on a digital canvas and directly matching their drawings to data. SketchStory~\cite{lee2013sketchstory} integrates real-time free-writing capabilities with interactive data charts, allowing presenters to move and resize data charts by touching the screen, easing and speeding up the creation of personalized and expressive data charts. Although all these tools can help create infographics, most tools can only transform specific data types into specific forms of visual charts, with line charts and bar charts being the majority. Designers still need to reintegrate the design elements and lay them out to form complete infographics.

{\textbf{ML/AI-supported tool:}}
%The AI-supported tools for creating infographics solve the problem that authoring tools cannot generate complete infographics. 
Lu et al.~\cite{lu2020exploring} built an infographic visual flow search tool, VIF-Explorer, by analyzing many infographics and extracting the Visual Information Flow (VIF) of these images. 
%This tool can search infographics based on selected or drawn VIF patterns, which improves the efficiency of infographic search and facilitates the creation of narrative infographics. 
However, this software can only analyze simple infographics. Complex or nonstandard infographics with creative elements are challenging to identify and characterize. Infographics Wizard~\cite{tyagi2021user} can generate infographics \rv{with} complex layouts. The tool first recommends VIF layouts based on the given information, then provides recommendations for visual \rb{group} (VG) designs, and finally generates connections between VGs to complete the infographics. Visme~\cite{visme2013}, Infogram~\cite{infogram2012} and \rv{Canva~\cite{CANVA2018} are examples of more commercial types of software.} These web-based tools allow users to drag and drop various images and graphic elements to create infographics of the highest quality. Additionally, an infographic's colors have a significant impact on the audience's first impression~\cite{harrison2015infographic, naparin2017infographics}. InfoColorizer \cite{yuan2021infocolorizer} allows users to employ color palettes to create data-driven infographics.

% These web-based platforms provide many images and graphic materials to help create high-quality infographics with dragging and dropping interactions. In addition, the colors of an infographic largely affect the audience's primary impression \cite{harrison2015infographic, naparin2017infographics}. InfoColorizer \cite{yuan2021infocolorizer} allows users to invoke color palettes when creating data-driven infographics. Users can try out different infographic layouts and get corresponding palette recommendations to refine their designs. %\rv{In addition, there are commercial tools like Canva~\cite{CANVA2018} for creating infographics quickly. It integrates many images, templates, illustrations, and other visual elements to generate infographics by directly entering text or inserting images automatically.} %The text and images can be laid out in any way you like, using them in various scenarios.

In short, ML/AI-supported tools for infographics aim to identify existing infographic layouts and color encodings and match them to corresponding infographic recommendations. While it could offer more design options and save efforts for designers, the existing ML/AI-supported tools are not intelligent enough to make creative and unique infographics similar to those created by designers who use authoring tools.%The infographics generated by AI-supported tools are stronger than authoring tools in overall and narrative aspects. However, it does not imply that AI-supported tools are better than authoring tools. Because the existing AI-supported tools are weak in identifying creative visual elements, the generated infographics are more modest, while the authoring tools could support making more creative and unique infographics.

{\textbf{ML/AI-generator tool:}}
Text-to-Viz \cite{cui2019text} \rb{generates} infographics by natural language techniques with predefined schemes in two steps: \rv{semantic parsing (identifying how this information is described \rb{by casual} users) and visual generation (layout, descriptions, graphics, and colors).} \rb{However, the tool is limited in three aspects: the generability problem, which only supports proportion facts; infographics expressiveness, which is based on predesigned styles; and expression ambiguity, which the current model cannot understand.}
Qian et al.\cite{qian2020retrieve} proposed Retrieve-Then-Adapt to automatically generate infographics by simulating Internet design works so that it can create richer designs.
%Qian et al. \cite{qian2020retrieve} proposed Retrieve-Then-Adapt to automatically generate infographics by simulating online examples in two main steps: retrieval (indexing online instances based on visual elements) and matching (replacing them with personal user data). Since this system creates infographics by matching information from the Internet, it can create richer content than Text-to-Viz \cite{cui2019text}.
Chen et al.~\cite{chen2019towards} proposed a similar solution in that it helps users turn existing timeline infographics into re-editable templates. \rv{In the deconstruction phase, a multitask deep neural network is used to parse the global and local information on the timeline; in the reconstruction phase, the infographic is then extended into an editable template by a channel technique.}
These approaches identify and visualize accurate information and ensure that the final generated infographic elements are organized harmoniously.

%Authoring tools for creating infographics and AI-supported tools require users to have an initial idea of the final infographics design, which can be challenging for inexperienced users. These two types of tools are mainly targeted at experienced users. On the other hand, AI-generator tools are more user-friendly for users without design experience. These cutting-edge technical studies help generate visualizations from data insights and design aesthetics through automated methods, effectively increasing the efficiency of creating infographics by significantly reducing the complexity of the creation process.

\textbf{Summary:}
\rv{Different types of tools have different focuses for infographic design. Design spaces of infographics mainly introduce the key components of a good infographics. For authoring tools, the focus is on how to bind images with data. ML/AI-supported tools and ML/AI-generator tools identify the layout of existing infographics and apply or recommend it to new infographics. Creating infographics with authoring tools and ML/AI-supported tools requires users to know what the final infographics look like, which can be challenging for amateurs. %These two types of tools are aimed at expert and intermediate users, who have a certain level of expertise in creating infographics. 
ML/AI-generator tools are more friendly to amateur users. These tools help users generate visualizations from data insights and design aesthetics by using an automated approach that reduces the complexity of the creative process and effectively increases productivity.}
% While ML/AI-generator tools are more friendly to amateurs, these cutting-edge technical studies help users generate visualizations from data insights and design aesthetics through automated methods. By significantly reducing the complexity of the creation process, they help amateurs create various infographics automatically, lowering the threshold for users to create visualizations and effectively increasing productivity.

% identify the design elements of existing infographics and apply them to new datasets with recommended visualizations, where AI-generator tools requires no user intervention in the creation process.%focus more on users without expertise. 
Although a great deal of research has been conducted, much work is still required in this category. The first direction is to adapt current tools to more visualization \rv{genres}. Existing tools for converting standard statistical charts into infographics support only simple chart conversions \cite{cui2021mixed, wang2018infonice}. A more comprehensive visual corpus needs to be built to support a wider variety of visualization \rv{genres} in future work. %In addition, infographics automation tools are generally designed for specific types of data, and better fault tolerance for complex data is also needed, and appropriate feedback moderation is necessary.
%How to enhance the interpretability of machine learning models is another challenge for automatically generating infographics \cite{zhu2020survey}.
The second direction is to offer more advanced extraction and editing functions to existing infographics. Some tools can identify design elements from existing infographics, but only support simple visual charts~\cite{cui2019text, qian2020retrieve, chen2019towards}. Meanwhile, the extraction of artistic effects in infographics is still relatively weak and intelligent algorithms can be applied to tackle this problem. Moreover, editing functions can be added to infographic identification tools directly to reduce the effort of switching between software.
The third direction is to enhance research on intelligent algorithms. Many rule-based algorithms are applied in current tools (e.g., color selection \cite{yuan2021infocolorizer} and icon selection). The quality of infographics generated by visualization systems can be further improved using more advanced machine learning or deep learning approaches. % \rv{There are also tools like Text to viz that have the problem of expression ambiguity due to their use of natural language algorithms.}
% \rv{The fourth direction is to integrate natural language algorithms from smart tools to productivity software so that users can benefit from automatically generated infographics while working on presentations and reports. Once the system detects a statement that can be enhanced with an infographic, a message will pop up asking the user if they want to use the recommendation chart. In this way, a wider range of users can be reached \cite{cui2019text}.}
%Fourth, add interactions to infographics. Most of the existing infographics are presented statically. Interaction can be added to the infographic to create interaction, allowing users to get more information and increase their interest.

\section{Timeline \& Storyline}

\begin{figure*}[htp]%[hbt!]
\setlength{\belowcaptionskip}{-0.1cm}
    \centering
    \includegraphics[width=16.5cm]{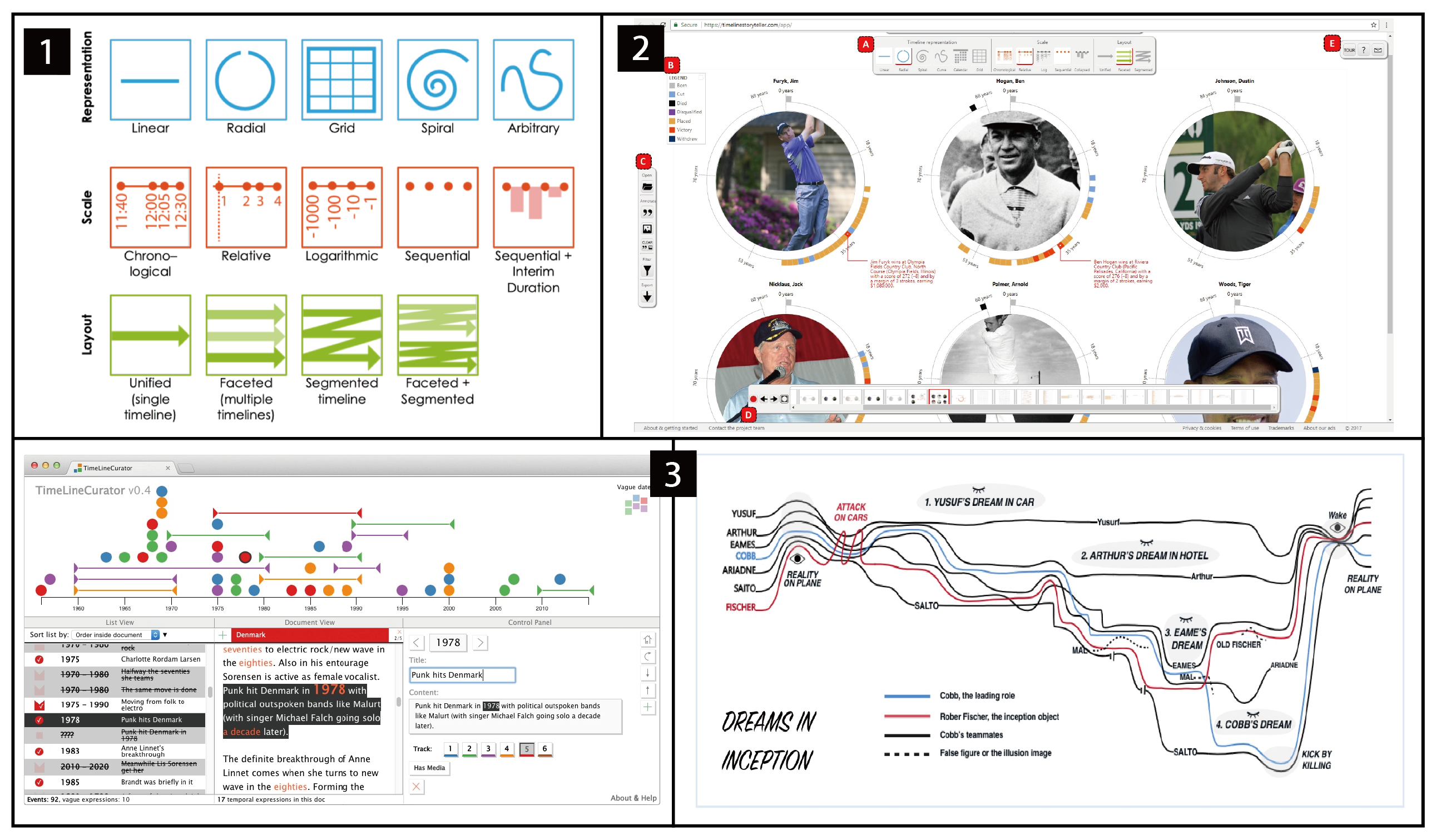}
    \caption{Selected examples of timeline's design spaces and tools. (1) Design space: Brehmer et al. \cite{brehmer2016timelines} proposed that storytelling with a timeline encompasses three levels of design space: representation, scale, and layout. (2) Authoring tool: Timeline Storyteller's \cite{brehmer2019timeline} working viewport, where the timeline canvas spans the entire browser window. (3) ML/AI-supported tool — Left: the working window of TimeLineCurator \cite{fulda2015timelinecurator}, a browser-based authoring tool. The diagram depicts a chronology of Scandinavian pop music, with each hue denoting a different nation. Right: Example of a storyline visualization created using PlotThread \cite{tang2020plotthread}. The layouts are developed collaboratively by AI agents and designers, while styles and visual labels are manually modified to enhance the narrative.}
    %\vspace{-2mm}
   % \label{fig:figures2.11} %交叉引用的标签
\end{figure*}

\begin{wrapfigure}{l}{0.055\textwidth}
    \includegraphics[width=0.055\textwidth]{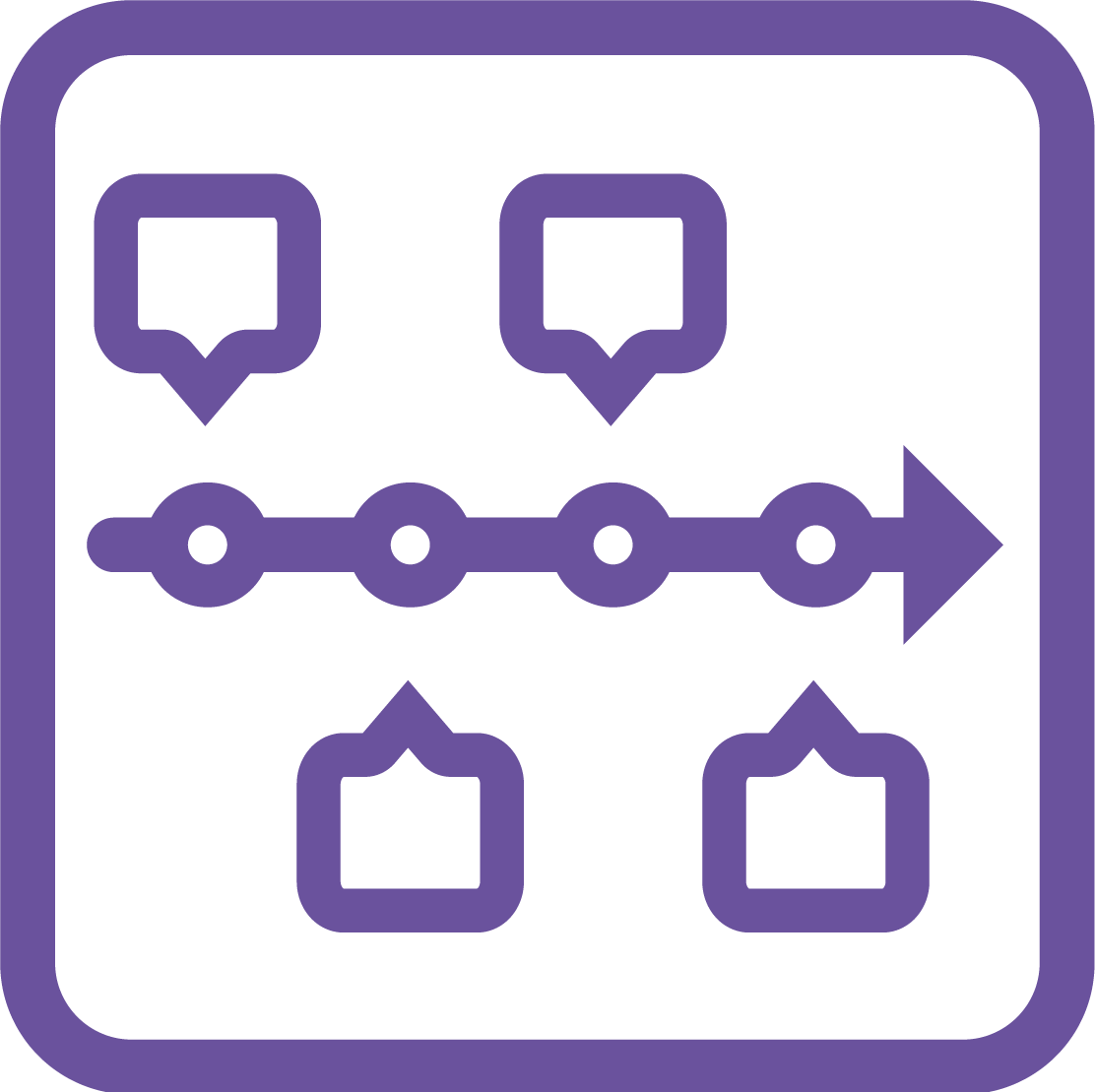}
\end{wrapfigure} 

%概念
Both Timeline and Storyline are used to describe sequences of events \cite{brehmer2016timelines}. %Aigner et al. \cite{aigner2011visualization} defines interval event data as distinguished from timely or continuous quantitative time-series data.
The most typical timeline has events arranged horizontally according to their timestamps and a horizontal axis used to represent time progression from left to right \cite{nguyen2016timesets}. %Simple timeline will elaborate on the type of event, the number of events, and the order in which they occurred; a detailed timeline will indicate when, how long they lasted, and whether the events overlapped \cite{brehmer2016timelines}. 
%A storyline is an effective means of presenting plot development and character interaction \cite{tang2020plotthread}. 
In a storyline visualization, the narrative unfolds from left to right; each person is represented as a line. When two people interact at the exact moment, their two lines intersect \cite{tang2018istoryline, di2020storyline}. As their presentations share many resemblances, timelines and storylines are jointly discussed in this section.

{\textbf{Design space:}}
Brehmer et al. \cite{brehmer2016timelines} proposed that storytelling with timelines contains three levels of representation (e.g., linear, radial, and grid), scale (e.g., relative and logarithmic), and layout (e.g., unified and faceted). Moreover, by combining these three levels, 20 timeline design options were identified to match the narrative style. Lan et al. \cite{lan2021understanding} identified six narrative sequencing patterns (chronology, trace-back, trailer, recurrence, halfway-back and anchor). The study results showed that nonlinear narratives are more likely to increase user engagement and that nonlinear narratives enable stories to be more expressive without hindering comprehension. Bach et al. \cite{bach2015time} proposed the concept of time curves for nonlinear narrative visualization. %The time curve technique is a self-similarity-based method for visualizing temporal data that employs multi dimensional scaling (MDS), which allows embedding points in time in a two-dimensional space. 
The aims of their work were to provide a general method for producing straightforward visual summaries for a variety of temporal datasets. 
The researchers describe the visual patterns that time curves often display (i.e., cluster, transition, cycle, U-turn, outlier, oscillation, and alternation) and how to interpret them. Similarly, Kim et al. \cite{kim2017visualizing} suggested the use of story curves to analyze and convey nonlinear narratives in film. Story curves in this style may be used to establish the general ordering of events by comparing the order of events in a film to their actual chronological order.
%The researchers present the visual patterns (cluster, transition, cycle, U-Turn, outlier, oscillation, and alternation) that time curves often exhibit and how to interpret these patterns. Similarly, Kim et al.\cite{kim2017visualizing} proposed story curves, which can be used to explore and convey non-linear narratives in films. By showing the sequence of events in a film and comparing them to the actual chronological order, story curves in this form can sort out the overall ranking in which events occur.

However, storyline visualization is usually limited in that participants cannot belong to two different groups simultaneously. %\rv{That is, a line can only be used to represent one time dimension. If a person is represented as a line, then based on the properties of the line, the person can only exist in one time dimension. For example, if the line represents "reading" and the time dimension is one year, then the person can only read and cannot do any other activity during the time dimension of one year.} 
\rv{As a participant is represented as a line, multiple lines bundled together at a time point usually indicate that they belong to the same group at that time. However, when the participant belongs to different groups simultaneously, for example, in co-author relationships, 
the participant's line of thinking is difficult to align with that of the co-authors.}
%This constraint is a severe problem for some application scenarios, for example, when the time point of the storyline corresponds to a relatively long time interval (e.g., one year), which leads to the fact that the participant can only perform one thing a year in the timeline.%改这一段话
To solve this problem, Di Giacomo et al. \cite{di2020storyline} proposed a model that aims to present participants with a tree diagram rather than a line diagram. In addition, several researchers have proposed a series of design guidelines regarding the timelines' aesthetics and readability, which can be roughly divided into three categories:
(1) attempt to keep straight lines to minimize line crossings \cite{tanahashi2012design, gronemann2016crossing, tang2018istoryline},
(2) the same set of lines should appear next to each other, and
(3) a certain distance should exist between lines \cite{tang2018istoryline}.
These design \rv{guidelines} are proposed to provide a theoretical basis for creating timelines, which can be used to guide users to better create timelines in authoring tools.

{\textbf{Authoring tool:}}
Creating timelines can be a time-consuming task for novices; consequently, researchers have developed several authoring tools for creating timelines \cite{webalon2011tiki, dukes2010dipity, TimelineJS2013, TimelineSetter2011}. % Examples include Tiki-Toki (interactive multimedia timeline creation application) \cite{webalon2011tiki}, Dipity  (online timeline application service) \cite{dukes2010dipity}, TimelineJS \cite{TimelineJS2013}, and TimelineSetter~ \cite{TimelineSetter2011}. 
Two of the most commonly used tools are TimelineJS~\cite{TimelineJS2013} and TimelineSetter~\cite{TimelineSetter2011}. Both tools can automatically generate a visual timeline by filling in dates and titles, describing events in Google spreadsheets, and linking to corresponding images, videos, and other media. The generated timeline can also be demonstrated in the form of slides~\cite{TimelineJS2013, TimelineSetter2011}. 
\rv{Although these tools are increasingly popular, they lack certain capabilities. For instance, they cannot generate timelines for nonlinear storylines or more complicated timeline layouts. On the basis of Genette's \cite{genette1983narrative} research on the sequence of events in a story, Kim et al. \cite{kim2017visualizing} built Story Explorer, a tool that allows users to organize the chronology of scenes in a movie script and to utilize story curves to explore the film’s nonlinear narrative. } 

\rv{However, several of the abovementioned tools can only create linear timelines. Before Timeline Storyteller\cite{brehmer2019timeline} was developed, designers who wanted to convey expressive stories by using special timeline layouts (matrices, spirals, etc.) usually applied time-consuming manual approaches or programming implementations. However, timelines created in using this method often lacked guidance in balancing the perception and narrative effects, resulting in being difficult to understand~\cite{kashan2012Timeline}.
To solve this problem, Brehmer et al. proposed a timeline design space \cite{brehmer2016timelines} and further developed tools~\cite{brehmer2019timeline} %\rb{\st{from three dimensions: representation (radial, grid, etc.), scale (chronological, relative, etc.) and layout (unified, faceted, etc.) to guide the creation of timelines, which}} 
that would easily allow users to create nonlinear forms of timelines.}

%The story curve is the core of the Story Explorer \cite{kim2017visualizing}, which visualizes events (scenes) as points in a two-dimensional diagram based on their order in the narrative (horizontally, left to right) and their chronological order in the story (vertically, top to bottom). Story Explorer \cite{kim2017visualizing} can also encode additional story information, such as characters (different line colors represent different characters and the thickness of the line segments represents the number of characters), locations (bands surrounding the characters), and periods of the day (represented using the vertical background in the background). 
%Brehmer et al.\cite{brehmer2019timeline} constructed Timeline Storyteller, a tool for creating event sequence data based on three levels of design space\rv{:} representation, scale, and layout, which can be used to create some unique timeline layouts (matrix, spiral, etc).

Although these authoring tools have lowered the threshold for users to create timelines, several challenges at the layout and visual encoding level still need to be addressed. For example, \rv{when designers need to finish hundreds or thousands of timelines, it becomes difficult to meet both the aesthetics and readability principles of the timeline design. It is also time-consuming and technically difficult for designers to manually adjust the layout to avoid line crossings and overlaps.}
%these tools cannot provide a clear result when designers need to design for hundreds of timelines \cite{liu2013storyflow}. And the large number of lines generated can cause visual inconsistencies. 

{\textbf{ML/AI-supported tool:}}
{Some ML/AI-supported tools in the timeline visualization domain solve the abovementioned problems. TimeSets \cite{nguyen2016timesets} uses the ``gestalt principles'' of proximity and uniformity of association to group together the relevant events and the use of backdrop colors to visually link collections' activities. %It shows how events in a collection change over time and how they relate to other collections. 
The tool addresses the visual inconsistency caused by too many lines. 
StoryFlow \cite{liu2013storyflow} uses a new hybrid optimization strategy that combines discrete (sorting and aligning line entities to create the initial layout) and continuous (optimizing the layout based on convex quadratic optimization) optimization methods to quickly create timelines with aesthetic and readable properties. %Efficient algorithms support rich real-time interactions (e.g., bundling, deletion, straightening), and users can edit story layouts.
However, this approach is insufficient in effectively supporting advanced design preferences, such as changing the general trend of lines~\cite{tang2018istoryline}. 
Tang et al. \cite{tang2018istoryline} created iStoryline to create more meaningful storyline visualizations that satisfy the needs of designers. This tool integrates user interactions into an optimization algorithm that allows users to easily create story visualizations by modifying the automatically generated layouts according to their preferences. %Moreover, the tool effectively incorporates hand-drawn effects. Studies have also shown that automatically generated storylines cannot match those with hand-drawn effects in terms of expressiveness. The automatically generated approach cannot cover rich narrative elements, including plot, tone, etc. 

While iStoryline's \cite{tang2018istoryline} interactions focus on modifying local areas, customizing the overall layout is time-consuming and the optimization process is unpredictable, which requires repeated trials to optimize the results. To improve the user experience, PlotThread \cite{tang2020plotthread} integrates AI agents into the authoring process. The AI agent can decompose a given storyline into a series of segments, allowing the user to understand the state of the intermediate layout and predict the following action. % through which a collaborative design of the timeline can be facilitated. 
In addition, Ellipsis \cite{satyanarayan2014authoring} and TimelineCurator \cite{fulda2015timelinecurator} are both timeline authoring tools focused on the field of journalism. Ellipsis\cite{satyanarayan2014authoring} blends a domain-specific language for narrative development with a graphical user interface framework. TimelineCurator \cite{fulda2015timelinecurator} can process unstructured documents with temporal text by using natural language and subsequently extract the temporal text from them along the way. These tools significantly facilitate the management and processing of documents containing timelines.}

\textbf{Summary:} 
\textls[-5]{\rv{Timelines and storylines are used to depict event progressions. Researchers focus on timeline aesthetics and narrative impact in timeline \& storyline design. Users can manually design timelines for particular scenes (i.e., movie narration) or use authoring tools (i.e., matrices and spirals). ML/AI-supported tools leverage intelligent algorithms to assist users in creating narratives by sorting temporal sequences and text information from unstructured raw data. It also enhances the aesthetics and usability of timelines and \rb{makes} writing tools more efficient. We observed that a significant amount of text information can be easily processed using ML/AI-supported tools, whereas a limited amount of text and a particular type of timeline can be created using authoring tools.}}
%Timelines and storylines can visually show the occurrence and progress of events. At the design space level of timeline \& storyline, researchers pay more attention to timeline aesthetics and narrative effect. Authoring tools allow users to interact directly with the interface, and users can manually create timelines for specific scenes (such as movie narration) and specific forms (such as matrices and spirals). ML/AI- Supported Tools combines the AI algorithm to provide some suggestions in the process of creating storylines, such as helping users sort out time text and reducing the time wasted by users repeatedly exploring unstructured documents. It not only has both aesthetic effect and readability, but also improves the efficiency of creating a timeline, which is an excellent improvement for authoring tools. Based on the above analysis, We found that information with a large amount of time text can be created through AI Supported Tools, while for a small amount of text and a special form of timeline needs to be created, it can be created using the authoring tool.

\rv{Following the research directions indicated by the existing studies, we believe the following directions may be studied in the future.
The first direction is to explore the need for special forms of timelines. Although Brehmer et al. \cite{brehmer2016timelines, brehmer2019timeline} proposed six forms (i.e., linear, radial, spiral, curved, calendar, and grid) of timeline representation, \rb{their study} mainly focused on two forms, linear and radial. Moreover, the representations of these particular timelines determined by the researchers have not been verified in terms of user acceptance and communication effectiveness. Future work needs to validate these \rb{representations} via formal experiments and implement more real-world applications of such new forms \rb{of} timelines.
%Brehmer et al.\cite{brehmer2016timelines, brehmer2019timeline} proposed six forms of timeline representation: linear, radial, spiral, curved, calendar, and grid. However, current research has focused on two forms: linear and radial. The application scenarios of other forms and their acceptance by users also need further study.  %虽然有一些这样的工作，但是在实际应用中的用户接受程度/传播有效性还没有被证实，已有的工作都缺乏正规的测试和验证，这些理论和验证都需要在实践中测试。
The second direction is that the existing authoring tools often overlap timelines when creating content with multiple temporal texts, and the subjective merging of timelines for aesthetic reasons results in the loss of information. In the future, we also need to strengthen the research in this area, ensuring the integrity of information while achieving the aesthetic goal. \rb{In the realm of timelines and storylines, ML/AI generator tools are still in their developmental stages. While current ML/AI-supported tools can assist users in creating timelines, they are primarily utilized for localized adjustments and fall short in terms of fulfilling the demands of the complete content creation process. The future holds immense potential for the research and development of advanced ML/AI generator tools for timelines.}}

\section{Data Comics} %20210325版本

\begin{figure*}[htp]%[hbt!]
\setlength{\belowcaptionskip}{-0.1cm}
    \centering
    \includegraphics[width=16.5cm]{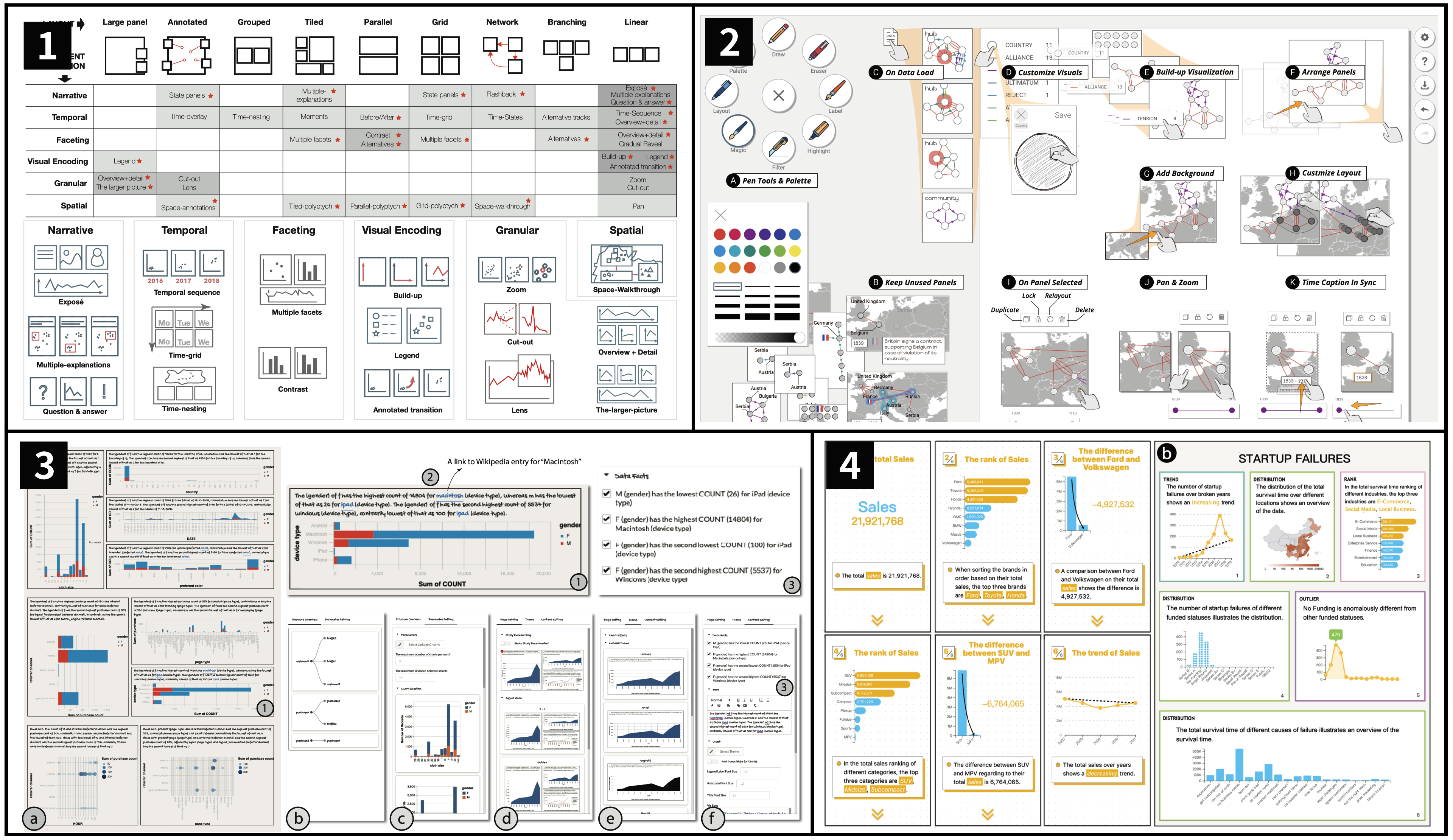}
    \caption{Selected examples of data comics' design spaces and tools. (1) Design space for data comics design patterns and illustrations of some examples in data comics~\cite{bach2018design}. (2) Authoring tool: DataToon's working viewport, which can create dynamic web data cartoons through pen-touch interaction~\cite{kim2019datatoon}. (3) ML/AI-supported tool: Chartstory's working viewport, which automates the analysis, layout, and creation of captions for data comics that tell tales using data \cite{zhao2021chartstory}. (4) ML/AI-generator tool: Calliope \cite{shi2020calliope} automatically generates visual data tales from spreadsheets and includes a story generator and editor.}
   % \label{fig:figures2.11} %交叉引用的标签
   %\vspace{-2mm}
\end{figure*}

\begin{wrapfigure}{l}{0.055\textwidth}
    \includegraphics[width=0.055\textwidth]{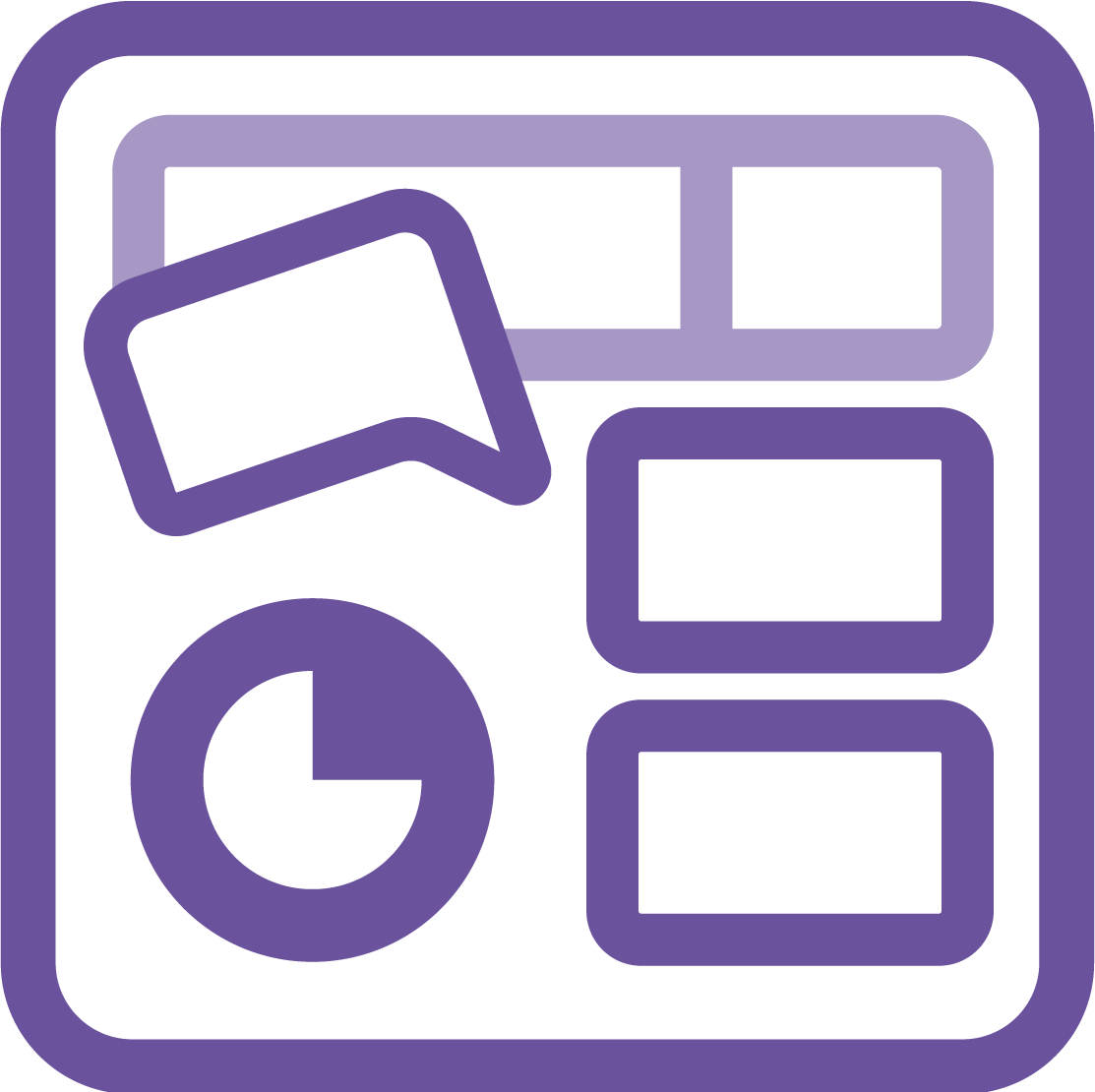}
\end{wrapfigure} 

%数据漫画设计模式的设计空间以及数据漫画中一些示例的插图

%概念
Data comics are an emerging form of narrative visualization \cite{segel2010narrative} that focuses on the variation of data information and the visual presentation of data sequences \cite{bach2017emerging}. Different from traditional comics, data comics must contain data-driven content, allowing multiple visualizations to be juxtaposed in a single panel in a comic strip layout, with annotations and visual decorations \cite{zhao2015data}. \rv{Data comics complement the linearity of narratives that are inherently imposed by movies and live presentations while offering the flexibility of two-dimensional spatial arrangements in infographics and annotated charts~\cite{bach2018design}.}

{\textbf{Design space:}}
Comics are a static format that is great for ideation and storyboards \cite{wang2019teaching}. Given that the technical barriers are low, comic creation can be shared and distributed in various formats, such as scientific papers, conference posters, slideshows, blogs, etc. \rv{The sequential nature of data comics and the tight integration of text and graphical information have great potential to explain complex data and to promote visualization and data literacy \cite{wang2019comparing}.}
%The sequential nature of data comics and the close combination of textual and graphic information allows for the successful implementation of detailed descriptions of different types of visualizations. both for simple charts (bar charts, line charts) and relatively complex visualizations (parallel coordinate charts, tree charts, etc.), which can be understood and used by the audience using the data comics format.Data comics have great potential to explain complex data and promote visualization and data literacy \cite{wang2019comparing}. 
Data comics have the potential to transform the manner we envision and produce infographics and presentations because they can convert storytelling approaches from one medium to another \cite{bach2018design}. Furthermore, data comics are incredibly flexible and communicative. They can be used to integrate graphic elements of comic properties with textual explanations and deliver visual content that requires memorization and quick navigation \cite{wang2020data, zhao2019understanding}. 

\rb{Although data comics have many advantages, creating good data comics is a complex task. Designers must consider many tradeoffs, such as balancing repetition and highlighting, and the results rely significantly on the expertise of designers.}
%Although data comics have many advantages, researchers have suggested that data comics have potential problems such as excessive content repetition and view sequencing due to subjective feedback. 
%\rb{\st{Moreover, due to their leisure and entertaining nature, data comics are rarely applied to sensitive or serious topics and scenarios.}} %Excessive visual repetition creates redundancy, which makes the picture look confusing and distracts the viewer. 
Zhao et al.~\cite{zhao2019understanding} addressed the issue of data comics view ordering by examining the narrative mechanism of comic strips. The order of the data comic panels must be shown to help recall details in data comics. %The whole order is more important than the order of a small range(i.e., two parallel thematic panels can be changed without affecting the whole storyline). 
%We propose leveraging data comics as study reports to provide an open and glanceable view of studies by tightly integrating text and images, illustrating design decisions and key insights visually, resulting in visual narratives that can be compelling to non-scientists and researchers alike. 
%Wang et al. \cite{wang2020data} proposed design guidelines for problematic points such as over-repetition of data comic content and view orders, balancing repetitive content with emphasized information and balancing sequences with overviews.
\rv{Wang et al. \cite{wang2020data} conducted a user study to compare data comics and infographics in terms of the degree of clarity of reading order and the degree of integration of text and images. The findings demonstrated that complicated spatiotemporal data are difficult to depict using infographics, while it is possible to present with data comics. \rb{The participants enjoyed reading data comics in the experiment and regarded them as more entertaining and more effective at retaining their attention.}}

%The participants enjoyed reading data comics in the experiment scenario and regarded such data comics are more entertaining and easier to keep user attention.} 

% Meanwhile, the researchers proposed corresponding design guidelines for the problematic points of data comics, such as balancing repetitive content with emphasized information, and sequences with overviews.
\rv{In another study, Zhao et al.~\cite{zhao2015data} compared data comics with PowerPoint~\cite{powerpoint2016}. The results also showed that data comics are more attractive, more space-efficient, and more enjoyable to use than PowerPoint \cite{powerpoint2016}.}
%Wang et al. \cite{wang2020data} proposed using data comics as study reports. They compared the differences between infographics and data comics, where data comics are more accessible to present complex spatio-temporal data than infographics. The study also shows that users are more likely to enjoy reading data comics and are more likely to maintain their attention. % 审查委员指出：这句话的依据是什么？
% 整理这部分的内容
Moreover, as the narrative style of comics is usually linear, a possible approach is to transform data comics into data videos with appropriate tools. Meanwhile, comics can present specific moments in separate frames, allowing for a more focused presentation of individual data information \cite{zhao2019understanding}.

% The core components of data comics lie in data, visualization, and storytelling \cite{wang2019teaching}. 

To help people comprehend the art of data, visualization, narrative, and the necessity for efficient data-based communication, Bach et al. \cite{bach2018design} offered a collection of data comic design patterns. They also constructed six design patterns for data comics according to different associations and layout methods. %The researcher argues that designing data comics requires considering two dimensions: association methods and layout methods, and constructs six design patterns for data comics based on these dimensions (narrative patterns, temporal patterns, faceting patterns, visual encoding patterns, granular patterns, spatial patterns).
Some researchers further validated the usefulness of this design space in practical cases. For example, Hasan et al. \cite{hasan2022playing} created an interactive data comic in the form of a card game. Each comic panel becomes an individual card instead of being arranged in a fixed sequence; learners can form different storylines by combining them in different ways. Their research showed that transforming data comics into card games allows learners to grasp information quickly via interaction and encourages collaborative thinking among participants.

{\textbf{Authoring tool:}}
%\rb{\st{Design space for data comics provides basic guidelines for designers to create data comics.}} 
\rb{Researchers have developed} various tools to create data comics to enhance the potential user experience. DataToon \cite{kim2019datatoon} is a tool for creating dynamic web data comics that support ``pen+touch'' interactions. The tool allows quick exploration of data, rapid generation of visual stories with custom annotations, and interactive filtering of layout templates. However, displaying exploration data and presentation information on the same page can cause visual distractions. Kang et al. \cite{kang2021toonnote} solved this problem by proposing ToonNote. ToonNote provides two view modes: notebook view, which adopts the format of a traditional computing notebook to conduct data analysis, and comic layout, which focuses on visual storytelling. %\rb{\st{It also enables users to switch between the two view modes, thus solving the visual distraction caused by code and other unnecessary information.}}

\rv{Suh et al.\cite{suh2022codetoon} developed CodeToon, a tool that supports the comic creation process by adopting two mechanisms. One is to facilitate the conception of code-related stories via metaphorical recommendations; the other is to generate comics from stories automatically. Both mechanisms allow users to add codes or select code examples provided by the tool, generate a story, and automatically produce comics. The tool allows users to quickly and easily create high-quality coding strips.} %(A comic strip with its associated code is known as a coding strip. To aid in the teaching and learning of programming concepts, languages, and practices, it offers programming concepts in both real and abstract contexts and representations.) 
%\rb{\st{that convey a salient connection between codes and comics. But such tools tool still require users to have some coding skills.}}}
To enhance the user experience of data comics, Wang et al. \cite{wang2021interactive} proposed a lightweight declarative scripting language, Comic Script, which \rb{supports adding interactivity to static comics.} %to support users in personalizing their operations in static comics. %Users can add or remove panels from a predefined panel layout to support branching, change perspective, get details on demand, and provide the ability to modify or interact with data. 
% This approach breaks away from the original model of only being able to create linear or immutable story narratives to enable non-linear narratives for personalized layouts, exploring levels of detail, and other potential user experiences. 
\rv{Their work allowed them to overcome the original narrative mode, which can only create linear or unchangeable stories, realize nonlinear narratives, offer more personalized layouts, and explore the level of detail and other potential user experiences.}

{\textbf{ML/AI-supported tool:}}
\textls[-5]{ChartStory \cite{zhao2021chartstory} is a tool that automatically converts a collection of charts into a data comic format. It divides charts into clusters of story segments by identifying narrative segments and then reorganizing the segments to generate a story. Users can further refine the generated data comics via interaction.}%, resulting in a data comic with a comic-style format that can be used to convey a data-driven narrative.

%we present ChartStory for crafting data stories from a collection of user-created charts, using a style akin to comic panels to imply the underlying sequence and logic of data-driven narratives. Our approach is to operationalize established design principles into an advanced pipeline which characterizes charts by their properties and similarity, and recommends ways to partition, layout, and caption story pieces to serve a narrative.ChartStory also augments this pipeline with intuitive user interactions for visual refinement of generated data comics
{\textbf{ML/AI-generator tool:}}
%The narrative is driven by the data facts and enhanced by the charts \cite{zhao2021chartstory}. 
\rv{Fact sheets present multiple data facts via visualization in a juxtaposed format that is highly similar to data comics. In a fact sheet, a data story is constructed from several facts and numerical or statistical findings produced from data \cite{wang2019datashot}. Although some comic elements are missing in fact sheets, we still categorize them in this category because \rb{they} can be easily extended to data comics by adding some comic-style decorations.}
% The factsheet combines multiple data facts in a visual and informational graphic format to tell a data story. 
%Its layout form is similar and can be easily extended to data comics; however, the decoration element is missing in the factsheet. We consider the factsheet as a form of data comics in this survey. 
%\rv{Its layout form is similar to data comics. Although the factsheet is missing the decoration element, it can be extended in data comics, so the factsheet is considered a form of data comics.}
Both DataShot~\cite{wang2019datashot} and Calliope \cite{shi2020calliope} can automatically generate fact sheets. DataShot~\cite{wang2019datashot} transforms tabular data into fact sheets  by adopting a three-step process of fact extraction, fact combination, and visual synthesis. \rv{This tool can effectively reduce the difficulty of data exploration, create information presentations and enhance the readability of data by means of expressive visual design.} 
This method is extended by Calliope~\cite{shi2020calliope} to automatically construct visual data stories from spreadsheets and use the Monte-Carlo-tree search technique to investigate tale fragments and portray them in a logical manner.
\rv{Calliope can generate coherent visual data stories in which logical connections can be maintained between the preceding and following segments. In this manner, the threshold for creating data stories can be effectively decreased.}

\textbf{Summary:} %\rv{Data comics combine the accessible layout of infographics in two dimensions with the linear ordering of videos and presentations and can be used to tell data stories.} 
Although in its infancy now, data comics have gained much attention in recent years. % It can be combined with different visualization types in a more engaging way to help users explain complex data and facilitate visual data exploration \cite{bach2018design}.
\rv{According to some preliminary studies~\cite{zhao2015data}, data comics perform better than slideshows and infographics in terms of spatial efficiency and reader enjoyment. However, a more detailed evaluation with a larger number of participants needs to be conducted to validate its usage and effectiveness in practice. \rb{Moreover, while data comics possess a leisurely and entertaining nature, they are occasionally applied in serious and sensitive contexts. For instance, at Charité in Berlin, comics are utilized on a regular basis to educate patients before heart surgery, showcasing the practical applicability of this medium beyond mere research settings \cite{brand2019medical}.}}

%Moreover, due to their casual and entertaining nature, data comics are rarely applied to sensitive or serious topics and scenarios.

\rv{Almost all the existing tools for creating data comics support basic data exploration and analysis. While authoring tools can reduce the difficulty of creating data comics, they are targeted at users who have a certain level of visualization creation skills, which is not user-friendly to amateurs who want to create data comics from scratch.
ML/AI-supported tools and ML/AI-generator tools for creating data comics integrate the ability to analyze data, visualize the analyzed content, and present the information in a narrative format. The difference between the two types of tools is that ML/AI-generator tools can automatically analyze data and arrange the data insights into comic-style narratives directly. By contrast, ML/AI-supported tools require users to select valuable insights or manually layout the panels of data comics.}
%Combined with the summary above, we see potential research directions for data comics in the following aspects.
%First, there will be much visual repetition in the data comic. The redundancy created will create confusion and distraction for the audience, so it is easiest for the audience to accept the number of comic values and the number of characters of text in the data comic \cite{wang2019comparing}.

%We believe that data comics can be widely used in the future. There are many opportunities for research and development in this genre. 
\rv{Reflecting on the collected work presented above, we think that the data comics can be studied in the following aspects in the future.}
\rv{First, the forms of comics vary to a large degree, and current research has ignored how different data types are suitable for which kind of design style and narrative strategy  \cite{wang2019comparing} and which style of data comics users prefer under what circumstances.} % For example, The position of the comic on the screen, the size of the images and text, the font style, the colors, etc. need more research on how to make it more acceptable to users and which type of users it is more suitable for.
%First, there are diverse forms of comics, various design styles and narrative strategies need to be studied\cite{wang2019comparing}. 
\rv{Second, the redundant and non-data related visual elements in data comics can sometimes be confusing and distracting to viewers, imploring the necessity to explore how the number of comics grids, the amount of text, the layout, and the color scheme can be designed to be more acceptable by users.}
Third, although \rv{data comics are a static medium for presenting data, a possible approach is to investigate how to include interactive features to promote user understanding and engagement \cite{zhao2019understanding}.}
%Third, since there will be a lot of visual repetition in the data comic, the resulting redundancy will create confusion and distract the viewers, so it is also necessary to explore the number value of comics and the number of characters of text in the data comic are more acceptable to the users\cite{wang2019comparing}. 

%Fourth, more research is needed on the position of comics in the design space, the size of images and text on the page, the style of fonts and colors, etc., to make them more acceptable to the audience and more suitable for them. 
%Forth, more research is needed on the location of the cartoon in the design space, the size of the picture and text on the page, the style and color of the font, and how to make it easier for the audience to accept and suit.

\section{Scrollytelling \& Slideshow} %20220325版本

\begin{figure*}[htp]%[hbt!]
\setlength{\belowcaptionskip}{-0.1cm}
    \centering
    \includegraphics[width=16.5cm]{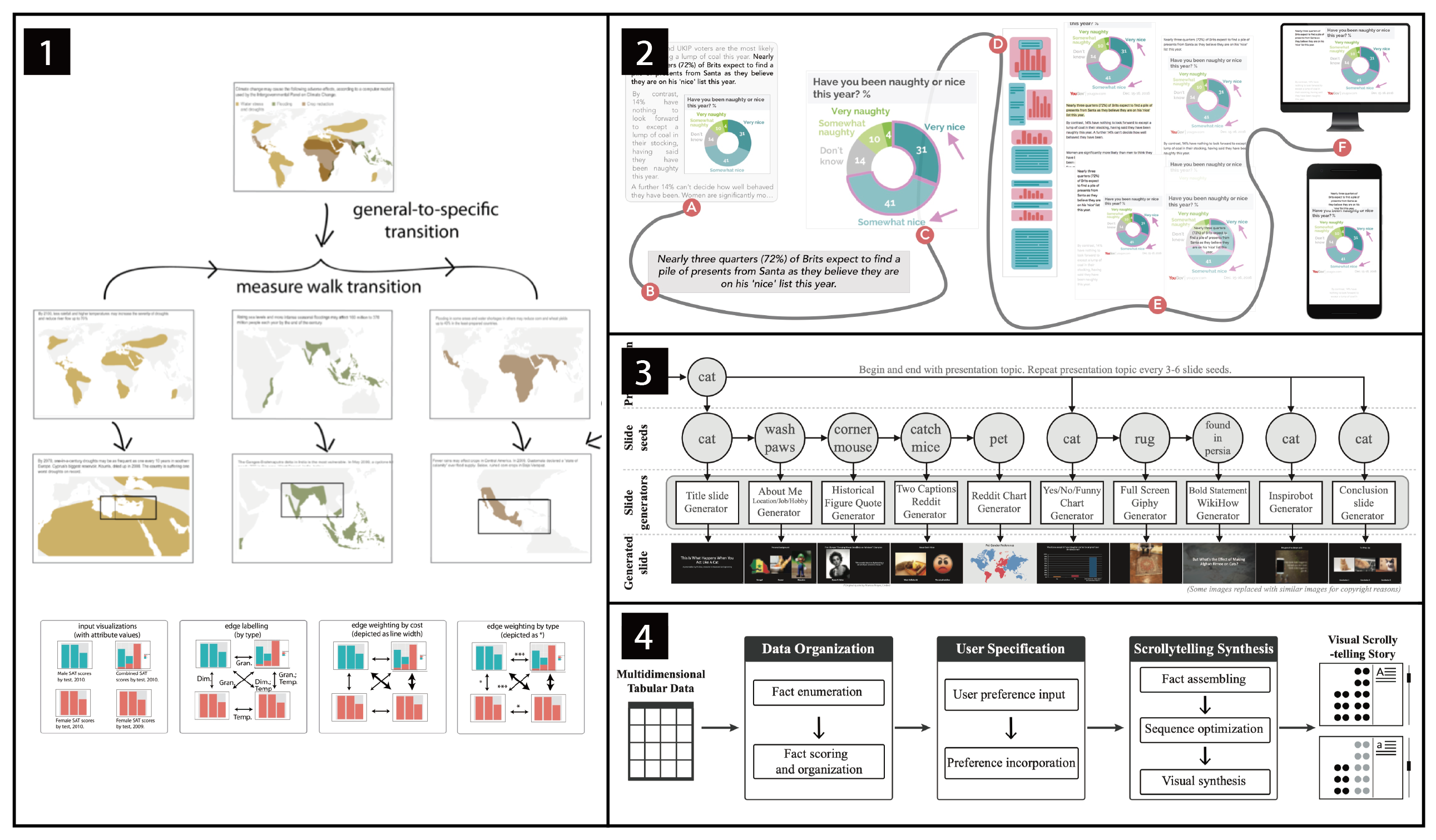}
    \caption{Selected examples of scrollytelling design spaces and tools. (1) Design space: Outlines how to use automatic sequencing in design systems to guide non-designers in making structured decisions when creating narrative visualizations~\cite{hullman2013deeper}.
    %(a) New York Times used a slideshow to cover the climate in Copenhagen, describing a complex subject through simplicity. (b)visualizations depict nodes in a graph-based method . %Edges (potential transitions) are identified and weighted according to user preferences using a cost function and type weightings (denoted by * symbols). 
    (2) Authoring tool: uses text-chart links to transform static data-driven articles containing text and charts into dynamic content~\cite{sultanum2021leveraging}. (3) ML/AI-supported tool: Tedric system workflow, which can be used to train presentation skills, reduce barriers to impromptu speaking and generate slideshow based on audience suggestions~\cite{winters2019automatically}. (4) ML/AI-generator tool: A method for automatically generating scrollytelling visualizations~\cite{lu2021automatic}.} %First, data facts are extracted from a multidimensional tabular dataset, then the system accepts, interprets, and integrates the input information, and finally, the facts are assembled and sorted to generate stories considering the factual connections and the scrolling design pattern 
   % \label{fig:figures2.11} %交叉引用的标签
   %\vspace{-2mm}
\end{figure*}

\begin{wrapfigure}{l}{0.055\textwidth}
    \includegraphics[width=0.055\textwidth]{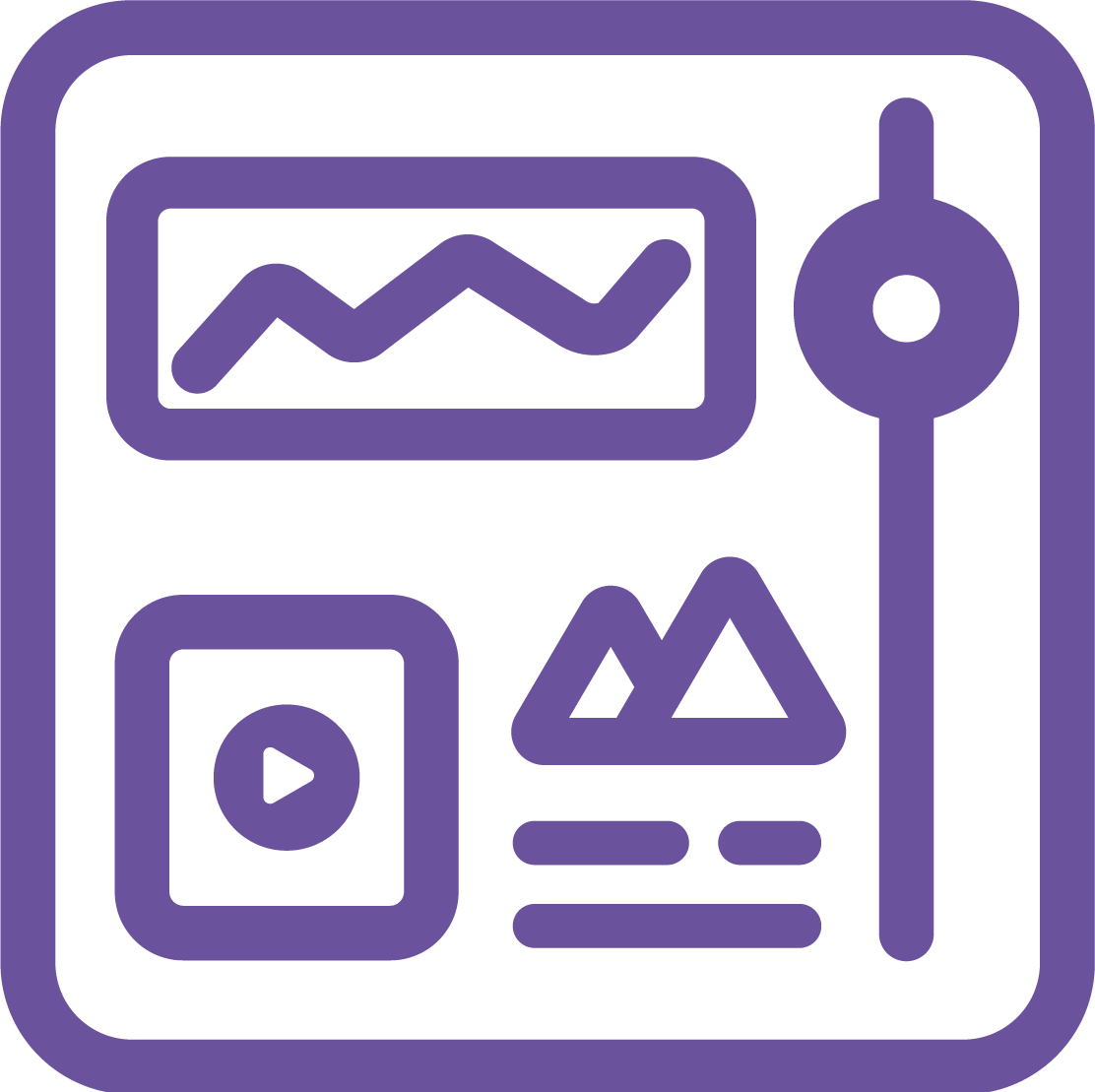}
\end{wrapfigure} 

%概念
The term ``scrollytelling'' is a combination of ``storytelling'' and ``scrolling.'' It is a scrolling-based visual narrative form that is widely used in data-driven articles \cite{seyser2018scrollytelling}. Scrollytelling articles usually start with a full-screen photo or video and scroll by considering the next part of the content \cite{seyser2018scrollytelling}. A similar form of visual narrative to scrollytelling is the slideshow~\cite{roth2021cartographic, mckenna2014design}. % a digital image display technique that takes the viewer through the stages of the story one by one.
% Slideshow is one of the seven genres of narrative visualization according to Segal and Heer's classification ~\cite{segel2010narrative}. The scrollytelling narrative is formed by scrolling while the slideshow narrative is triggered by clicking, pressing, or sliding~\cite{roth2021cartographic}.
\rv{Mckenna et al.\cite{mckenna2017visual} noted that many recent websites integrate buttons and sliders, demonstrating that the distinction between the stepper and the scroller depends on \rb{whether the user input is clicking the stepper or scrolling the slider.}
In addition, in terms of story layout, pages often appear as slideshows or hybrids\rb{that combine features of both slideshows and steppers, with different animations and scrolling. They resemble both steppers and scrollers,} in which the latter form supports scrollytelling. As the slideshow form and the hybrid form can be interconverted, we jointly studied scrollytelling and slideshow.}

{\textbf{Design space:}}
% \rb{\st{The advantage of scrollytelling is that it narrates the story in a linear scrolling fashion, and its interactive format fits well with the way we interact with computers. This format helps the user present the story fluidly}} \cite{seyser2018scrollytelling,lu2021automatic}.
Scrollytelling articles are usually text-centered and use multimedia elements such as images and videos to assist narrative storytelling~\cite{godulla2017digitale}.
% Scrollytelling articles usually start with a photo or video as wide as the screen, and the user moves to the next element by scrolling (usually vertical) \cite{lu2021automatic}. Scrollytelling articles are usually text-centric. Text is the main constituent element and an essential part of connecting other multimedia elements (pictures, videos, etc.) \cite{godulla2017digitale}. In image-centric scrollytelling content, the text is a necessary complement to photos, graphics, and animations. In text-centric articles, multimedia elements appear either on the left or right side of the screen page or in a full-screen form in paragraphs. Multimedia elements interweave background information about the core content with the text, supporting the narrative structure \cite{seyser2018scrollytelling}.
Various transition styles between pages can be triggered by scrolling. The choice of transition styles is usually determined by the relationship between facts (e.g., comparative, similar, and sequential). 
\rv{Scrollytelling can be used as visual cues, such as highlighting facts in visualization, to direct attention or to indicate stages to assist browsing~\cite{lu2021automatic}.}
%If non-related data items or completely different fact types are presented sequentially, the transition is often abrupt. The scrolling interaction can be used as a step to help navigate the story, to present the story fluidly, or as a trigger to direct the reader's attention. Other interactions such as clicking cannot serve the above goals because they can cause the reader to feel interrupted while reading the story \cite{lu2021automatic}. 

%\rv{A slideshow is composed of a collection of slide similar to scrollytelling, represents information in a structured layout, while a slideshow is composed of a collection of slides.}
A slideshow is composed of a collection of slides instead of continuous content in scrollytelling. %Different media types can be displayed, such as images, tables, audio, video, text, etc \cite{elias2018towards}.
Elias et al. \cite{elias2018towards} reviewed the elements that comprise a slideshow presentation, identifying six typical elements: slide title, text box, image, embedded content, equations, and tables to ensure accessibility. Hullman et al. \cite{hullman2013deeper} analyzed 42 narrative visualizations in the form of slides and investigated how the choice of order affects narrative visualization. 
For slideshows, the narrative is told by discrete clicking, tapping, keying, or swiping dynamic slideshows, allowing the designer to control the storytelling pace. In addition, users can add or remove pages to the slideshows according to their needs and can exit the presentation page at any time. Slide layouts can show continuous progress between slides or support nonlinear breaks in the narrative~\cite{roth2021cartographic}. %Usually, the number of pages is best from 6 to 19. 
However, when readers have to navigate too many pages, they may eventually suffer from boredom, but too few pages also hinder the user from remembering the story. Therefore, the story’s length in the slides must be accurately established \cite{lu2021automatic}. 

{\textbf{Authoring tool:}}
Scrollytelling is a challenging task. Idyll \cite{conlen2018idyll} provides a ``scroller'' component for building scrolling narratives, allowing users to control document style, layout, and control pages by clicking or scrolling. Sultanum et al. \cite{sultanum2021leveraging} explored a data-driven approach to article story creation that separates semantic, textual, and graphical links and story layout forms. \rv{On this basis, researchers developed VizFlow \cite{sultanum2021leveraging}, a tool for creating dynamic data-driven articles. With a text-chart linking strategy, VizFlow allows users to create dynamic layouts for static data-driven articles.}

\rv{Users have more options or tools to create slideshows compared with scrollytelling. The most popular ones are PowerPoint \cite{powerpoint2016}, Keynote \cite{keynote2003} \rb{and Google Slides \cite{googleslides2006}.}}
This type of software aims at helping users manually create a set of slideshows that contain text, images, and other multimedia content. Providing abundant design templates allows users to focus on the information they want to present rather than spending plenty of time in the visual layout \cite{bocklandt2021sandslide}. 
% \rb{\st{However, different software have their own output formats, resulting in poor compatibility. Bocklandt et al. %\cite{bocklandt2021sandslide}addressed this problem by developing SandSlide, a system that automatically converts slides in PDF format into editable PowerPoint files. Sandslide adheres to the default slide templates and therefore takes full advantage of the slide editor's flexible layout capabilities}}.

% The researchers built VizFlow, a concept authoring tool for creating dynamic data-driven articles based on this approach. \rv{Using VizFlow\cite{sultanum2021leveraging} to guide different storytelling layouts, static data-driven articles (with text and supporting charts) can be transformed into dynamic articles.}

{\textbf{ML/AI-supported tool:}}
\rv{Users often employ slideshows for presentations or speeches. However, it usually takes considerable time and effort to create slideshows before the presentation, and for impromptu speech, users cannot create slideshows in such a short time. Tedric \cite{winters2019automatically} is a tool to construct a coherent slideshow from a single subject idea. This tool blends a semantic word web with text and picture data sources to produce a slideshow that matches the subject. The user studies conducted by the authors demonstrated that the use of the tool significantly reduces the barriers to impromptu speech and saves users much time.}

% Tedric \cite{winters2019automatically} is a slideshow generation system that efficiently creates slideshows on topics suggested by the audience. The system can help train presentation skills and reduce the barriers to delivering impromptu presentations, saving presenters much time.

{\textbf{ML/AI-generator tool:}}
{\rv{Leake et al.\cite{leake2020generating} developed a system that converts text into speech by recognizing specific words in each sentence and automatically selects relevant images to transform these texts into audiovisual slides.}
%These tools mentioned above are poor at supporting the automatic ordering and transition of scrollytelling, which are crucial for logic formulation, viewing experience, and understanding of scrollytelling stories \cite{hullman2013deeper}. 
Lu et al. \cite{lu2021automatic} proposed a method for automatically generating scrollytelling visualizations. The method begins by listing the data facts for a given dataset, scores the facts and arranges them into stories, and then produces visualizations, transitions, and text descriptions for the scrolling display. However, as the existing work in this category is mostly prototypes, the practical usage of ML/AI-generator tools for scrollytelling has not yet been proved.}
% This framework is similar to DataShot \cite{wang2019datashot} and Calliope \cite{shi2020calliope} that are tools to create data comics. Although the presentations of data comics and scrollytelling are not exactly the same, the generated content of data comics can be easily transformed into scrollytelling.

\textbf{Summary:}
\textls[+8]{Scrollytelling is a scrolling view of content, an interaction that is consistent with our everyday behavior of browsing web pages and articles on mobile devices. A slideshow is another common display that is a step-based display. \rv{Although we often encounter the two forms of narratives in daily practice, academic research \rb{on} slideshow and scrollytelling is generally lacking.} %Both are easily accessible to audiences as narrative visualizations. While slideshows are frequently applied in daily practice, there is a lack of research in this genre.
First, as mentioned in the timeline chapter, nonlinear narratives are more likely to engage users, and scrollytelling and slideshows can use both linear and nonlinear ways of presenting information. \rb{Scrollytelling and slideshows allow the audiences to explore different paths by referring to the content based on their own interests and needs. Instead of following a predetermined linear sequence, the audience can select their own journey by referring to the information by clicking on links, making selections, or following different branches of the narrative. This approach gives the audience more control over the pace and order of information, allowing them to focus on the aspects that are most relevant or meaningful to them.} 
Future work can investigate whether other nonlinear narrative structures are also suitable for scrollytelling or slideshow.
Second, existing research has focused on different media combinations, such as images, text, and video, with minimal research on data visualization and intelligent tools. \rv{In particular, slideshow creation tools are % \rb{\st{mainly focused on essential productivity software, which is}} 
inexplicitly designed to create narrative visualizations. Therefore, future research can investigate the real needs and design requirements for narrative visualization, thus providing more \rb{support} to create data-driven scrollytelling and slideshows.}} % \rb{\st{Third, there is a lack of research on how scrollytelling and slideshow can be seamlessly switched to different display devices, and research on this type of tool needs to be strengthened.}}

% Therefore, future studies could strengthen the research in AI-supported and AI-generator to assist users in automating the creation of data-driven scrollytelling and slideshows.

% Idyll Studio: A Structured Editor for Authoring Interactive & Data-Driven Articles\cite{conlen2021idyll}
\section{Data Video} %20210325版本

\begin{figure*}[htp]%[hbt!]
    \centering
    \includegraphics[width=16.5cm]{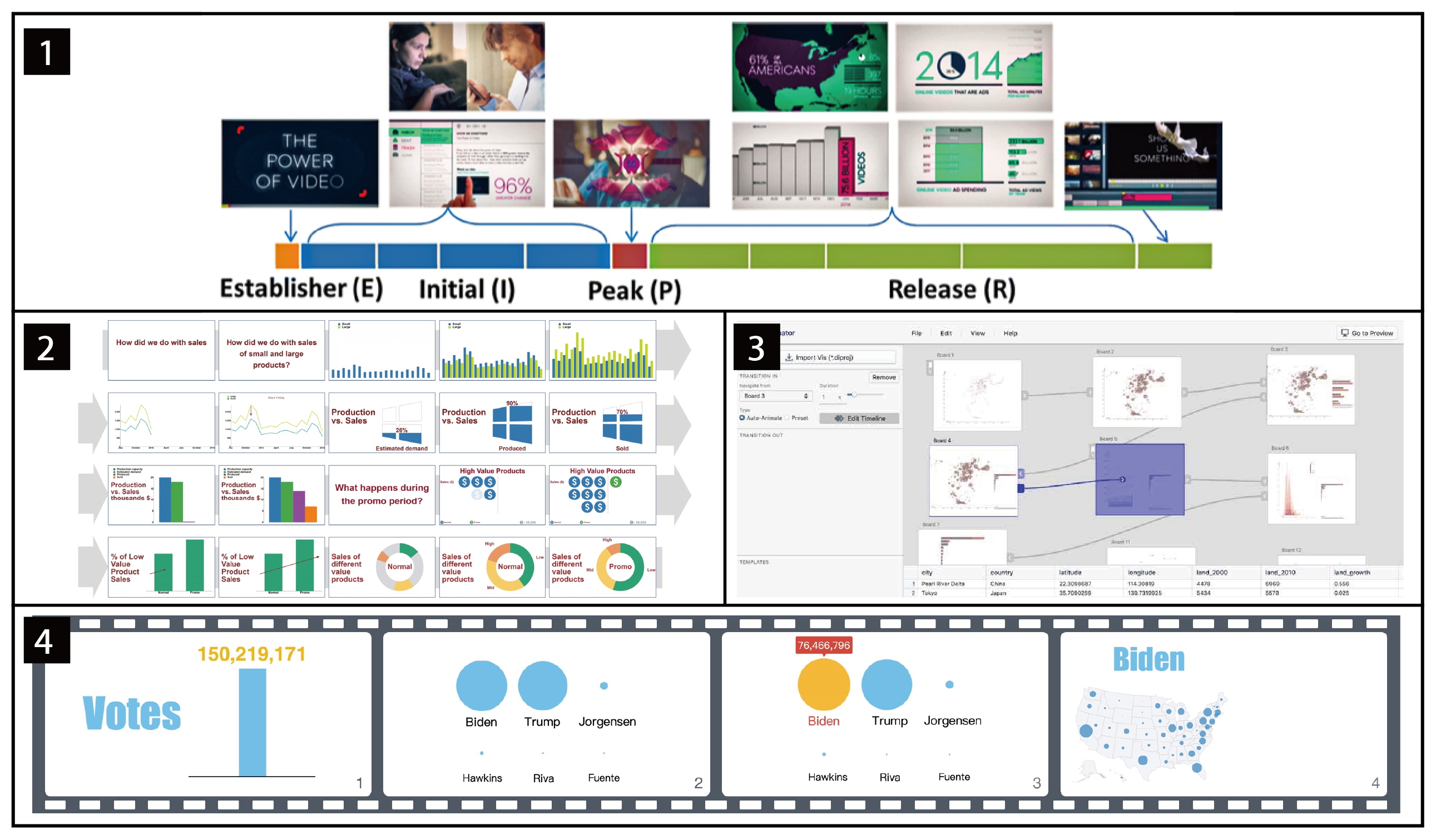}
    \caption{Selected examples of data video's design spaces and tools. (1) Design space: Amini et al. \cite{amini2015understanding} states that E+I+PR+ in data video is the most balanced narrative structure. (2) Authoring tool: example of a data-driven video generated using DataClip for financial analysis \cite{amini2016authoring}. (3) ML/AI-supported tool: Data Animator's storyboard editing work window. It is able to segment complicated animations by stacking keyframes and using data parameters to stagger the start time and modify the pace of animated objects in the timeline view~\cite{thompson2021data}. (4)  ML/AI-generator tool: Autoclips automatically generates keyframes for data video based on a series of data facts \cite{shi2021autoclips}.}
   % \label{fig:figures2.11} %交叉引用的标签
  % \vspace{-2mm}
\end{figure*}

\begin{wrapfigure}{l}{0.055\textwidth}
    \includegraphics[width=0.055\textwidth]{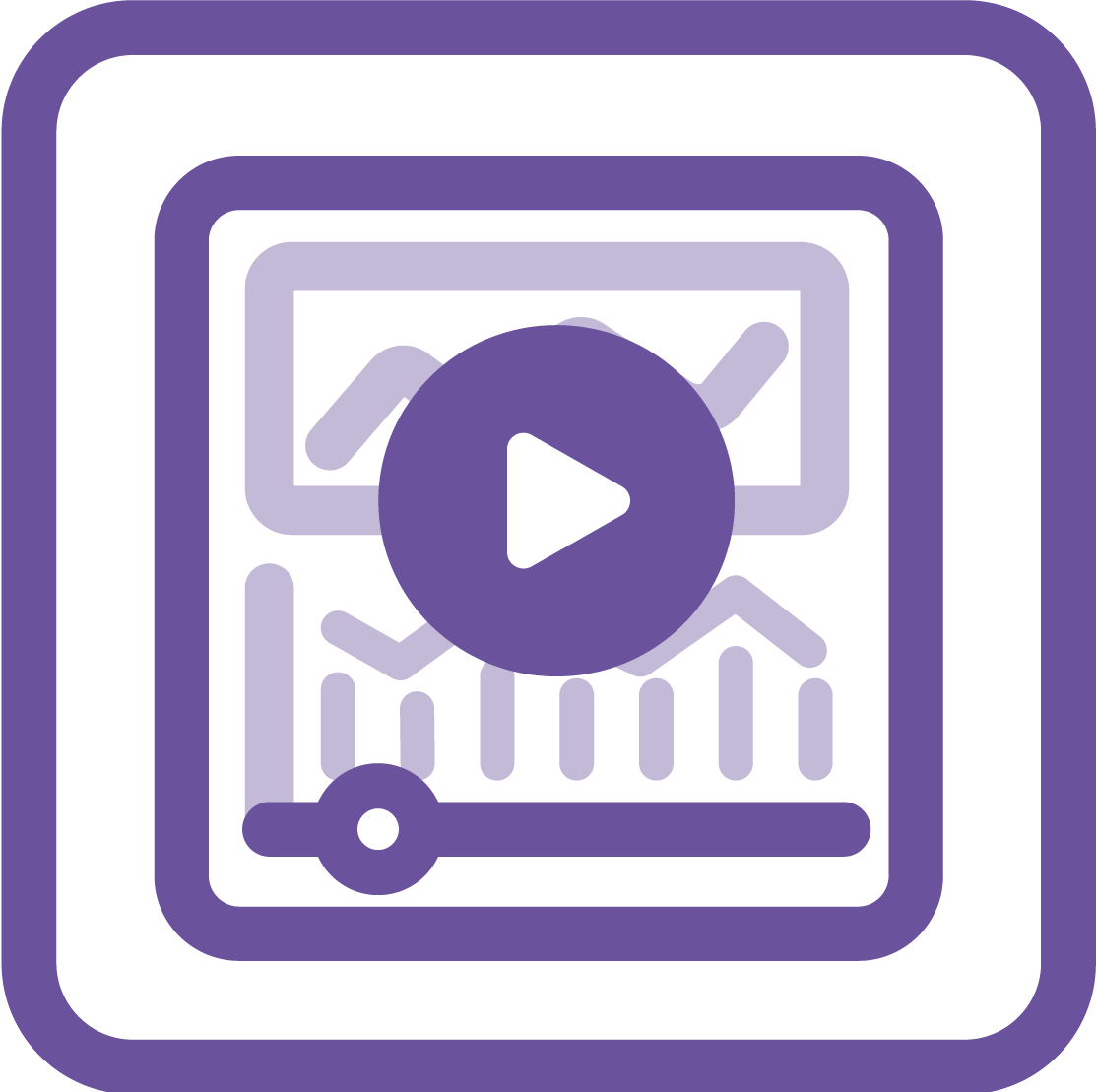}
\end{wrapfigure} 

%概念
Data video is a narrative visualization type \cite{segel2010narrative} that combines data visualization with motion graphics and tells data-driven stories. Data videos can present viewers with diverse visual information in a short period, and therefore, they are widely used in disseminating data information \cite{amini2015understanding, amini2016authoring}.
%The popularity of data video can be attributed to the diversity of narrative structure forms and the wide range of visuals they present to viewers in a short period of time. As a result, researchers believe that data video have a substantial impact, making them an exciting form for studying narrative visualization \cite{amini2015understanding, amini2016authoring}.

\textbf{Design space:}
Researchers have primarily focused on understanding, creating, and disseminating data videos. Amini et al.~\cite{amini2015understanding} first proposed a visual narrative structure theory, in which the narrative structure of data video can be divided into four roles: establisher (E), initial (I), peak (P), and release (R). On this basis, Cao et al. \cite{cao2020examining} presented a more extensive taxonomy of data video, including four narrative structures, five main genres, and six narrative qualities. Users can quickly find specific types of data videos with the help of this classification. These studies provide a solid foundation for designers to create data videos.
%This detailed categorization can bring benefits to data video in many practical activities. For example, individuals can easily find a specific type of data video, and developers can choose the most appropriate type of data video for their work. Moreover, the classification approach can help users construct their narrative content and guide them to model their data and eventually identify data video. The above work analyzes the structure and composition of data video and provides a solid foundation for understanding them.
%The opening of a data video has a significant impact on user acceptance. A good opening will attract users to keep watching, while a lousy opening will make them abandon watching the video. 
Xu et al. \cite{xu2022wow} considered data videos' opening narrative and visual presentation design. They proposed six cinematic opening styles (symbolism and metaphor, camera eye, big bang, old footage, and ending first styles) and 28 design guidelines for the six styles.%, which is essential in data video because the data itself changes and secondly because time is a fundamental property of data sets.

Visual narratives in data videos are usually performed using animation because animation can represent temporal changes and enhance the comprehension and user engagement of data stories~\cite{thompson2020understanding}. 
Shi et al. \cite{shi2021communicating} analyzed 43 animation techniques for narrative visualizations and categorized eight narrative strategies (e.g., emphasis, suspense, and comparison) to construct a design space. \rv{This design space helps describe how data videos are produced and how various components are integrated with visual narrative strategies. Such design spaces provide users with useful design suggestions; thus, they weaken the barriers to designing and producing expressive data videos.}
% that systematically describes these patterns. This design space can be used to reduce barriers to producing data videos and stimulate audience creativity in designing expressive data videos. 
By examining animated data charts, Tang et al. \cite{tang2020design} created a design space for data videos with five dimensions: data, motion, layout, duration, and narrative. Moreover, they proposed 20 design guidelines based on these dimensions. %For example, they select the data that matches the video intently and determine the duration of the visualization based on complexity. 
In addition, other researchers have conducted studies on how to increase the effectiveness of data video communication. Sallam et al. \cite{sallam2022towards} found that for a problem with no clear solution, a better option is to present it in a data video because the audience may feel high levels of negative emotions. 
%The researchers also noted that different populations react differently to the same message. It is essential to be cautious about using messages with bad connotations in data video when dealing with people with neurotic tendencies, prone to negative emotions, and challenging to persuade using data video. %Also, this type of person can be irritated by more complex information. 
%To improve video services and reduce the complexity of data video, Wang et al. \cite{wang2016animated} proposed non-linear time mapping (i.e., making the interesting parts the focus of the animation and compressing the parts that are not of interest to the user into the background) and foreshadowing (i.e., drawing and reminding viewers of the events in the animation) to improve the video service. 
To improve the quality and reduce the complexity of data video, Wang et al.~\cite{wang2016animated} proposed nonlinear time mapping and foreshadowing. The ``foreshadowing'' technique, as described by researchers, is only applicable to animated stacked images. Li et al. \cite{li2020improving} expanded on this by formally defining visual foreshadowing, a technique that addresses the problem of time-consuming videos that ignore the viewer's attention. %Viewers to maintain their attention throughout the process. 
%Animated visualizations with visual foreshadowing are attractive in displaying temporal data changes. It can guide viewers to cues for what is to be displayed in the generated animated visualization to provide an effective narrative visualization. 
%As a shorter form of data videos, data-GIFs show only the focused information of data changes in less than 15 seconds. Data-GIFs can represent information in a more focused manner, easier to understand and more easily disseminated than data videos. 
Shu et al. \cite{shu2020makes} examined the Data-GIFs design space and offered recommendations.

{\textbf{Authoring tool:}}
Producing data videos is time-consuming because it requires collaboration between people from different backgrounds (e.g., data analysts to generate data and insights, scripters to write narratives, and motion designers and graphics experts to produce video assets). Each element may depend on one or more particular software tools \cite{amini2016authoring}. DataClips \cite{amini2016authoring} provides a set of data clip libraries that allow \rv{amateurs} to combine data-driven clips to form longer sequences of data videos.
Lan et al. \cite{lan2021kineticharts} developed Kineticharts, a chart animation scheme for conveying emotions, based on the animations provided in DataClips~\cite{amini2016authoring}. Compared with DataClips, Kineticharts \cite{lan2021kineticharts} can enhance the emotional engagement of users by improving the presentation of the story without hindering users' understanding of the data. In addition, Chen et al. \cite{chen2021augmenting} %focused on data visualization of ball videos in sports, and through extensive analysis of sports videos, derived element-level (what the components are) and clip-level (how the components are organized) as a design space for enhanced sports videos (data type, visual type, data hierarchy, narrative order). Furthermore, the researchers
developed VisCommentator, a tool for analyzing ball sports videos in sports programs, facilitating the creation of  enhanced sports analysis videos through data insights and visualization suggestions.

{\textbf{ML/AI-supported tool:}}
Researchers have designed and developed tools to transform static visualizations into dynamic versions. Data Animator \cite{thompson2021data} utilizes the Data Illustrator \cite{liu2018data} framework to match two static visual objects and generate automatic transitions by default. %Designers have the flexibility to interpret and adjust the matching results through a visual interface. 
The tool also supports dividing complex animations into segments by layering keyframes, utilizing data characteristics to stagger the start time, and adjusting the pace of animated objects through a timeline view. Similarly, InfoMotion~\cite{wang2021animated} can build data films by extracting the graphical attributes of infographics, understanding its underlying information structure, and adding animation effects to the visual pieces of the infographic in chronological sequence. \rv{As InfoMotion~\cite{wang2021animated} is built into PowerPoint~\cite{powerpoint2016} as a plug-in, it can automatically link a variety of built-in animation effects to the visual parts of slides, which is excellent for speeding up data video production. This tool \cite{wang2021animated} is also easier to create data videos than Data Animator \cite{thompson2021data} because it can only use data in Data Illustrator \cite{liu2018data} format.}
% InfoMotion \cite{wang2021animated} is integrated into PowerPoint\cite{powerpoint2016} as an add-in, making it easier than Data Animator \cite{thompson2021data}, which can only use data in Data Illustrator \cite{liu2018data} format. InfoMotion \cite{wang2021animated} can automatically attach various built-in animation effects to the visual elements of a slide show, which is useful for quickly creating data video.
In addition, while Gemini2 \cite{kim2021gemini} and Cast \cite{ge2021cast} are not dedicated tools for creating data videos, both tools can build keyframes for charts. Gemini2\cite{kim2021gemini} focuses on helping users create animations by referring to keyframe suggestions. Similarly, Cast \cite{ge2021cast} allows users to manipulate directly to change the parameters of animation effects (e.g., animation type and jogging function) and refine animation specifications (e.g., adjusting keyframes to play across tracks and adjusting delays) by providing a GUI interface.
\rv{ML/AI-supported tools for creating data videos focus on how to identify existing infographic elements and convert them into dynamic video clips while authoring tools provide a library of existing data clips that users can use directly.}

{\textbf{ML/AI-generator tool:}}
While the abovementioned technologies ease the design process, data videos are still difficult to create because users must select which visualizations and animations to utilize and how to assemble a cohesive video.
The aforementioned problem was solved by the emergence of AutoClips \cite{shi2021autoclips}, a tool that automatically makes a data video from a series of data facts. This tool \cite{shi2021autoclips} saves users the time of analyzing the data and the obstacles of creating animations by using video motion software, significantly reducing the complexity of creating data videos. However, AutoClips normally only supports tabular data and tends to favor datasets with diverse column types. %The difficulty of creating more complex data content still exists.

\textbf{Summary:}
Data video has become popular owing to the growth of social media platforms. Research on data video has also received much attention in recent years. At the design space level, researchers have explored the understanding, creation, and dissemination of data videos to help better understand the components of data videos. These design spaces and guidelines provide the theoretical basis for developing authoring tools, ML/AI-supported tools, and ML/AI-generator tools. \rv{Authoring tools simplify the creation of data videos by offering a library of existing data clips that can be replicated. ML/AI-supported tools focus on how to identify existing static visualization elements and convert them into dynamic videos. ML/AI-generator tools can automatically generate data videos directly from input data.}

%\rv{Several researchers have discussed the duration of data videos to solve the problem of audience attention, but there are still some problems that need to be further studied in future work.}
However, certain issues still need further investigation.
%The data video design space focuses on the presentation of narrative structure and animation. Based on the proposed design spaces and guidelines, the theoretical basis for developing authoring tools, ML/AI-supported tools, and AI-generator tools can be provided. However, there are still some issues that need to be further studied.
\rv{First, existing automatic tools for creating data videos are still limited to a few visualization genres and input formats. For example, AutoClips \cite{shi2021autoclips} only supports tabular data, limiting the visual display possibilities of data video. More tools are needed to handle various data types, such as spatial-temporal data and textual data, which are essential for constructing diverse data narratives. Second, researchers also a need to study how the speed, continuity, and smoothness of animations in data videos, the transitions between charts and graphs, and the embellishment effects added to the videos would affect the understanding and overall experience of readers \cite{lan2021kineticharts}.} % Third, by applying advanced natural language algorithms, ML/AI supported or generator tools could recommend or produce animated infographics directly from datasets and text input.
%Second, in order to enable users to create and share data videos, it is necessary to create a platform to collect that accepts numerous infographics as input \cite{wang2016animated}. 

%First, how the speed, continuity, and smoothness of the animations in the data video, the transitions between charts and graphs, and the embellishment effects that need to be added to the video will affect the user's attention.
%Third, different animated data graphics creation tools use different standards and formats. Currently, no common source file format conveys complete information about static data visualization. Research in this area could be enhanced in the future, for example, by developing a standard format or allowing the software to support the import of files in different formats, which would reduce the distress of users due to software format issues and frequent software switching \cite{thompson2021data}. It also enables non-expert users to create narrative stories for data-driven videos by providing a set of templates that suggest different styles based on different messages \cite{amini2016authoring}. 
% Fifth, due to the development of the Immersive domain, it is possible to study the validity of data video in 3D and VR technology domains to determine if they support narrative visualization \cite{cao2020examining}.

\section{Discussions and Future Work}\label{sec:discussion}
%Fully automated AI-generator is not the optimal solution, human-ai collaboration (人智融合-智能体）
%design thinking process, tools at different stages
%This paper constructs a two-dimensional coordinate based on prior research. The horizontal axis is the strength of the narrative. The vertical axis shows the complexity of visualization design. It focuses on the design space and tools (authoring tools, ML/AI-supported tools, and AI-generator tools) of seven narrative visualization types on the +x axis (dashboard, annotated chart, infographic, timeline, data comics, scrollytelling, data video). 
%In this section, we first outline the future research opportunities of the design space and tools at different automation levels for narrative visualization. Then, we discuss limitations of this survey.
In this section, we outline the current limitations and future research opportunities of design spaces and tools at different automation levels for narrative visualization. 

\vspace{1mm}
%\subsection{Future Research Opportunities} 
\begin{wrapfigure}{l}{0.055\textwidth}
  %\begin{center}
    \includegraphics[width=0.055\textwidth]{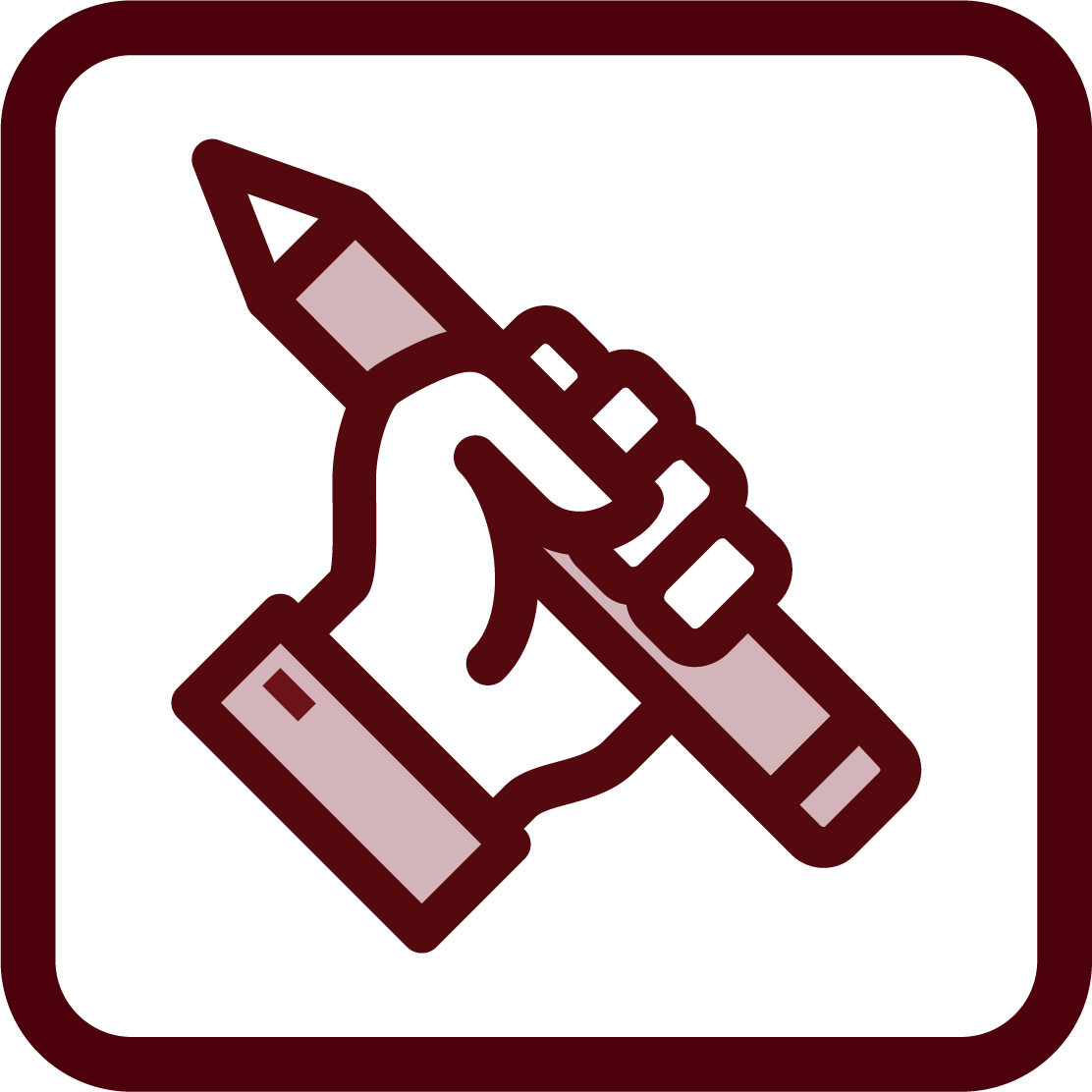}
  %\end{center}
\end{wrapfigure} 
\noindent
\textbf{Design space} \textls[+12]{aims to describe all the possible design aspects for various narrative genres. The summary of the design space allows us to capture some implicit knowledge of visual designers and practitioners~\cite{schulz2010design}. Most existing studies propose clear design guidelines in specific design scenarios~\cite{suprata2019data, bach2018design, tang2020design}. However, the design space articles on the different narrative genres vary in focus. For example, in annotated charts and infographic genres, the focus is on how to effectively create a correct and aesthetically appealing visualization. For timeline, data comics, scrollytelling, and data video genres, the focus is more on exploring the narrative structure. In particular, data video pays special attention to creating animations, while other genres focus more on static presentations. The major future research directions are listed below.}

\textbf{Simplify and validate the design space.} Creators can generally access many existing visualization design guidelines, but choosing the right guidelines is difficult for them. Moreover, \rb{design guidelines often fall short in explaining when it is more appropriate to use, and lack proper validation} \cite{tang2020design}. For example, researchers have proposed visualization design process frameworks \cite{sedlmair2012design, oppermann2020data}, but have not explained what scenarios and how to use these frameworks for visualization design. Therefore, a potential research direction is to validate the usage of various design spaces and classify them according to application domains. Amateurs may also be provided with an overview of design spaces to tackle specific design problems. %Design space research also requires to relate specific design problems and design purposes so that beginners can directly retrieve the corresponding design guidelines by searching design questions. This will improve both the utilization of design space and the efficiency in finding an appropriate reference. %However, the design space is only a solution for users, not a hard-and-fast rule. Users can use the visual design guide to design according to the actual situation, but the design guide should not be used as a standard to measure the design. Users can use the visual design guide to design according to the actual situation, but the design guide should not be used as a standard to measure the design. 
%Visual design guidelines are a starting point for designers, not an endpoint. If one lacks sufficient experience in visualization, then using visual design guidelines will, in most cases, make the design better. As a mature visualization designer, it is not necessary to follow the visualization guidelines dogmatically.
For more experienced designers, we could pay more attention to the subtle design guidelines that can improve the user experience and user perception in the visualization.

%rvq:1.there are some borrowed from literature and drama - pyramid/non-linear; 2.do TableuStory support narrative structure, what is lacking
\rv{\textbf{Explore new narrative structures.} The existing narrative structures are primarily derived from movies or other audio-visual content \cite{amini2015understanding}. Information on the application of narrative structures in novels and plays in narrative visualizations is generally lacking in the extant literature. Due to the different characteristics of various narrative genres, the choice of narrative structure can also be different. For example, a timeline mainly presents content in a linear narrative sequence \cite{nguyen2016timesets}, whereas in data videos, using a nonlinear narrative approach is more likely to engage the audience~\cite{lan2021understanding}. Therefore, a potential research direction is to explore different data types and which narrative structure is more suitable for different narrative genres.}

\textls[-5]{\rv{\textbf{Explore other narrative visualization genres.} Some traditional visualization genres that focus more on visual analytics are embracing narrative and storytelling concepts. Suprata \cite{suprata2019data} noted that adding narrative attributes to dashboards allows users to become more aware of their goals and how to take action next. Fernandez Nieto et al. \cite{fernandez2022beyond} enhanced teachers' guidance of the content by including narrative attributes in designing learning analytics dashboards. With more attention and practical applications of narrative attributes to traditional dashboards, narrative dashboards can be another future narrative genre.
Meanwhile, some new genres of visualizations have emerged, such as immersive visualization \cite{isenberg2018immersive} and data physicalization~\cite{karyda2020narrative}, which can also be developed with narrative characteristics.
Some researchers have demonstrated visualizations such as 3D scatter plots \cite{bach2017hologram}, parallel coordinates~\cite{butscher2018clusters}, and networks \cite{cordeil2016immersive} in immersive environments to help users interpret complex data and facilitate visual data exploration. Others have explored the narrative roles of data physicalization~\cite{karyda2020narrative}, which enabled the participants to develop meaningful narratives in the forms of physical data representations in their study. However, these studies are in their infancy. An emerging trend is the need to study new forms of narrative visualizations. For example, the potential research direction of narrative immersive visualization is a more in-depth exploration of data types, spatial layouts, and user interactions for narrative communication in the virtual environment. The physicalization of data encodes information in a perceptible form, allowing users to explore using all their senses and motor skills~\cite{hogan2017towards}. 
More research on narrative data physicalization is still needed in the future to better understand the design space, the process of producing data, and the benefits compared with flat visualization or virtual presentation~\cite{dragicevic2020data}.}}

\vspace{1mm}
\begin{wrapfigure}{l}{0.055\textwidth}
  %\begin{center}
    \includegraphics[width=0.055\textwidth]{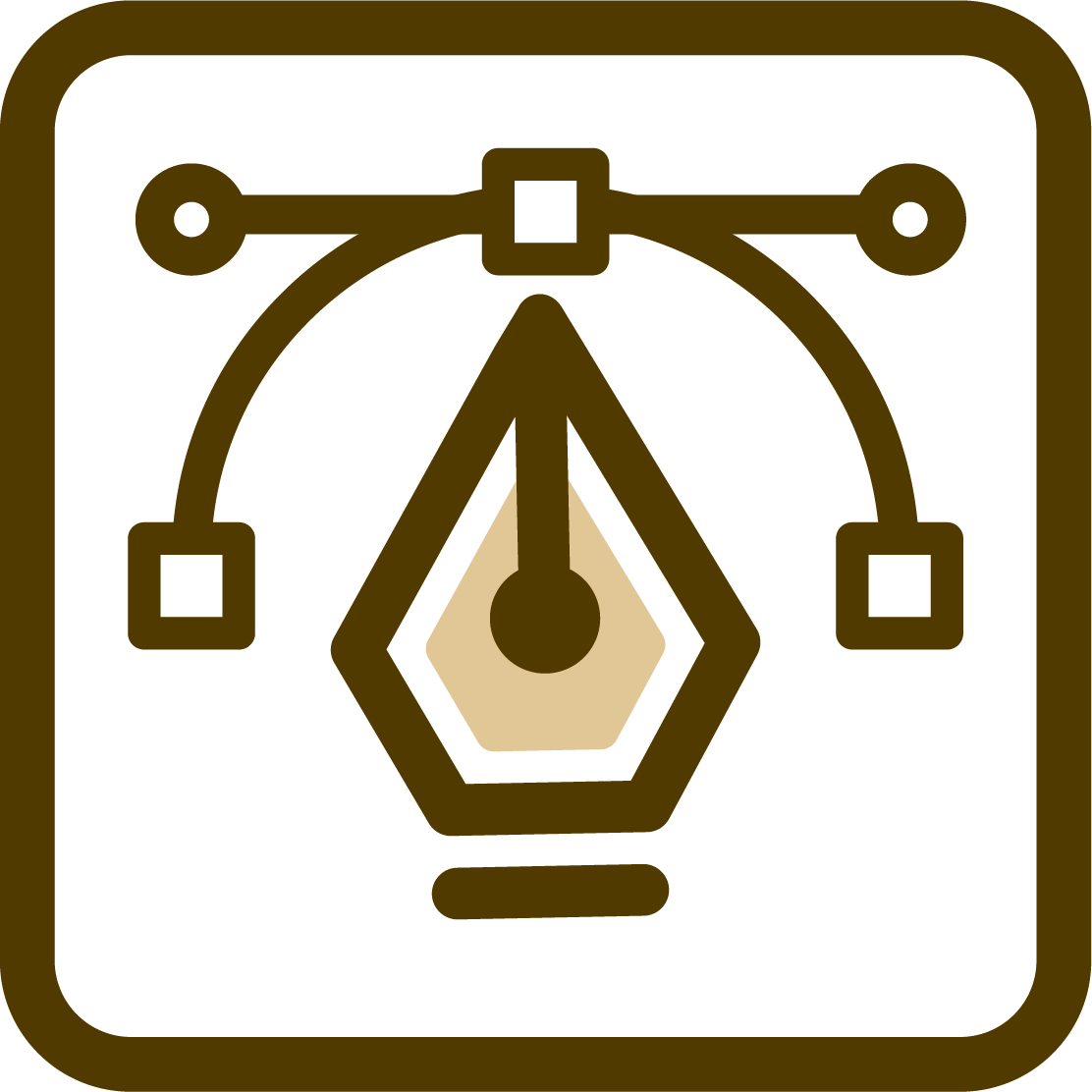}
  %\end{center}
\end{wrapfigure} 
\noindent
\textbf{Authoring tools} \textls[+3]{aims to facilitate the visualization creation process with controllable interactions. These tools include stand-alone applications~\cite{ ren2017chartaccent, kim2019datatoon}, web-based tools~\cite{TimelineJS2013, kim2016data}, and authoring tools that combine with office software~\cite{cui2021mixed, wang2018infonice}. The advantage of these tools is that users have enough control to create customized visualizations, including more complex visualizations that cannot be supported by automated tools. Although these authoring tools significantly improve the efficiency of creating narrative visualizations, most tools are aimed at users with a certain level of expertise. 
\rv{For example, authoring tools for infographics and timelines require users to have visual design skills, while data videos require users to have video editing skills. Future research may invest more efforts in the following directions.}}
% \rv{\textbf{Provide different models for authoring tools.} Tools are designed to fit the needs of different users and environments, the tool should also do the corresponding match. For example, in Adobe Illustrator, users are provided with Web mode, typography rules mode, essential functions, and other modes. By clicking on different methods, the work area of the page will change accordingly.} Researchers found significant differences in design goals and interaction styles between expert and amateurs. Experts use tools to support the detailed exploration of data, while amateurs mainly use them for communication\cite{elias2011exploration}. \rv{The same strategy can be used for narrative visual authoring tools with other workspace interfaces for different users, making the user interface simple on the one hand and avoiding the psychological burden of too many functions for the user on the other.}

% Researchers found significant differences in design goals and interaction styles between expert and novice users. Experts use tools to support the detailed exploration of data, while novice users mainly use them for communication \cite{elias2011exploration}. Novice users could be resistant to authoring tools with complex interfaces and functions and may feel at a loss to start. Therefore, in future research, expert mode and novice mode can be developed so that different users can switch between the two modes to fit different needs.%to reduce the difficulty of learning authoring tools for novice users.

% Offer more flexible interaction for authoring tools 
\rv{\textbf{Develop flexible interfaces for authoring tools.}  %In the design field, there are many tools to help produce creative design works \cite{kang2021metamap}. 
Among narrative authoring tools, a few tools can be used to freely draw creative patterns on a 
 screen, including DataSelfies \cite{kim2019dataselfie}, DataInk~\cite{xia2018dataink} and SketchStory~\cite{lee2013sketchstory} for infographics and DataToon~\cite{kim2019datatoon} for data comics. However, other narrative visualization tools are relatively lacking. More flexible interface methods should be provided to help designers achieve more creative ideas and more artistic effects in creating various narrative visualizations.}%research on the integration of data visualization and creative design should be strengthened to support more artistic designers to participate in the scenario development of narrative visualization.}}

\textbf{Develop more interactive visualizations}. \rv{Among the existing narrative visualization genres, only scrollytelling and slideshow have strong interactive properties,}\rb{
while annotated charts, infographics, data comics, and data videos are mostly static visualizations that lack interactive functionality. However, studies have proven that by providing interactivity\cite{wang2021interactive} and adding interesting~\cite{diakopoulos2011playable, dunlap2016getting} and emotional factors \cite{lan2021smile} to the visualization, users are more likely to memorize the information. Therefore, in future research, a possible approach is to explore more narrative genres of interactive visualizations and to add interesting and emotional elements.}

\vspace{1mm}
\begin{wrapfigure}{l}{0.055\textwidth}
  %\begin{center}
    \includegraphics[width=0.055\textwidth]{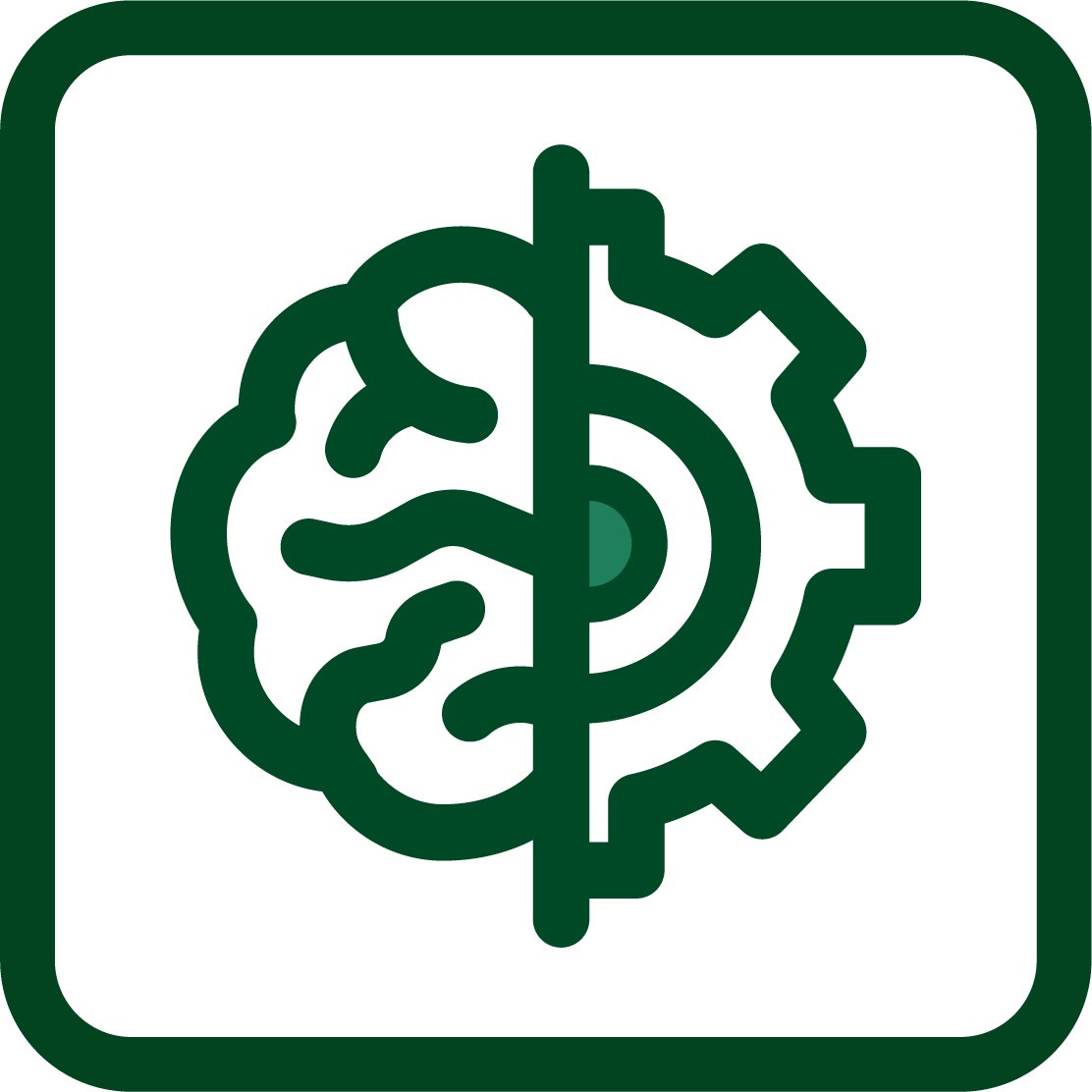}
  %\end{center}
\end{wrapfigure} 
\noindent % 修改到这里了
\textbf{ML/AI-supported tools} are designed to assist users in visualization creation by applying intelligent algorithms and techniques. Such tools can provide recommendations or guide the user via the creation process. ML/AI-supported tools for narrative visualizations can serve a wider range of users than authoring tools. For example, \rv{designers who lack data analytic skills} can easily create data comics with the data analysis capabilities of ML/AI-supported tools; data analysts \rv{who lack design skills} can use ML/AI-supported tools to create more aesthetically pleasing timelines or data videos.

\rv{However, the automatic goals and functions of current ML/AI-supported tools for different narrative visualization types are different. For example, tools for annotated charts, infographics, and data comics have the auxiliary function of identifying and parsing visualizations. Among them, the purpose of annotated chart recognition is to add annotations to facilitate comprehension of the visual story; the purpose of infographic recognition is to create new visualizations based on the original visual styles; and the purpose of data comics recognition is to transform visual content into the comic layout. ML/AI-supported tools for timelines focus on placing timeline text and optimizing visual aesthetic effects, while data videos pay more attention to the creation of animation.}
By summarizing existing research in ML/AI-supported tools for narrative visualization, the following directions can be studied.

% \textls[-5]{\textbf{Adapt to more visualization \rv{genres}.} The existing ML/AI-supported tools for narrative visualization, such as annotated charts, infographic, data comics, only support simple statistical charts and data formats, excluding more complex input data and visualization \rv{genres}. In future research, a more comprehensive library of visual resources can be created to allow ML/AI-supported tools to support more narrative visualization \rv{genres}. Besides, we need to study applying larger-scale and more complex data in ML/AI-supported tools to facilitate broader application scenarios~\cite{zhu2020survey}.}

\rv{\textbf{Enhance the research of annotation tools.} Although the annotated chart is one specific genre in narrative visualization, annotations are important for any visualization. It can help users understand the visualization and help data analysts review the past analysis process \cite{shrinivasan2009connecting}. However, the ability to add annotations to visualizations is lacking in ML/AI-supported tools for different narrative visualizations \cite{thompson2021data}. Moreover, the annotations generated by most existing tools can only explain the statistical information on a single chart \cite{coelho2020infomages, elias2012annotating}. Studies are generally lacking in terms of applying intelligent techniques to extract context information to build visualizations with narrative structures.}

%\textbf{Develop more tools to detect visualization problems.} Many tools can detect problems in creating visual charts in the traditional visualization domain, such as VisuaLint \cite{hopkins2020visualint}, Vizlinter \cite{chen2021vizlinter}, and Draco \cite{moritz2018formalizing}. In the narrative visualization domain, only annotated charts have similar detection tools, but they only support the detection of line charts \cite{fan2022annotating}. Such tools are urgently needed for other types of narrative visualizations. Because many practitioners lack expertise in visualization, they often create misleading visualizations that distort the underlying data \cite{choi2021toward}; there are also researchers who, in their teaching practice, find that students often have problems introducing data and news narrative stages in the process of creating data comics \cite{wang2019teaching}. Some timelines also have problems that affect readability \cite{kumar1998metadata, stab2010sematime}. Therefore, a potential research direction is to develop tools for detecting data problems and violations of visual design rules to help users create better visualizations and improve the accuracy of visual content \cite{lin2020dziban}.

\textls[+15]{\textbf{Improve the reusability of existing visualizations.} In practice, the majority of charts are saved as bitmap pictures. Although they are simple to spread and use, they are difficult to modify. VisCode \cite{zhang2020viscode} and Chartem \cite{fu2020chartem} can store and hide the original data information inside the picture of a chart. However, only rudimentary visual charts are supported by these tools. Therefore, tools to support the recognition and reprocessing of more complex visual charts and more diverse narrative genres must be developed.}
By improving the reusability of existing narrative visualizations, \rv{amateurs} are able to create more visual stories efficiently and effectively~\cite{thompson2021data}. %这段话介绍的还是对复杂图表进行识别

%Text2vis的challenge没有说清楚，内容和后面的Generator相似
%\textls[-3]{\rv{\textbf{Integrate natural language algorithms.} The use of natural language algorithms to facilitate the development of tools is mentioned in various genres of narrative visualization. For example, natural language algorithms are used in annotated charts to detect incorrect chart annotations and incorrect information on the screen\cite{latif2021kori}. TimelineCurator \cite{fulda2015timelinecurator} can process unstructured documents with temporal text using natural language and extract temporal text from them. Text-to-viz \cite{cui2019text} recommends and generates infographics directly from natural language statements. However, the current natural language algorithms still face the accuracy problem due to the ambiguity of expressions and users' actual needs. There is a need to study more advanced algorithms to tackle this problem in future research.}}

\rv{\textbf{Facilitate the adaptability of different software.} Some existing tools are integrated with office software. For example, all the features of the DataComicsJS~\cite{zhao2015data} tool can be replicated in presentation tools (e.g., Microsoft PowerPoint~\cite{powerpoint2016}) and drawing tools (e.g., Adobe Illustrator \rb{\cite{adobeIllustrator}}). Chartreuse~\cite{cui2021mixed} and InfoNice~\cite{wang2018infonice} are also both integrated into Microsoft Office software in the form of plug-ins. %Compared to different stand-alone tools, office software has a much larger user base. 
After incorporating the natural language algorithms of intelligent tools into productivity software, the corresponding functions can work in the background. For example, once a statement is detected that can be enhanced with visualization, a message can pop up to ask the user if he or she wants to use the recommended chart~\cite{cui2019text}. In this manner, ML/AI-supported tools could reach a wider audience.} 

%there is a lack of standard for better evaluation of automated outputs
%\textbf{Evaluate the outputs of automated tools.} Although there are many evaluation efforts on output visualization, most of them are evaluated for a specific type of visualization. Moreover, there is still a great challenge to define a comprehensive algorithm to quantify the evaluation of automated visualization tools. Evaluation of visualization tools can focus on three levels, which are (1) measurement of the diversity of input datasets (diversity of data types, formats, and content), (2) the degree of automation of the tools in generating visualizations (qualitative or quantitative measurements to reveal how these techniques become smarter), and (3) the quality of the output visualizations (are the data accurate, do they match user perception, and whether it is aesthetically pleasing) \cite{zhu2020survey}.
\vspace{1mm}
\begin{wrapfigure}{l}{0.055\textwidth}
  %\begin{center}
    \includegraphics[width=0.055\textwidth]{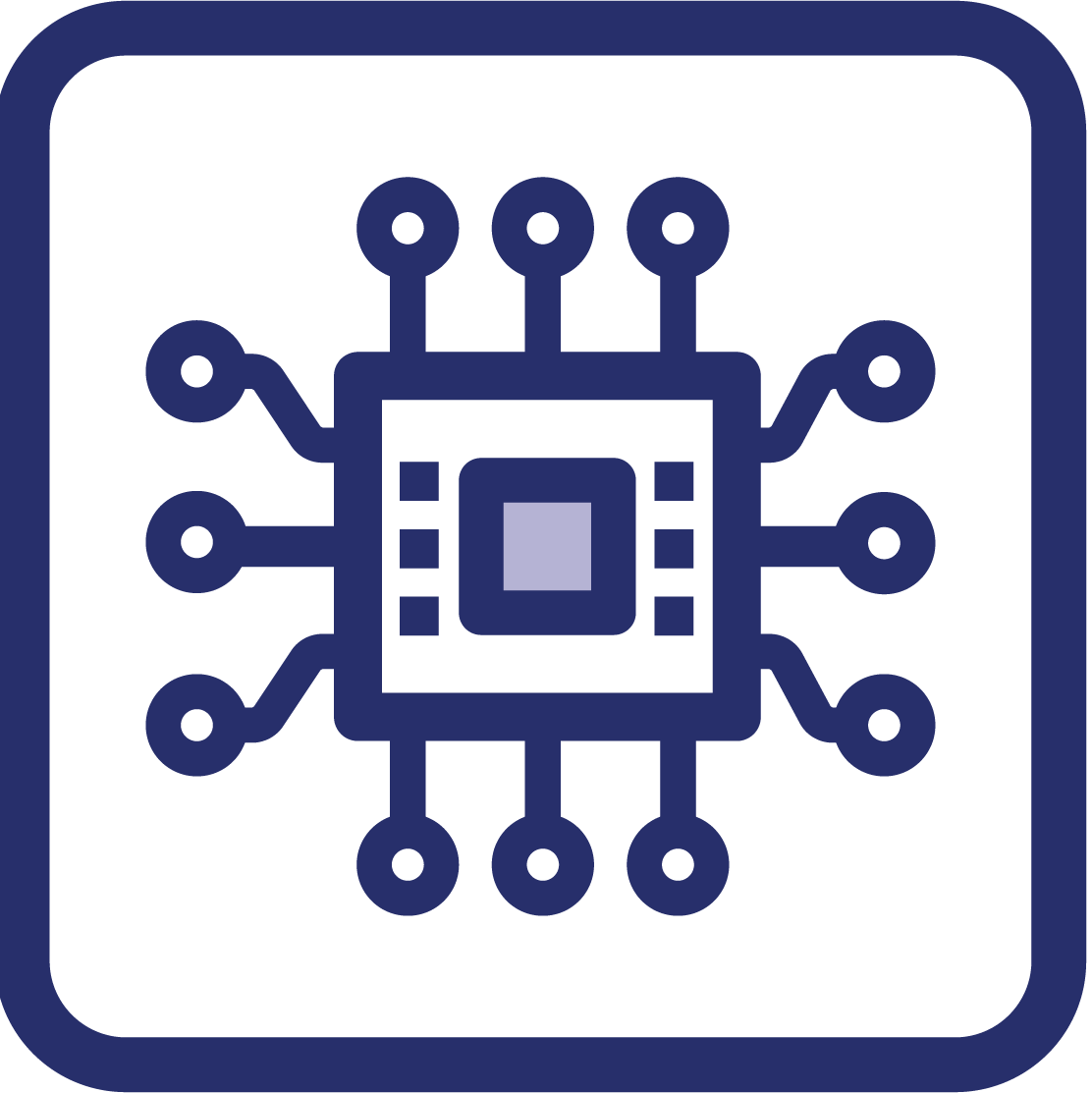}
  %\end{center}
\end{wrapfigure} 
\noindent
\textbf{ML/AI-generator tools} are more intelligent than the previous three types of tools in that they require minimal or no user involvement in the entire creation process. These tools automate the analysis of data and directly generate a complete narrative visualization without user intervention. ML/AI-generator tools mostly target \rv{amateurs}. %Most AI-generator tools help users generate visualizations by automating the process of gaining insight into data and then following design aesthetics. In addition, these tools have a distinct advantage over the previous tools in that they can directly analyze data and automatically generate visual data stories with narrative properties.
%Especially for data co mics, scrollytelling, and data videos can directly combine data fragments with telling a data story. 
%A literature review \cite{这里需要加reference} shows that AI-generator tools have entered a rapid development phase after 2014.
The development of such tools has gradually increased in the past decade. % We believe there would be more research and tools in the near future. in the pre-creation, creation, and post-creation processes in the future. 
As visual communication becomes increasingly important in our daily life, we believe that such tools can play an important role in the creation of narrative visualization. The following directions can be studied in terms of understanding user intent to improve accuracy and efficiency.

\textbf{Apply natural language techniques to understand and predict user design intent.} On the one hand, we can better understand user intent by using natural language interaction in the design process. Once the interaction data and design outcomes are collected, we can use them to train models to predict and recommend potential interaction or design options. On the other hand, recent advances in natural language processing can generate realistic images and artworks directly from textual descriptions~\cite{dibia2019data2vis}. This situation may also drive visualization academics to use cutting-edge approaches to aid in the automation of visual story tools. %and to assist users in creating visualizations more efficiently. One potential approach is first to explore how users describe or express a visualization for a given data set; then apply knowledge of the user's utterances describing the visualization to the three elements of natural language description, data sequence, and visualization specification; and finally, use this data to train a model that learns to generate visualizations based on natural language descriptions. Such a model can extend the expressive power of existing visualization authoring systems and further enhance visualization tools' intelligence \cite{dibia2019data2vis}.

\rv{\textbf{Improve data analysis capability to identify user design intent.} Among the current six genres of narrative visualizations, ML/AI-generator tools that can be used for timelines are generally lacking. Even though certain intelligent tools can be used to create a timeline, they only modify the local area. Completing the entire creation remains time-consuming. While several ML/AI-generator tools for other genres, such as Autoclips \cite{shi2021autoclips}, can analyze the data and extract essential parameters from the dataset, the final output is not satisfactory when facing different datasets, different contexts, or more complex data types. Therefore, the ability of ML/AI-generator tools to analyze complex data in the future must be improved. In addition, a possible direction is to study how to input the user’s creative intent into the automation process and at which point in the creation process; in this manner, the user’s intent can be fully grasped to achieve the most satisfying outcome.}

\textbf{Develop narrative recommendation tools to clarify design intent.} In statistical charts, researchers have developed many visual recommendation systems such as Voyager \cite{heer2007voyagers} and SeeDB~\cite{vartak2015seedb}. However, research on such tools, specifically for narrative visualization, is lacking. This situation can be explained by recommendation methods being based on data characteristics or design guidelines rather than the user’s design intent. One potential research direction is to develop a narrative visualization recommendation platform, where both the design process and design outcomes of visualizations are stored on the platform. By analyzing the collected information, we could identify the user’s design intent with the help of machine learning algorithms~\cite{luo2018deepeye}. Such recommendation tools can provide users with abundant design ideas and recommendations in the pre-creation stage of narrative visualizations.

%data-driven models具体指什么 chen19toward提到什么
%\rv{\textbf{Use data-driven models to enhance ML/AI generator tools.} Due to the rapid development of data mining techniques, it is possible to interpret charts and even infographics using data-driven models rather than rule-based approaches \cite{chen2019towards}. Many existing ML/AI-generator tools are rule-based, which limits the creation of more creative visualizations. In the future, rule-based ML/AI-generator tools can be fused with data-driven models. The data-driven tools can use different types of visualizations to train the models and thus extend them to a wider range of scenarios.}

%For all the work in the four levels of automation summarized above, xxx Although different tools are classified into different narrative visualization genres in this study, it does not mean that these tools can only be used to produce one type of narrative visualization. %For example, DataShot\cite{wang2019datashot} can create both data comics and generate infographics.
%(更换例子)Some other tools are also compatible or adaptable across different narrative visualization \rv{genres}. However, there is a learning curve of varying degrees for amateurs of either visualization tool. 

This study outlines four narrative visualization tools at the automation level. Furthermore, although the different tools are divided into different narrative visualization genres in this study, it does not mean that these tools can only create one genre of narrative visualization. The tools present certain compatibility across different genres of narrative visualizations. However, novice users need to undergo a learning curve to varying degrees for either visualization tool type.
\rv{Moreover, these visualization tools can only tell users how the data have changed, without explaining why it has changed, suggesting that the user will still need to analyze the reasons for the data change results.} 
Moreover, a one-size-fits-all tool to handle all scenarios to address different users and goals does not exist. Therefore, all four levels of automation have their unique values and are worth further investigation, from pure manual design following design theories to the ultimate ML/AI-generated tools that support automation in the whole visualization creation pipeline. Furthermore, with the development of AI technology and the need to create and share data visualization by \rv{amateurs}, ML/AI-supported tools and ML/AI-generated tools are becoming more popular in both research and various application domains.
%, with the progress of machine learning technology, two types of tools, ML/AI-supported tools, and AI-generator tools, can gradually fill the problems encountered so far, and these two types of tools will be the main visualization creation tools in the future. 
%These two types of tools are the primary trend for future visualization creation. 
In particular, with both human participation and machine assistance, ML/AI-supported tools provide a better user experience than authoring tools and support more diverse design opportunities than ML/AI-generator tools. More efforts can be invested in such human-centered ML/AI-supported narrative visualization tools in the future.

\section{Conclusion}\label{sec:conclusion}
In this study, we systematically reviewed \rv{105} papers and tools to study how automation can progressively engage in visualization design and narrative processes \rv{to help users create narrative \rb{visualizations} more easily, effectively, and efficiently. We have summarized six genres of narrative visualization (i.e., annotated charts, infographics, timeline \& storyline, data comics, scrollytelling \& slideshow, and data videos) based on previous research, and four types of tools (i.e., design space, authoring tool, ML/AI-supported tool, ML/AI-generator tool) based on the intelligence and automation level of the tools. }\rb{This study enables users to comprehend the explicit and implicit design elements of various narrative visualization genres, facilitating the selection of appropriate tools for visual storytelling. However, our survey excluded scientific visualization. In the field of scientific visualization, narrative visualization has been applied in scenarios such as climate or medical condition narratives \cite{ma2011scientific}. We believe that more research and tools in scientific visualization storytelling can be performed and developed in the future. We further discuss new research challenges and outline potential directions for future research and implementation.}

\section*{Acknowledgments}
{This work was supported in part by the NSFC 62002267, 62072338, 62061136003, NSF Shanghai 23ZR1464700, and Shanghai Education Development Foundation ``Chen-Guang Project'' (21CGA75). We would like to thank anonymous reviewers for their constructive feedback.}

%\newpage
\bibliographystyle{unsrt} %按引用的先后进行排列参考文献
%\bibliography{cas-refs}
%\bibliographystyle{abbrv}
%\bibliographystyle{abbrv-doi}
%\bibliographystyle{abbrv-doi-narrow}
%\bibliographystyle{abbrv-doi-hyperref}
%\bibliographystyle{abbrv-doi-hyperref-narrow}

%\bibliographystyle{IEEEtran}
\bibliography{template}

\begin{thebibliography}{100}

\bibitem{kosara2013storytelling}
Robert Kosara and Jock Mackinlay.
\newblock Storytelling: The next step for visualization.
\newblock {\em Computer}, 46(5):44--50, 2013.

\bibitem{hullman2011visualization}
Jessica Hullman and Nick Diakopoulos.
\newblock Visualization rhetoric: Framing effects in narrative visualization.
\newblock {\em IEEE Transactions on Visualization and Computer Graphics},
  17(12):2231--2240, 2011.

\bibitem{wang2021survey}
Qianwen Wang, Zhutian Chen, Yong Wang, and Huamin Qu.
\newblock A survey on ml4vis: Applying machine learning advances to data
  visualization.
\newblock {\em IEEE Transactions on Visualization and Computer Graphics}, 2021.

\bibitem{wu2021ai4vis}
Aoyu Wu, Yun Wang, Xinhuan Shu, Dominik Moritz, Weiwei Cui, Haidong Zhang,
  Dongmei Zhang, and Huamin Qu.
\newblock Ai4vis: Survey on artificial intelligence approaches for data
  visualization.
\newblock {\em IEEE Transactions on Visualization and Computer Graphics}, 2021.

\bibitem{zhu2020survey}
Sujia Zhu, Guodao Sun, Qi~Jiang, Meng Zha, and Ronghua Liang.
\newblock A survey on automatic infographics and visualization recommendations.
\newblock {\em Visual Informatics}, 4(3):24--40, 2020.

\bibitem{lee2015more}
Bongshin Lee, Nathalie~Henry Riche, Petra Isenberg, and Sheelagh Carpendale.
\newblock More than telling a story: Transforming data into visually shared
  stories.
\newblock {\em IEEE Computer Graphics and Applications}, 35(5):84--90, 2015.

\bibitem{borkin2013makes}
Michelle~A Borkin, Azalea~A Vo, Zoya Bylinskii, Phillip Isola, Shashank
  Sunkavalli, Aude Oliva, and Hanspeter Pfister.
\newblock What makes a visualization memorable?
\newblock {\em IEEE Transactions on Visualization and Computer Graphics},
  19(12):2306--2315, 2013.

\bibitem{borkin2015beyond}
Michelle~A Borkin, Zoya Bylinskii, Nam~Wook Kim, Constance~May Bainbridge,
  Chelsea~S Yeh, Daniel Borkin, Hanspeter Pfister, and Aude Oliva.
\newblock Beyond memorability: Visualization recognition and recall.
\newblock {\em IEEE Transactions on Visualization and Computer Graphics},
  22(1):519--528, 2015.

\bibitem{kong2017internal}
Ha-Kyung Kong, Zhicheng Liu, and Karrie Karahalios.
\newblock Internal and external visual cue preferences for visualizations in
  presentations.
\newblock {\em Computer Graphics Forum}, 36(3):515--525, 2017.

\bibitem{ren2017chartaccent}
Donghao Ren, Matthew Brehmer, Bongshin Lee, Tobias H{\"o}llerer, and Eun~Kyoung
  Choe.
\newblock Chartaccent: Annotation for data-driven storytelling.
\newblock In {\em 2017 IEEE Pacific Visualization Symposium (PacificVis)},
  pages 230--239. IEEE, 2017.

\bibitem{chen2010touch2annotate}
Yang Chen, Jing Yang, Scott Barlowe, and Dong~H Jeong.
\newblock Touch2annotate: Generating better annotations with less human effort
  on multi-touch interfaces.
\newblock {\em CHI'10 Extended Abstracts on Human Factors in Computing
  Systems}, pages 3703--3708, 2010.

\bibitem{chen2010click2annotate}
Yang Chen, Scott Barlowe, and Jing Yang.
\newblock Click2annotate: Automated insight externalization with rich
  semantics.
\newblock In {\em 2010 IEEE Symposium on Visual Analytics Science and
  Technology}, pages 155--162. IEEE, 2010.

\bibitem{kandogan2012just}
Eser Kandogan.
\newblock Just-in-time annotation of clusters, outliers, and trends in
  point-based data visualizations.
\newblock In {\em 2012 IEEE Conference on Visual Analytics Science and
  Technology (VAST)}, pages 73--82. IEEE, 2012.

\bibitem{bryan2016temporal}
Chris Bryan, Kwan-Liu Ma, and Jonathan Woodring.
\newblock Temporal summary images: An approach to narrative visualization via
  interactive annotation generation and placement.
\newblock {\em IEEE Transactions on Visualization and Computer Graphics},
  23(1):511--520, 2016.

\bibitem{latif2021kori}
Shahid Latif, Zheng Zhou, Yoon Kim, Fabian Beck, and Nam~Wook Kim.
\newblock Kori: Interactive synthesis of text and charts in data documents.
\newblock {\em IEEE Transactions on Visualization and Computer Graphics},
  28(1):184--194, 2021.

\bibitem{fan2022annotating}
Arlen Fan, Yuxin Ma, Michelle Mancenido, and Ross Maciejewski.
\newblock Annotating line charts for addressing deception.
\newblock In {\em CHI Conference on Human Factors in Computing Systems}, pages
  1--12, 2022.

\bibitem{kong2012graphical}
Nicholas Kong and Maneesh Agrawala.
\newblock Graphical overlays: Using layered elements to aid chart reading.
\newblock {\em IEEE Transactions on Visualization and Computer Graphics},
  18(12):2631--2638, 2012.

\bibitem{srinivasan2018augmenting}
Arjun Srinivasan, Steven~M Drucker, Alex Endert, and John Stasko.
\newblock Augmenting visualizations with interactive data facts to facilitate
  interpretation and communication.
\newblock {\em IEEE Transactions on Visualization and Computer Graphics},
  25(1):672--681, 2018.

\bibitem{subramonyam2018smartcues}
Hariharan Subramonyam and Eytan Adar.
\newblock Smartcues: a multitouch query approach for details-on-demand through
  dynamically computed overlays.
\newblock {\em IEEE Transactions on Visualization and Computer Graphics},
  25(1):597--607, 2018.

\bibitem{hullman2013contextifier}
Jessica Hullman, Nicholas Diakopoulos, and Eytan Adar.
\newblock Contextifier: automatic generation of annotated stock visualizations.
\newblock In {\em Proceedings of the SIGCHI Conference on Human Factors in
  Computing Systems}, pages 2707--2716, 2013.

\bibitem{liu2020autocaption}
Can Liu, Liwenhan Xie, Yun Han, Datong Wei, and Xiaoru Yuan.
\newblock Autocaption: An approach to generate natural language description
  from visualization automatically.
\newblock In {\em 2020 IEEE Pacific Visualization Symposium (PacificVis)},
  pages 191--195. IEEE, 2020.

\bibitem{cmeciu2016beyond}
Camelia Cmeciu, Madalina Manolache, and Alexandra Bardan.
\newblock Beyond the narrative visualization of infographics on european
  issues.
\newblock {\em Studies in Media and Communication}, 4(2):54--69, 2016.

\bibitem{harrison2015infographic}
Lane Harrison, Katharina Reinecke, and Remco Chang.
\newblock Infographic aesthetics: Designing for the first impression.
\newblock In {\em Proceedings of the 33rd Annual ACM Conference on Human
  Factors in Computing Systems}, pages 1187--1190, 2015.

\bibitem{lyra2016infographics}
Kamila~T Lyra, Seiji Isotani, Rachel~CD Reis, Leonardo~B Marques, La{\'\i}s~Z
  Pedro, Patr{\'\i}cia~A Jaques, and Ig~Ibert Bitencourt.
\newblock Infographics or graphics+ text: Which material is best for robust
  learning?
\newblock In {\em 2016 IEEE 16th International Conference on Advanced Learning
  Technologies (Icalt)}, pages 366--370. IEEE, 2016.

\bibitem{lan2021smile}
Xingyu Lan, Yang Shi, Yueyao Zhang, and Nan Cao.
\newblock Smile or scowl? looking at infographic design through the affective
  lens.
\newblock {\em IEEE Transactions on Visualization and Computer Graphics},
  27(6):2796--2807, 2021.

\bibitem{diakopoulos2011playable}
Nicholas Diakopoulos, Funda Kivran-Swaine, and Mor Naaman.
\newblock Playable data: characterizing the design space of game-y
  infographics.
\newblock In {\em Proceedings of the SIGCHI Conference on Human Factors in
  Computing Systems}, pages 1717--1726, 2011.

\bibitem{dunlap2016getting}
Joanna~C Dunlap and Patrick~R Lowenthal.
\newblock Getting graphic about infographics: design lessons learned from
  popular infographics.
\newblock {\em Journal of Visual Literacy}, 35(1):42--59, 2016.

\bibitem{kim2016data}
Nam~Wook Kim, Eston Schweickart, Zhicheng Liu, Mira Dontcheva, Wilmot Li, Jovan
  Popovic, and Hanspeter Pfister.
\newblock Data-driven guides: Supporting expressive design for information
  graphics.
\newblock {\em IEEE Transactions on Visualization and Computer Graphics},
  23(1):491--500, 2016.

\bibitem{cui2021mixed}
Weiwei Cui, Jinpeng Wang, He~Huang, Yun Wang, Chin-Yew Lin, Haidong Zhang, and
  Dongmei Zhang.
\newblock A mixed-initiative approach to reusing infographie charts.
\newblock {\em IEEE Transactions on Visualization and Computer Graphics},
  28(1):173--183, 2021.

\bibitem{wang2018infonice}
Yun Wang, Haidong Zhang, He~Huang, Xi~Chen, Qiufeng Yin, Zhitao Hou, Dongmei
  Zhang, Qiong Luo, and Huamin Qu.
\newblock Infonice: Easy creation of information graphics.
\newblock In {\em Proceedings of the 2018 CHI Conference on Human Factors in
  Computing Systems}, pages 1--12, 2018.

\bibitem{zhang2020dataquilt}
Jiayi~Eris Zhang, Nicole Sultanum, Anastasia Bezerianos, and Fanny Chevalier.
\newblock Dataquilt: Extracting visual elements from images to craft pictorial
  visualizations.
\newblock In {\em Proceedings of the 2020 CHI Conference on Human Factors in
  Computing Systems}, pages 1--13, 2020.

\bibitem{coelho2020infomages}
Darius Coelho and Klaus Mueller.
\newblock Infomages: Embedding data into thematic images.
\newblock {\em Computer Graphics Forum}, 39(3):593--606, 2020.

\bibitem{kim2019dataselfie}
Nam~Wook Kim, Hyejin Im, Nathalie Henry~Riche, Alicia Wang, Krzysztof Gajos,
  and Hanspeter Pfister.
\newblock Dataselfie: Empowering people to design personalized visuals to
  represent their data.
\newblock In {\em Proceedings of the 2019 CHI Conference on Human Factors in
  Computing Systems}, pages 1--12, 2019.

\bibitem{xia2018dataink}
Haijun Xia, Nathalie Henry~Riche, Fanny Chevalier, Bruno De~Araujo, and Daniel
  Wigdor.
\newblock Dataink: Direct and creative data-oriented drawing.
\newblock In {\em Proceedings of the 2018 CHI Conference on Human Factors in
  Computing Systems}, pages 1--13, 2018.

\bibitem{lee2013sketchstory}
Bongshin Lee, Rubaiat~Habib Kazi, and Greg Smith.
\newblock Sketchstory: Telling more engaging stories with data through freeform
  sketching.
\newblock {\em IEEE Transactions on Visualization and Computer Graphics},
  19(12):2416--2425, 2013.

\bibitem{lu2020exploring}
Min Lu, Chufeng Wang, Joel Lanir, Nanxuan Zhao, Hanspeter Pfister, Daniel
  Cohen-Or, and Hui Huang.
\newblock Exploring visual information flows in infographics.
\newblock In {\em Proceedings of the 2020 CHI Conference on Human Factors in
  Computing Systems}, pages 1--12, 2020.

\bibitem{tyagi2021user}
Anjul Tyagi, Jian Zhao, Pushkar Patel, Swasti Khurana, and Klaus Mueller.
\newblock User-centric semi-automated infographics authoring and
  recommendation.
\newblock {\em arXiv preprint arXiv:2108.11914}, 2021.

\bibitem{yuan2021infocolorizer}
Linping Yuan, Ziqi Zhou, Jian Zhao, Yiqiu Guo, Fan Du, and Huamin Qu.
\newblock Infocolorizer: Interactive recommendation of color palettes for
  infographics.
\newblock {\em IEEE Transactions on Visualization and Computer Graphics}, pages
  1--16, 2021.

\bibitem{cui2019text}
Weiwei Cui, Xiaoyu Zhang, Yun Wang, He~Huang, Bei Chen, Lei Fang, Haidong
  Zhang, Jian-Guan Lou, and Dongmei Zhang.
\newblock Text-to-viz: Automatic generation of infographics from
  proportion-related natural language statements.
\newblock {\em IEEE Transactions on Visualization and Computer Graphics},
  26(1):906--916, 2019.

\bibitem{qian2020retrieve}
Chunyao Qian, Shizhao Sun, Weiwei Cui, Jian-Guang Lou, Haidong Zhang, and
  Dongmei Zhang.
\newblock Retrieve-then-adapt: Example-based automatic generation for
  proportion-related infographics.
\newblock {\em IEEE Transactions on Visualization and Computer Graphics},
  27(2):443--452, 2020.

\bibitem{chen2019towards}
Zhutian Chen, Yun Wang, Qianwen Wang, Yong Wang, and Huamin Qu.
\newblock Towards automated infographic design: Deep learning-based
  auto-extraction of extensible timeline.
\newblock {\em IEEE Transactions on Visualization and Computer Graphics},
  26(1):917--926, 2019.

\bibitem{brehmer2016timelines}
Matthew Brehmer, Bongshin Lee, Benjamin Bach, Nathalie~Henry Riche, and Tamara
  Munzner.
\newblock Timelines revisited: A design space and considerations for expressive
  storytelling.
\newblock {\em IEEE Transactions on Visualization and Computer Graphics},
  23(9):2151--2164, 2016.

\bibitem{lan2021understanding}
Xingyu Lan, Xinyue Xu, and Nan Cao.
\newblock Understanding narrative linearity for telling expressive
  time-oriented stories.
\newblock In {\em Proceedings of the 2021 CHI Conference on Human Factors in
  Computing Systems}, pages 1--13, 2021.

\bibitem{bach2015time}
Benjamin Bach, Conglei Shi, Nicolas Heulot, Tara Madhyastha, Tom Grabowski, and
  Pierre Dragicevic.
\newblock Time curves: Folding time to visualize patterns of temporal evolution
  in data.
\newblock {\em IEEE Transactions on Visualization and Computer Graphics},
  22(1):559--568, 2015.

\bibitem{di2020storyline}
Emilio Di~Giacomo, Walter Didimo, Giuseppe Liotta, Fabrizio Montecchiani, and
  Alessandra Tappini.
\newblock Storyline visualizations with ubiquitous actors.
\newblock In {\em Graph Drawing}, pages 324--332, 2020.

\bibitem{tanahashi2012design}
Yuzuru Tanahashi and Kwan-Liu Ma.
\newblock Design considerations for optimizing storyline visualizations.
\newblock {\em IEEE Transactions on Visualization and Computer Graphics},
  18(12):2679--2688, 2012.

\bibitem{gronemann2016crossing}
Martin Gronemann, Michael J{\"u}nger, Frauke Liers, and Francesco Mambelli.
\newblock Crossing minimization in storyline visualization.
\newblock In {\em International Symposium on Graph Drawing and Network
  Visualization}, pages 367--381. Springer, 2016.

\bibitem{kim2017visualizing}
Nam~Wook Kim, Benjamin Bach, Hyejin Im, Sasha Schriber, Markus Gross, and
  Hanspeter Pfister.
\newblock Visualizing nonlinear narratives with story curves.
\newblock {\em IEEE Transactions on Visualization and Computer Graphics},
  24(1):595--604, 2017.

\bibitem{brehmer2019timeline}
Matthew Brehmer, Bongshin Lee, Nathalie~Henry Riche, David Tittsworth, Kate
  Lytvynets, Darren Edge, and Christopher~M. White.
\newblock Timeline storyteller: The design \& deployment of an interactive
  authoring tool for expressive timeline narratives.
\newblock In {\em Computation+Journalism Symposium}, pages 1--5, February 2019.

\bibitem{nguyen2016timesets}
Phong~H Nguyen, Kai Xu, Rick Walker, and BL~William Wong.
\newblock Timesets: Timeline visualization with set relations.
\newblock {\em Information Visualization}, 15(3):253--269, 2016.

\bibitem{liu2013storyflow}
Shixia Liu, Yingcai Wu, Enxun Wei, Mengchen Liu, and Yang Liu.
\newblock Storyflow: Tracking the evolution of stories.
\newblock {\em IEEE Transactions on Visualization and Computer Graphics},
  19(12):2436--2445, 2013.

\bibitem{tang2018istoryline}
Tan Tang, Sadia Rubab, Jiewen Lai, Weiwei Cui, Lingyun Yu, and Yingcai Wu.
\newblock istoryline: Effective convergence to hand-drawn storylines.
\newblock {\em IEEE Transactions on Visualization and Computer Graphics},
  25(1):769--778, 2018.

\bibitem{tang2020plotthread}
Tan Tang, Renzhong Li, Xinke Wu, Shuhan Liu, Johannes Knittel, Steffen Koch,
  Lingyun Yu, Peiran Ren, Thomas Ertl, and Yingcai Wu.
\newblock Plotthread: Creating expressive storyline visualizations using
  reinforcement learning.
\newblock {\em IEEE Transactions on Visualization and Computer Graphics},
  27(2):294--303, 2020.

\bibitem{satyanarayan2014authoring}
Arvind Satyanarayan and Jeffrey Heer.
\newblock Authoring narrative visualizations with ellipsis.
\newblock {\em Computer Graphics Forum}, 33(3):361--370, 2014.

\bibitem{fulda2015timelinecurator}
Johanna Fulda, Matthew Brehmer, and Tamara Munzner.
\newblock Timelinecurator: Interactive authoring of visual timelines from
  unstructured text.
\newblock {\em IEEE Transactions on Visualization and Computer Graphics},
  22(1):300--309, 2015.

\bibitem{zhao2015data}
Zhenpeng Zhao, Rachael Marr, and Niklas Elmqvist.
\newblock Data comics: Sequential art for data-driven storytelling.
\newblock {\em Technical Report, Univ. of Maryland}, 2015.

\bibitem{wang2019teaching}
Zezhong Wang, Harvey Dingwall, and Benjamin Bach.
\newblock Teaching data visualization and storytelling with data comic
  workshops.
\newblock In {\em Extended Abstracts of the 2019 CHI Conference on Human
  Factors in Computing Systems}, pages 1--9, 2019.

\bibitem{bach2017emerging}
Benjamin Bach, Nathalie~Henry Riche, Sheelagh Carpendale, and Hanspeter
  Pfister.
\newblock The emerging genre of data comics.
\newblock {\em IEEE Computer Graphics and Applications}, 37(3):6--13, 2017.

\bibitem{wang2019comparing}
Zezhong Wang, Shunming Wang, Matteo Farinella, Dave Murray-Rust, Nathalie
  Henry~Riche, and Benjamin Bach.
\newblock Comparing effectiveness and engagement of data comics and
  infographics.
\newblock In {\em Proceedings of the 2019 CHI Conference on Human Factors in
  Computing Systems}, pages 1--12, 2019.

\bibitem{bach2018design}
Benjamin Bach, Zezhong Wang, Matteo Farinella, Dave Murray-Rust, and Nathalie
  Henry~Riche.
\newblock Design patterns for data comics.
\newblock In {\em Proceedings of the 2018 CHI conference on human factors in
  computing systems}, pages 1--12, 2018.

\bibitem{wang2020data}
Zezhong Wang, Jacob Ritchie, Jingtao Zhou, Fanny Chevalier, and Benjamin Bach.
\newblock Data comics for reporting controlled user studies in human-computer
  interaction.
\newblock {\em IEEE Transactions on Visualization and Computer Graphics},
  27(2):967--977, 2020.

\bibitem{zhao2019understanding}
Zhenpeng Zhao, Rachael Marr, Jason Shaffer, and Niklas Elmqvist.
\newblock Understanding partitioning and sequence in data-driven storytelling.
\newblock In {\em International Conference on Information}, pages 327--338.
  Springer, 2019.

\bibitem{hasan2022playing}
Md~Tanvir Hasan, Annika Wolff, Antti Knutas, Anne P{\"a}ssil{\"a}, and Lasse
  Kantola.
\newblock Playing games through interactive data comics to explore water
  quality in a lake: A case study exploring the use of a data-driven
  storytelling method in co-design.
\newblock In {\em CHI Conference on Human Factors in Computing Systems Extended
  Abstracts}, pages 1--7, 2022.

\bibitem{kim2019datatoon}
Nam~Wook Kim, Nathalie Henry~Riche, Benjamin Bach, Guanpeng Xu, Matthew
  Brehmer, Ken Hinckley, Michel Pahud, Haijun Xia, Michael~J McGuffin, and
  Hanspeter Pfister.
\newblock Datatoon: Drawing dynamic network comics with pen+ touch interaction.
\newblock In {\em Proceedings of the 2019 CHI Conference on Human Factors in
  Computing Systems}, pages 1--12, 2019.

\bibitem{kang2021toonnote}
DaYe Kang, Tony Ho, Nicolai Marquardt, Bilge Mutlu, and Andrea Bianchi.
\newblock Toonnote: Improving communication in computational notebooks using
  interactive data comics.
\newblock In {\em Proceedings of the 2021 CHI Conference on Human Factors in
  Computing Systems}, pages 1--14, 2021.

\bibitem{wang2021interactive}
Z.~Wang, H.~Romat, F.~Chevalier, N.~H. Riche, and B.~Bach.
\newblock Interactive data comics.
\newblock In {\em VIS 2021}, 2021.

\bibitem{suh2022codetoon}
Sangho Suh, Jian Zhao, and Edith Law.
\newblock Codetoon: Story ideation, auto comic generation, and structure
  mapping for code-driven storytelling.
\newblock {\em arXiv preprint arXiv:2208.12981}, 2022.

\bibitem{zhao2021chartstory}
Jian Zhao, Shenyu Xu, Senthil Chandrasegaran, Chris Bryan, Fan Du, Aditi
  Mishra, Xin Qian, Yiran Li, and Kwan-Liu Ma.
\newblock Chartstory: Automated partitioning, layout, and captioning of charts
  into comic-style narratives.
\newblock {\em arXiv preprint arXiv:2103.03996}, 2021.

\bibitem{wang2019datashot}
Yun Wang, Zhida Sun, Haidong Zhang, Weiwei Cui, Ke~Xu, Xiaojuan Ma, and Dongmei
  Zhang.
\newblock Datashot: Automatic generation of fact sheets from tabular data.
\newblock {\em IEEE Transactions on Visualization and Computer Graphics},
  26(1):895--905, 2019.

\bibitem{shi2020calliope}
Danqing Shi, Xinyue Xu, Fuling Sun, Yang Shi, and Nan Cao.
\newblock Calliope: Automatic visual data story generation from a spreadsheet.
\newblock {\em IEEE Transactions on Visualization and Computer Graphics},
  27(2):453--463, 2020.

\bibitem{seyser2018scrollytelling}
Doris Seyser and Michael Zeiller.
\newblock Scrollytelling--an analysis of visual storytelling in online
  journalism.
\newblock In {\em 2018 22nd International Conference Information Visualisation
  (IV)}, pages 401--406. IEEE, 2018.

\bibitem{godulla2017digitale}
Alexander Godulla and Cornelia Wolf.
\newblock {\em Digitale Langformen im Journalismus und Corporate Publishing}.
\newblock Springer, 2017.

\bibitem{elias2018towards}
Mirette Elias, Abi James, Steffen Lohmann, S{\"o}ren Auer, and Mike Wald.
\newblock Towards an open authoring tool for accessible slide presentations.
\newblock In {\em International Conference on Computers Helping People with
  Special Needs}, pages 172--180. Springer, 2018.

\bibitem{hullman2013deeper}
Jessica Hullman, Steven Drucker, Nathalie~Henry Riche, Bongshin Lee, Danyel
  Fisher, and Eytan Adar.
\newblock A deeper understanding of sequence in narrative visualization.
\newblock {\em IEEE Transactions on visualization and computer graphics},
  19(12):2406--2415, 2013.

\bibitem{roth2021cartographic}
Robert~E Roth.
\newblock Cartographic design as visual storytelling: Synthesis and review of
  map-based narratives, genres, and tropes.
\newblock {\em The Cartographic Journal}, 58(1):83--114, 2021.

\bibitem{conlen2018idyll}
Matthew Conlen and Jeffrey Heer.
\newblock Idyll: A markup language for authoring and publishing interactive
  articles on the web.
\newblock In {\em Proceedings of the 31st Annual ACM Symposium on User
  Interface Software and Technology}, pages 977--989, 2018.

\bibitem{sultanum2021leveraging}
Nicole Sultanum, Fanny Chevalier, Zoya Bylinskii, and Zhicheng Liu.
\newblock Leveraging text-chart links to support authoring of data-driven
  articles with vizflow.
\newblock In {\em Proceedings of the 2021 CHI Conference on Human Factors in
  Computing Systems}, pages 1--17, 2021.

\bibitem{winters2019automatically}
Thomas Winters and Kory~W Mathewson.
\newblock Automatically generating engaging presentation slide decks.
\newblock In {\em International Conference on Computational Intelligence in
  Music, Sound, Art and Design (Part of EvoStar)}, pages 127--141. Springer,
  2019.

\bibitem{lu2021automatic}
Junhua Lu, Wei Chen, Hui Ye, Jie Wang, Honghui Mei, Yuhui Gu, Yingcai Wu,
  Xiaolong~Luke Zhang, and Kwan-Liu Ma.
\newblock Automatic generation of unit visualization-based scrollytelling for
  impromptu data facts delivery.
\newblock In {\em 2021 IEEE 14th Pacific Visualization Symposium (PacificVis)},
  pages 21--30. IEEE, 2021.

\bibitem{amini2015understanding}
Fereshteh Amini, Nathalie Henry~Riche, Bongshin Lee, Christophe Hurter, and
  Pourang Irani.
\newblock Understanding data videos: Looking at narrative visualization through
  the cinematography lens.
\newblock In {\em Proceedings of the 33rd Annual ACM Conference on Human
  Factors in Computing Systems}, pages 1459--1468, 2015.

\bibitem{cao2020examining}
Ruochen Cao, Subrata Dey, Andrew Cunningham, James Walsh, Ross~T Smith,
  Joanne~E Zucco, and Bruce~H Thomas.
\newblock Examining the use of narrative constructs in data videos.
\newblock {\em Visual Informatics}, 4(1):8--22, 2020.

\bibitem{xu2022wow}
Xian Xu, Leni Yang, David Yip, Mingming Fan, Zheng Wei, and Huamin Qu.
\newblock From ‘wow’to ‘why’: Guidelines for creating the opening of a
  data video with cinematic styles.
\newblock In {\em CHI Conference on Human Factors in Computing Systems}, pages
  1--20, 2022.

\bibitem{thompson2020understanding}
John Thompson, Zhicheng Liu, Wilmot Li, and John Stasko.
\newblock Understanding the design space and authoring paradigms for animated
  data graphics.
\newblock {\em Computer Graphics Forum}, 39(3):207--218, 2020.

\bibitem{sallam2022towards}
Samar Sallam, Yumiko Sakamoto, Jason Leboe-McGowan, Celine Latulipe, and
  Pourang Irani.
\newblock Towards design guidelines for effective health-related data videos:
  An empirical investigation of affect, personality, and video content.
\newblock In {\em CHI Conference on Human Factors in Computing Systems}, pages
  1--22, 2022.

\bibitem{wang2016animated}
Yun Wang, Zhutian Chen, Quan Li, Xiaojuan Ma, Qiong Luo, and Huamin Qu.
\newblock Animated narrative visualization for video clickstream data.
\newblock In {\em SIGGRAPH Asia 2016 Symposium on Visualization}, pages 1--8.
  ACM, 2016.

\bibitem{li2020improving}
Wenchao Li, Yun Wang, Haidong Zhang, and Huamin Qu.
\newblock Improving engagement of animated visualization with visual
  foreshadowing.
\newblock In {\em 2020 IEEE Visualization Conference (VIS)}, pages 141--145.
  IEEE, 2020.

\bibitem{shu2020makes}
Xinhuan Shu, Aoyu Wu, Junxiu Tang, Benjamin Bach, Yingcai Wu, and Huamin Qu.
\newblock What makes a data-gif understandable?
\newblock {\em IEEE Transactions on Visualization and Computer Graphics},
  27(2):1492--1502, 2020.

\bibitem{shi2021communicating}
Yang Shi, Xingyu Lan, Jingwen Li, Zhaorui Li, and Nan Cao.
\newblock Communicating with motion: A design space for animated visual
  narratives in data videos.
\newblock In {\em Proceedings of the 2021 CHI Conference on Human Factors in
  Computing Systems}, pages 1--13, 2021.

\bibitem{tang2020design}
Tan Tang, Junxiu Tang, Jiayi Hong, Lingyun Yu, Peiran Ren, and Yingcai Wu.
\newblock Design guidelines for augmenting short-form videos using animated
  data visualizations.
\newblock {\em Journal of Visualization}, 23(4):707--720, 2020.

\bibitem{amini2016authoring}
Fereshteh Amini, Nathalie~Henry Riche, Bongshin Lee, Andres Monroy-Hernandez,
  and Pourang Irani.
\newblock Authoring data-driven videos with dataclips.
\newblock {\em IEEE Transactions on Visualization and Computer Graphics},
  23(1):501--510, 2016.

\bibitem{lan2021kineticharts}
Xingyu Lan, Yang Shi, Yanqiu Wu, Xiaohan Jiao, and Nan Cao.
\newblock Kineticharts: Augmenting affective expressiveness of charts in data
  stories with animation design.
\newblock {\em IEEE Transactions on Visualization and Computer Graphics},
  28(1):933--943, 2021.

\bibitem{chen2021augmenting}
Zhutian Chen, Shuainan Ye, Xiangtong Chu, Haijun Xia, Hui Zhang, Huamin Qu, and
  Yingcai Wu.
\newblock Augmenting sports videos with viscommentator.
\newblock {\em IEEE Transactions on Visualization and Computer Graphics}, 2021.

\bibitem{thompson2021data}
John~R Thompson, Zhicheng Liu, and John Stasko.
\newblock Data animator: Authoring expressive animated data graphics.
\newblock In {\em Proceedings of the 2021 CHI Conference on Human Factors in
  Computing Systems}, pages 1--18, 2021.

\bibitem{wang2021animated}
Yun Wang, Yi~Gao, Ray Huang, Weiwei Cui, Haidong Zhang, and Dongmei Zhang.
\newblock Animated presentation of static infographics with infomotion.
\newblock In {\em Computer Graphics Forum}, volume~40, pages 507--518. Wiley
  Online Library, 2021.

\bibitem{kim2021gemini}
Younghoon Kim and Jeffrey Heer.
\newblock Gemini 2: Generating keyframe-oriented animated transitions between
  statistical graphics.
\newblock In {\em 2021 IEEE Visualization Conference (VIS)}, pages 201--205.
  IEEE, 2021.

\bibitem{ge2021cast}
Tong Ge, Bongshin Lee, and Yunhai Wang.
\newblock Cast: Authoring data-driven chart animations.
\newblock In {\em Proceedings of the 2021 CHI Conference on Human Factors in
  Computing Systems}, pages 1--15, 2021.

\bibitem{shi2021autoclips}
Danqing Shi, Fuling Sun, Xinyue Xu, Xingyu Lan, David Gotz, and Nan Cao.
\newblock Autoclips: An automatic approach to video generation from data facts.
\newblock {\em Computer Graphics Forum}, 40(3):495--505, 2021.

\bibitem{segel2010narrative}
Edward Segel and Jeffrey Heer.
\newblock Narrative visualization: Telling stories with data.
\newblock {\em IEEE Transactions on Visualization and Computer Graphics},
  16(6):1139--1148, 2010.

\bibitem{tong2018storytelling}
Chao Tong, Richard Roberts, Rita Borgo, Sean Walton, Robert~S Laramee, Kodzo
  Wegba, Aidong Lu, Yun Wang, Huamin Qu, Qiong Luo, and Ma~Xiaojuan.
\newblock Storytelling and visualization: An extended survey.
\newblock {\em Information}, 9(3):65, 2018.

\bibitem{botero2010expanding}
Andrea Botero, Kari-Hans Kommonen, and Sanna Marttila.
\newblock Expanding design space: Design-in-use activities and strategies.
\newblock In {\em Aalto University publication series Doctoral Dissertations},
  2010.

\bibitem{fischer2006meta}
Gerhard Fischer and Elisa Giaccardi.
\newblock Meta-design: A framework for the future of end-user development.
\newblock In {\em End user development}, pages 427--457. Springer, 2006.

\bibitem{westerlund2005design}
Bo~Westerlund.
\newblock Design space conceptual tool--grasping the design process.
\newblock In {\em Nordes}, 2005.

\bibitem{schulz2010explorative}
Hans-J{\"o}rg Schulz.
\newblock {\em Explorative graph visualization}.
\newblock PhD thesis, University of Rostock, 2010.

\bibitem{schulz2010design}
Hans-Jorg Schulz, Steffen Hadlak, and Heidrun Schumann.
\newblock The design space of implicit hierarchy visualization: A survey.
\newblock {\em IEEE Transactions on Visualization and Computer Graphics},
  17(4):393--411, 2010.

\bibitem{zikas2020immersive}
Paul Zikas, George Papagiannakis, Nick Lydatakis, Steve Kateros, Stavroula
  Ntoa, Ilia Adami, and Constantine Stephanidis.
\newblock Immersive visual scripting based on vr software design patterns for
  experiential training.
\newblock {\em The Visual Computer}, 36(10):1965--1977, 2020.

\bibitem{brath2018automated}
Richard Brath and Martin Matusiak.
\newblock Automated annotations.
\newblock In {\em An IEEE VIS Workshop on Visualization for Communication
  (VisComm)}, 2018.

\bibitem{marshall1997annotation}
Catherine~C Marshall.
\newblock Annotation: from paper books to the digital library.
\newblock In {\em Proceedings of the second ACM International Conference on
  Digital Libraries}, pages 131--140, 1997.

\bibitem{gomez2017ggplot2}
Virgilio G{\'o}mez-Rubio.
\newblock ggplot2-elegant graphics for data analysis.
\newblock {\em Journal of Statistical Software}, 77:1--3, 2017.

\bibitem{bostock2011d3}
Michael Bostock, Vadim Ogievetsky, and Jeffrey Heer.
\newblock D$^3$ data-driven documents.
\newblock {\em IEEE Transactions on Visualization and Computer Graphics},
  17(12):2301--2309, 2011.

\bibitem{tableau2006}
Tableau.
\newblock \url{https://www.tableau.com/}, 2006.
\newblock Accessed: 2022-02-14.

\bibitem{lee2021viral}
Crystal Lee, Tanya Yang, Gabrielle~D Inchoco, Graham~M Jones, and Arvind
  Satyanarayan.
\newblock Viral visualizations: How coronavirus skeptics use orthodox data
  practices to promote unorthodox science online.
\newblock In {\em Proceedings of the 2021 CHI conference on human factors in
  computing systems}, pages 1--18, 2021.

\bibitem{shrinivasan2009connecting}
Yedendra~B Shrinivasan, David Gotzy, and Jie Lu.
\newblock Connecting the dots in visual analysis.
\newblock In {\em 2009 IEEE Symposium on Visual Analytics Science and
  Technology}, pages 123--130. IEEE, 2009.

\bibitem{kittivorawong2020fast}
Chanwut Kittivorawong, Dominik Moritz, Kanit Wongsuphasawat, and Jeffrey Heer.
\newblock Fast and flexible overlap detection for chart labeling with occupancy
  bitmap.
\newblock In {\em 2020 IEEE Visualization Conference (VIS)}, pages 101--105.
  IEEE, 2020.

\bibitem{otten2015infographics}
Jennifer~J Otten, Karen Cheng, and Adam Drewnowski.
\newblock Infographics and public policy: using data visualization to convey
  complex information.
\newblock {\em Health Affairs}, 34(11):1901--1907, 2015.

\bibitem{naparin2017infographics}
Husni Naparin and A~Binti Saad.
\newblock Infographics in education: Review on infographics design.
\newblock {\em The International Journal of Multimedia \& Its Applications
  (IJMA)}, 9(4):5, 2017.

\bibitem{albers2014infographics}
Michael~J Albers.
\newblock Infographics: Horrid chartjunk or quality communication.
\newblock In {\em 2014 IEEE International Professional Communication Conference
  (IPCC)}, pages 1--4. IEEE, 2014.

\bibitem{adobeIllustrator}
Adobe~Systems Incorporated.
\newblock Adobe illustrator.
\newblock \url{https://www.adobe.com/products/illustrator.html}, 2023.
\newblock Online; Accessed: 2023-02-14.

\bibitem{sketch2010}
Bohemian Coding.
\newblock {Sketch - Professional Digital Design for Mac}.
\newblock \url{https://www.sketch.com/}, 2010.
\newblock Online; Accessed: 2023-02-14.

\bibitem{visme2013}
Visme.
\newblock \url{https://www.visme.co/make-infographics/}, 2013.
\newblock Accessed: 2022-01-07.

\bibitem{infogram2012}
Infogram.
\newblock \url{https://infogram.com/}, 2012.
\newblock Accessed: 2022-01-07.

\bibitem{CANVA2018}
Canva.
\newblock \url{https://www.canva.cn/create/}, 2018.
\newblock Accessed: 2022-01-07.

\bibitem{webalon2011tiki}
Webalon.
\newblock Tiki-toki.
\newblock \url{http://tiki-toki.com/}, 2011.
\newblock Accessed: 2023-02-14.

\bibitem{dukes2010dipity}
D~Dukes and BJ~Heinley.
\newblock Dipity.
\newblock \url{https://www.timetoast.com/timelines/dipity-online-timeline},
  2010.
\newblock Accessed: 2023-02-14.

\bibitem{TimelineJS2013}
Northwestern University~Knight Lab.
\newblock Timelinejs.
\newblock \url{http://timeline.knightlab.com/}, 2013.
\newblock Accessed: 2023-02-14.

\bibitem{TimelineSetter2011}
Al~Shaw, Jeff Larson, and Ben Welsh.
\newblock Timelinesetter.
\newblock \url{http:// propublica.github.io/timeline-setter/}, 2011.
\newblock Online; Accessed: 2023-02-14.

\bibitem{genette1983narrative}
G{\'e}rard Genette.
\newblock {\em Narrative discourse: An essay in method}, volume~3.
\newblock Cornell University Press, 1983.

\bibitem{kashan2012Timeline}
O~Kashan.
\newblock Timeline of the universe.
\newblock
  \url{https://www.informationisbeautifulawards.com/showcase/456-timeline-of-the-universe},
  2012.
\newblock Accessed: 2022-09-12.

\bibitem{powerpoint2016}
Microsoft.
\newblock Powerpoint.
\newblock \url{https://office.live.com/start/powerpoint.aspx}, 2016.
\newblock Accessed: 2022-02-14.

\bibitem{brand2019medical}
Anna Brand, Linde Gao, Alexandra Hamann, Claudia Crayen, Hannah Brand, Susan~M
  Squier, Karl Stangl, Friederike Kendel, and Verena Stangl.
\newblock Medical graphic narratives to improve patient comprehension and
  periprocedural anxiety before coronary angiography and percutaneous coronary
  intervention: a randomized trial.
\newblock {\em Annals of Internal Medicine}, 170(8):579--581, 2019.

\bibitem{mckenna2014design}
Sean McKenna, Dominika Mazur, James Agutter, and Miriah Meyer.
\newblock Design activity framework for visualization design.
\newblock {\em IEEE Transactions on Visualization and Computer Graphics},
  20(12):2191--2200, 2014.

\bibitem{mckenna2017visual}
Sean McKenna, Nathalie Henry~Riche, Bongshin Lee, Jeremy Boy, and Miriah Meyer.
\newblock Visual narrative flow: Exploring factors shaping data visualization
  story reading experiences.
\newblock {\em Computer Graphics Forum}, 36(3):377--387, 2017.

\bibitem{keynote2003}
Apple.
\newblock Keynote.
\newblock \url{https://www.apple.com/keynote/}, 2003.
\newblock Accessed: 2022-02-14.

\bibitem{googleslides2006}
Google.
\newblock Google slides.
\newblock \url{https://www.google.com/slides/about/}, 2006.
\newblock Accessed: 2022-02-14.

\bibitem{bocklandt2021sandslide}
Sieben Bocklandt, Gust Verbruggen, and Thomas Winters.
\newblock Sandslide: Automatic slideshow normalization.
\newblock In {\em International Conference on Document Analysis and
  Recognition}, pages 445--461. Springer, 2021.

\bibitem{leake2020generating}
Mackenzie Leake, Hijung~Valentina Shin, Joy~O Kim, and Maneesh Agrawala.
\newblock Generating audio-visual slideshows from text articles using word
  concreteness.
\newblock In {\em Proceedings of the 2020 CHI Conference on Human Factors in
  Computing Systems}, pages 1--11, 2020.

\bibitem{liu2018data}
Zhicheng Liu, John Thompson, Alan Wilson, Mira Dontcheva, James Delorey, Sam
  Grigg, Bernard Kerr, and John Stasko.
\newblock Data illustrator: Augmenting vector design tools with lazy data
  binding for expressive visualization authoring.
\newblock In {\em Proceedings of the 2018 CHI Conference on Human Factors in
  Computing Systems}, pages 1--13, 2018.

\bibitem{suprata2019data}
Ferdian Suprata.
\newblock Data storytelling with dashboard: Accelerating understanding through
  data visualization in financial technology company case study.
\newblock {\em Jurnal Metris}, 20(01):1--10, 2019.

\bibitem{sedlmair2012design}
Michael Sedlmair, Miriah Meyer, and Tamara Munzner.
\newblock Design study methodology: Reflections from the trenches and the
  stacks.
\newblock {\em IEEE Transactions on Visualization and Computer Graphics},
  18(12):2431--2440, 2012.

\bibitem{oppermann2020data}
Michael Oppermann and Tamara Munzner.
\newblock Data-first visualization design studies.
\newblock In {\em 2020 IEEE Workshop on Evaluation and Beyond-Methodological
  Approaches to Visualization (BELIV)}, pages 74--80. IEEE, 2020.

\bibitem{fernandez2022beyond}
Gloria~Milena Fernandez~Nieto, Kirsty Kitto, Simon Buckingham~Shum, and Roberto
  Martinez-Maldonado.
\newblock Beyond the learning analytics dashboard: Alternative ways to
  communicate student data insights combining visualisation, narrative and
  storytelling.
\newblock In {\em LAK22: 12th International Learning Analytics and Knowledge
  Conference}, pages 219--229, 2022.

\bibitem{isenberg2018immersive}
Petra Isenberg, Bongshin Lee, Huamin Qu, and Maxime Cordeil.
\newblock Immersive visual data stories.
\newblock In {\em Immersive Analytics}, pages 165--184. Springer, 2018.

\bibitem{karyda2020narrative}
Maria Karyda, Danielle Wilde, and Mette~Gislev Kj{\ae}rsgaard.
\newblock Narrative physicalization: supporting interactive engagement with
  personal data.
\newblock {\em IEEE Computer Graphics and Applications}, 41(1):74--86, 2020.

\bibitem{bach2017hologram}
Benjamin Bach, Ronell Sicat, Johanna Beyer, Maxime Cordeil, and Hanspeter
  Pfister.
\newblock The hologram in my hand: How effective is interactive exploration of
  3d visualizations in immersive tangible augmented reality?
\newblock {\em IEEE Transactions on Visualization and Computer Graphics},
  24(1):457--467, 2017.

\bibitem{butscher2018clusters}
Simon Butscher, Sebastian Hubenschmid, Jens M{\"u}ller, Johannes Fuchs, and
  Harald Reiterer.
\newblock Clusters, trends, and outliers: How immersive technologies can
  facilitate the collaborative analysis of multidimensional data.
\newblock In {\em Proceedings of the 2018 CHI Conference on Human Factors in
  Computing Systems}, pages 1--12, 2018.

\bibitem{cordeil2016immersive}
Maxime Cordeil, Tim Dwyer, Karsten Klein, Bireswar Laha, Kim Marriott, and
  Bruce~H Thomas.
\newblock Immersive collaborative analysis of network connectivity: Cave-style
  or head-mounted display?
\newblock {\em IEEE Transactions on Visualization and Computer Graphics},
  23(1):441--450, 2016.

\bibitem{hogan2017towards}
Trevor Hogan and Eva Hornecker.
\newblock Towards a design space for multisensory data representation.
\newblock {\em Interacting with Computers}, 29(2):147--167, 2017.

\bibitem{dragicevic2020data}
Pierre Dragicevic, Yvonne Jansen, and Andrew Vande~Moere.
\newblock Data physicalization.
\newblock {\em Handbook of Human Computer Interaction}, pages 1--51, 2020.

\bibitem{elias2012annotating}
Micheline Elias and Anastasia Bezerianos.
\newblock Annotating {BI} visualization dashboards: Needs \& challenges.
\newblock In {\em Proceedings of the SIGCHI Conference on Human Factors in
  Computing Systems}, pages 1641--1650, 2012.

\bibitem{zhang2020viscode}
Peiying Zhang, Chenhui Li, and Changbo Wang.
\newblock Viscode: Embedding information in visualization images using
  encoder-decoder network.
\newblock {\em IEEE Transactions on Visualization and Computer Graphics},
  27(2):326--336, 2020.

\bibitem{fu2020chartem}
Jiayun Fu, Bin Zhu, Weiwei Cui, Song Ge, Yun Wang, Haidong Zhang, He~Huang,
  Yuanyuan Tang, Dongmei Zhang, and Xiaojing Ma.
\newblock Chartem: Reviving chart images with data embedding.
\newblock {\em IEEE Transactions on Visualization and Computer Graphics},
  27(2):337--346, 2020.

\bibitem{dibia2019data2vis}
Victor Dibia and {\c{C}}a{\u{g}}atay Demiralp.
\newblock Data2vis: Automatic generation of data visualizations using
  sequence-to-sequence recurrent neural networks.
\newblock {\em IEEE Computer Graphics and Applications}, 39(5):33--46, 2019.

\bibitem{heer2007voyagers}
Jeffrey Heer, Fernanda~B Vi{\'e}gas, and Martin Wattenberg.
\newblock Voyagers and voyeurs: supporting asynchronous collaborative
  information visualization.
\newblock In {\em Proceedings of the SIGCHI Conference on Human Factors in
  Computing Systems}, pages 1029--1038, 2007.

\bibitem{vartak2015seedb}
Manasi Vartak, Sajjadur Rahman, Samuel Madden, Aditya~G. Parameswaran, and
  Neoklis Polyzotis.
\newblock Seedb: Efficient data-driven visualization recommendations to support
  visual analytics.
\newblock {\em Proceedings of the VLDB Endowment International Conference on
  Very Large Data Bases}, 8:2182 -- 2193, 2015.

\bibitem{luo2018deepeye}
Yuyu Luo, Xuedi Qin, Nan Tang, and Guoliang Li.
\newblock Deepeye: Towards automatic data visualization.
\newblock In {\em 2018 IEEE 34th International Conference on Data Engineering
  (ICDE)}, pages 101--112. IEEE, 2018.

\bibitem{ma2011scientific}
Kwan-Liu Ma, Isaac Liao, Jennifer Frazier, Helwig Hauser, and Helen-Nicole
  Kostis.
\newblock Scientific storytelling using visualization.
\newblock {\em IEEE Computer Graphics and Applications}, 32(1):12--19, 2011.

\end{thebibliography}
\newpage
\begin{IEEEbiography}[{\includegraphics[width=1in,height=1.25in,clip,keepaspectratio]{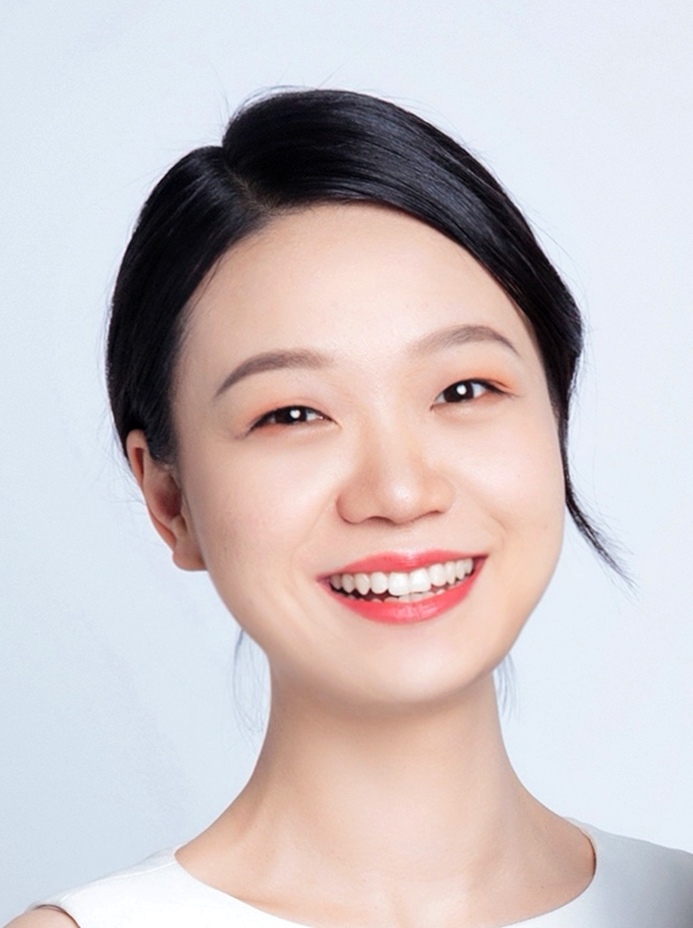}}]{Qing Chen} received her B.Eng degree from the Department of Computer Science, Zhejiang University and her Ph.D. degree from the Department of Computer Science and Engineering, Hong Kong University of Science and Technology (HKUST). After receiving her PhD degree, she worked as a postdoc at Inria and Ecole Polytechnique. She is currently an assistant professor at Tongji University. Her research interests include information visualization, visual analytics, human-computer interaction, online education, visual storytelling, intelligent healthcare and design. 
\end{IEEEbiography}
%\vspace{-4mm}
%\vskip 0pt plus -1fil

\begin{IEEEbiography}[{\includegraphics[width=1in,height=1.25in,clip,keepaspectratio]{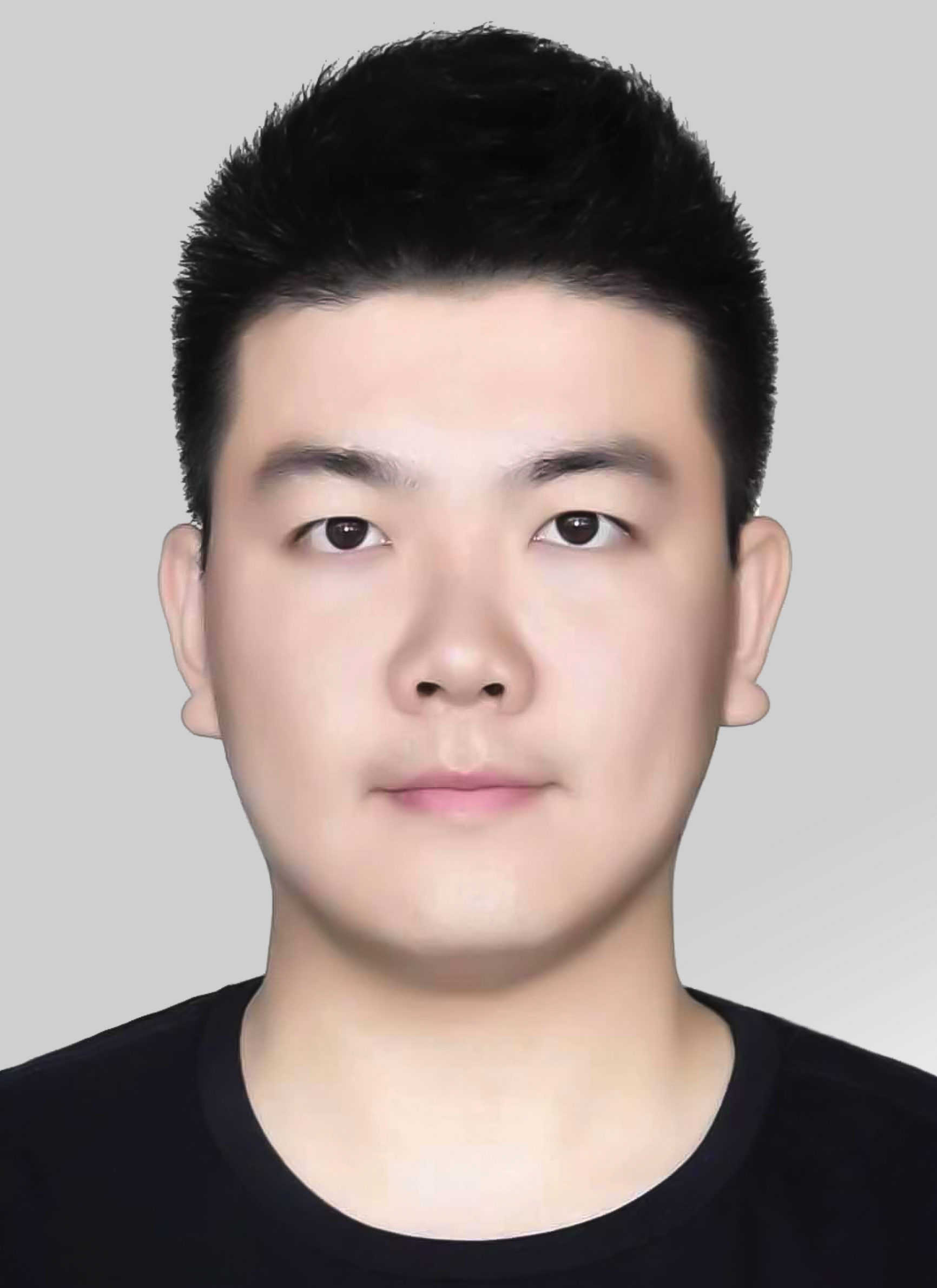}}]{Shixiong Cao} received his Master's degree in Design from Sangmyung University in South Korea in 2019, and subsequently obtained a Ph.D. degree from Sungkyunkwan University in South Korea in 2023. Currently, he works as a postdoctoral researcher at Tongji University, and his research interests include information design, narrative visualization design, and user experience design.

\end{IEEEbiography}
%\vspace{-4mm}
%\vskip 0pt plus -1fil

\begin{IEEEbiography}[{\includegraphics[width=1in,height=1.25in,clip,keepaspectratio]{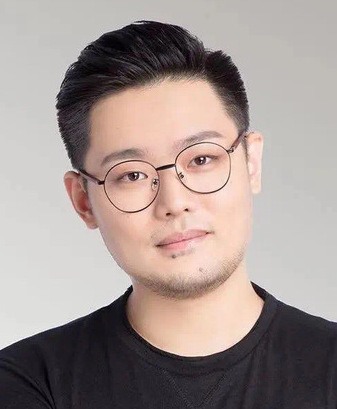}}]{Jiazhe Wang} received his Master's degree from the Department of Computer Science, University of Oxford. He is currently a data and front-end technologist in Ant Group, a core member of the data visualization team AntV. He is also a tech leader of the augmented analytics team for the internal BI product of Ant Group. His research interests include automated visualization, augmented analytics and narrative visualization.
\end{IEEEbiography}
%\vspace{-4mm}
%\vskip 0pt plus -1fil

\begin{IEEEbiography}[{\includegraphics[width=1in,height=1.25in,clip,keepaspectratio]{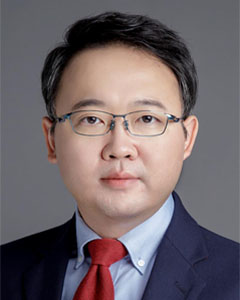}}]{Nan Cao} received his Ph.D. degree in Computer Science and Engineering from the Hong Kong University of Science and Technology (HKUST), Hong Kong, China in 2012. He is currently a professor at Tongji University and the Assistant Dean of the Tongji College of Design and Innovation. He also directs the Tongji Intelligent Big Data Visualization Lab (iDV$^x$ Lab) and conducts interdisciplinary research across multiple fields, including data visualization, human computer interaction, machine learning, and data mining. He was a research staff member at the IBM T.J. Watson Research Center, New York, NY, USA before joining the Tongji faculty in 2016.
\end{IEEEbiography}

\end{document}